\documentclass[aps,amsmath,amssymb,twocolumn,showpacs]{revtex4}
\usepackage{graphicx,color}
 \graphicspath{{./}{./figures/}}
\usepackage{dcolumn}% Align table columns on decimal point
\usepackage{bm}% bold math
\definecolor{labelkey}{cmyk}{.4,.2,0,0}

\newcommand{\be}{\begin{equation}}
\newcommand{\ee}{\end{equation}}
\newcommand{\bea}{\begin{eqnarray}}
\newcommand{\eea}{\end{eqnarray}}

\newcommand{\nn}{\nonumber }

\newcommand{\fig}[2]{\includegraphics[width=#1]{./figures/#2}}

\begin{document}

\title{Universality in the mean spatial shape of avalanches}
\author{Thimoth\'ee Thiery and Pierre Le Doussal} \affiliation{CNRS-Laboratoire
de Physique Th{\'e}orique de l'Ecole Normale Sup{\'e}rieure, 24 rue
Lhomond,75231 Cedex 05, Paris, France} 

\begin{abstract}

Quantifying the universality of avalanche observables beyond critical exponents
is of current great interest in theory and experiments.
Here, we improve the characterization of the spatio-temporal process
inside avalanches in the universality class of the depinning of elastic interfaces 
in random media. Surprisingly, at variance with the temporal shape, the spatial shape
of avalanches has not yet been predicted. 
In part this is due to
a lack of an analytically tractable definition: how should the shapes
be centered? Here we introduce such a definition, accessible in experiments, and study the 
{\it mean spatial shape of avalanches at fixed size centered around their starting point (seed).} 
We calculate the associated universal scaling functions, both in a mean-field model and
beyond. Notably, they are predicted to exhibit a cusp singularity near the seed. The results are in good agreement with a numerical simulation of an elastic line. 

\end{abstract}

\maketitle

Numerous slowly driven non-linear systems exhibit motion which is
not smooth in time but rather proceeds discontinuously via jumps extending 
over a broad range of space and time scales. Developing
predictive models of avalanche motion and understanding their universality, or lack thereof, 
has emerged as an outstanding challenge of modern statistical physics \cite{Sethna}. 
In condensed matter recent developments have led to distinguish two broad
classes, depending on the importance of plastic deformations. 
In systems such as dislocated solids, metallic glasses, 
granular media near jamming, plastic deformations play a crucial role and despite recent progresses 
a theoretical description is still under construction 
\cite{MullerWyart,RossoWyart2014,Robbins,ZapperiAlava}.
In many other situations the description by an elastic
interface driven in a disordered medium 
has proved relevant 
\cite{DSFisher1998,BlatterFeigelmanGeshkenbeinLarkinVinokur1994,NattermannScheidl2000,
GiamarchiLeDoussalBookYoung}. Examples are 
domain walls in soft magnets \cite{ZapperiCizeauDurinStanley1998, DurinZapperi2000}, fluid contact lines on rough surfaces \cite{Moulinet,LeDoussalWieseMoulinetRolley2009}, strike-slip faults in geophysics \cite{earthquakes}, fractures in brittle materials \cite{Ponson2009,Santucci2010,LaursonSantucciZapperi2010,BonamySantucciPonson2008} or imbibition fronts \cite{PlanetSantucciOrtÂn2009}. 
This class exhibits a dynamical phase transition - the so-called depinning transition -
accompanied by collective avalanche motion.
While the microscopic details of the dynamics are specific to each system, the large scale statistical properties of the avalanches are believed to be universal. The most studied quantities in this context are the critical exponents characterizing the scale-free probability distribution function (PDF) of avalanche total sizes $S$, $P(S) \sim S^{-\tau_S}$ and durations $T$, $P(T) \sim T^{-\tau_T}$. They are related to the roughness and dynamical exponents, $\zeta$ and $z$, defined at the depinning transition of the interface, using the scaling relations 
$S \sim \ell^{d +\zeta}$ and $T \sim \ell^z$ with $\ell$ the lateral extension of the avalanche.

Recent improvements in experimental techniques allow studies of avalanches with higher accuracy and to access new, finer quantities, with the aim of distinguishing more efficiently the different universality classes. 
This notably includes the direct imaging of the spatio-temporal process of the velocity field inside an avalanche $v(x,t)$ where $x$ denotes the internal coordinate of the ($d$-dimensional) interface and $t$ is the time since the beginning of the avalanche. A question of great interest is to understand whether and how scaling and universality extend to $v(x,t)$. 
\begin{figure}
\centerline{
\includegraphics[width=0.37\textwidth]{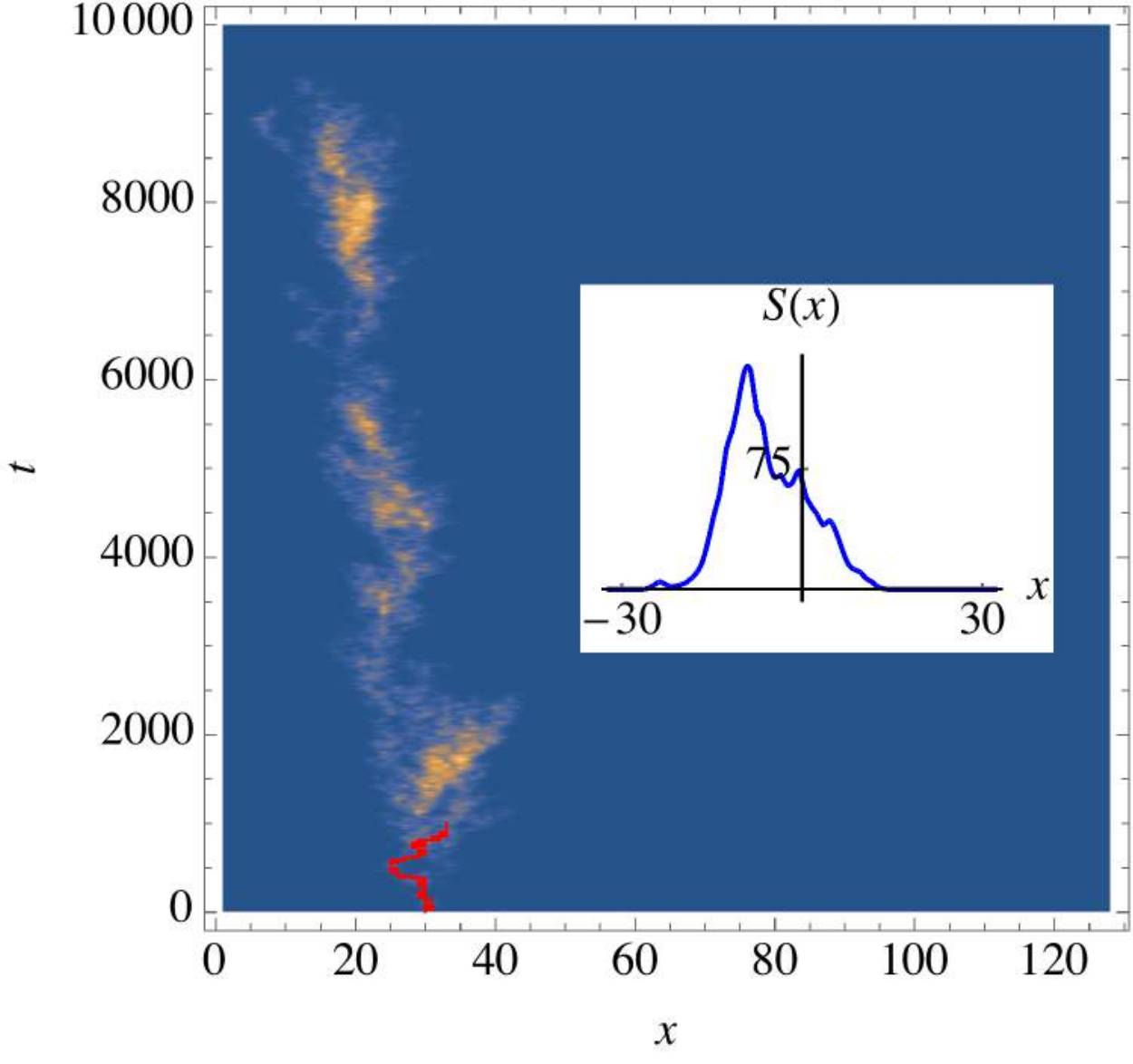} \includegraphics[width=0.06\textwidth]{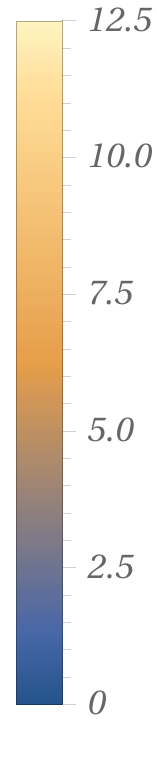} 
}
 \caption{Density plot of the velocity field $v(x,t)$ inside an avalanche of size $S=1760$ in the mean-field model (Brownian Force Model) for $d=1$ discretized with $N=128$ points. Time is given in machine-time unit. Line in red: backward
 path produced by the algorithm used to find the seed of the avalanche (see text). Inset: the spatial shape of this avalanche when centered around its starting point.
}
\label{fig:BranchingAvalanche}
\end{figure}

Until now the focus was 
on the center of mass velocity $ v_{\rm cm}(t) \sim \int d^dx ~ v(x,t)$ and the mean temporal shape at fixed duration $T$, $\langle v_{{\rm cm}}(t) \rangle_T$, where here $\langle \rangle_T$ denotes the statistical average over all avalanches of fixed duration $T$.
A scaling analysis suggests, through the sum rule $S= \int dt d^dx v(x,t)$, the existence of a scaling function $f^{\rm temp}_d(t )$ such that $\langle v_{{\rm cm}}(t) \rangle_T = T^{\gamma-1} f^{\rm temp}_d(t/T)$, where $\gamma = (d+\zeta)/z$. The universality of $f^{\rm temp}_d(t)$ was shown 
theoretically and studied experimentally in \cite{Assymetry,PapanikolaouBohnSommerDurinZapperiSethna2011,Laurson2013,DobrinevskiLeDoussalWiese2014,Durin-Doussal-Wiese-Shape}. The beautiful parabola-shape predicted at mean field level, 
$f_{\rm temp}(t) = t(1-t)$ (and $\gamma=2$), stimulated 
the excitement around this observable.

Though very interesting, this observable does not contain information on the remarkable spatial structure of avalanche processes (see for illustration Fig.~\ref{fig:BranchingAvalanche}). A characterization of even the
mean spatial shape of avalanches in terms of a simple scaling function is presently lacking. 
In this Letter we propose and calculate such a scaling function. We consider the mean shape of avalanches at fixed total size $S$,  for which a scaling analysis suggests (in real or in Fourier space $\langle S(q) \rangle_S = \int d^dx e^{iqx}  \langle S(x) \rangle_S$)
\bea \label{ScalingS}
&& \langle S(x) \rangle_S=  S^{1- \frac{d}{d+\zeta}}  f_d(\frac{x}{S^{\frac{1}{d+\zeta}}}) \ , \nn \\
&& \langle S(q) \rangle_S=  S  \tilde{f}_d(q S^{\frac{1}{d+\zeta}}) \ ,
\eea
where $S(x) = \int dt v(x,t)$ is the ``local size" at $x$,   $f_d(x)$ and $\tilde f_d(q)$ are radial scaling functions (hence $x$ and $q$ as arguments of the scaling functions always denote the norm of the vectors $x$ and $q$), normalized as $\int d^d x  f_d(x) = \tilde f_d (q=0) = 1$,
since $S= \int d^d x~ S(x)$. Here the local size at $x$, $S(x)$ is the local displacement of the interface between the beginning and the end of an avalanche at the point $x$, while the total size $S$ is the area swept by the interface during the avalanche. Note that these definitions are not complete: there are various ways of centering an avalanche. Our proposal is to study the spatial structure by {\it centering the avalanches on their starting points.} Hence in (\ref{ScalingS}) $\langle \rangle_S$ denotes the statistical average over all avalanches of fixed total size $S$ and starting point $x=0$. We call this procedure the {\it seed-centering} which %, though arbitrary, 
appears natural when one thinks of how an avalanche unfolds following a branching process (see Fig.~\ref{fig:BranchingAvalanche}). Furthermore, it permits analytical treatment and is thus appropriate to compare theory and 
experiments.

\medskip

We first calculate the above scaling functions at the level of mean-field. This requires
to go beyond the simplest mean-field toy model, the ABBM model 
\cite{AlessandroBeatriceBertottiMontorsi1990,AlessandroBeatriceBertottiMontorsi1990b}
which only describes the center of mass motion of the interface.
To this aim we consider the Brownian Force Model (BFM), recently
introduced as the relevant mean-field theory to describe spatial correlations
\cite{LeDoussalWiese2011b,DobrinevskiLeDoussalWiese2011b,LeDoussalWiese2012a,
ThieryLeDoussalWiese2015}. For this model, we even compute the
full {\it mean velocity-field inside a seed-centered avalanche of given size $S$} 
which in general obeys the scaling form
\bea \label{SpatioTemp2}
&& \langle v(x,t) \rangle_S = S^\frac{\zeta-z}{d+\zeta} F(t/S^{\frac{z}{d+\zeta}},x/S^{\frac{1}{d+\zeta}})  \ .
\eea

More generally, in this Letter we consider 
elastic interfaces in the quenched Edward-Wilkinson universality class with short ranged disorder.
In this context, the BFM is accurate for $d \geq d_c$, where $d_c$ is the upper critical
dimension of the depinning transition, $d_c=4$ for short-range (SR) elasticity and 
$d_c=2$ for the most common
long-range (LR) elasticity. In lower dimensions $d <d_c$, correlations play an important role.
To take them into account and study this more difficult case,
we use the Functional Renormalization Group (FRG) and calculate the 
scaling functions $f_d(x)$ and $\tilde{f}_d(q)$ perturbatively in $\epsilon=d_c-d$,
to one-loop, i.e. $O(\epsilon)$ accuracy 
(see \cite{Fisher86,Nattermann92, NarayanFisher92, ChauveDoussalWiese} for background on FRG, and \cite{DahmenSethna1996,LeDoussalWiese2008c,LeDoussalWiese2011b,LeDoussalWiese2012a} for its application to the study of avalanches). We show that the scaling ansatz 
(\ref{ScalingS}) holds and that the scaling functions contain only 
one non-universal scale $\ell_{\sigma}$ (which is discussed in details below)
\bea \label{ScalingS2}
f_d(x) = \frac{1}{\ell_\sigma^d} {\cal F}_d(\frac{x}{\ell_{\sigma}})  \quad , \quad  \tilde f_d(q) = \tilde{{\cal F}}_d(\ell_{\sigma} q ) \ ,
\eea
where ${\cal F}_d$ and $\tilde{{\cal F}}_d$ are fully universal and 
depend only on the space dimension $d$ and the universality class of the model
(i.e. range of elasticity and disorder). The precise model that is the starting point of our theoretical analysis (for elastic interfaces with short-ranged elasticity) is given in (\ref{eqm}). Our conclusions however apply in much greater generality and the details of the model are unimportant (once the range of elasticity and disorder correlation have been set). Indeed, since the scaling functions that we compute are universal and entirely determined by the properties of the FRG fixed point for models in the quenched Edward-Wilkinson universality class, any model in the same universality class leads to the same scaling functions. In the first part of the Letter we thus focus on stating our results, and report the discussion of the model and of the method to the second part. For a generic system, we expect scaling and universality to hold for avalanche of size $S$ in a scaling regime $S_{\rm min} \ll S \ll S_{\rm max}$. Note that in (\ref{ScalingS2}), the space variable $x$ is measured in units of $S^\frac{1}{d+\zeta}$ (see (\ref{ScalingS})). In the original units, the universality in the avalanche shape should hold for both small and large $x$ (compared to $S^\frac{1}{d+\zeta}$) as long as $x_{\rm \min} \ll  x \ll x_{\rm max}$ where $x_{{\rm min } / {\rm max }} \sim   S_{{\rm min } / {\rm max }}^{\frac{1}{d+\zeta}} $. 
%We now describe our main results,
%and report the discussion of the model and methods in the second part of the Letter. 
We will start by discussing the exact results obtained for the BFM (defined below, see (\ref{eqm})). These results are also of interests for the SR disorder universality class as the lowest order terms in the $\epsilon$ expansion (i.e. $O(\epsilon^0)$ terms) of the true universal scaling functions.

\begin{figure}
\centerline{
\includegraphics[width=0.25\textwidth]{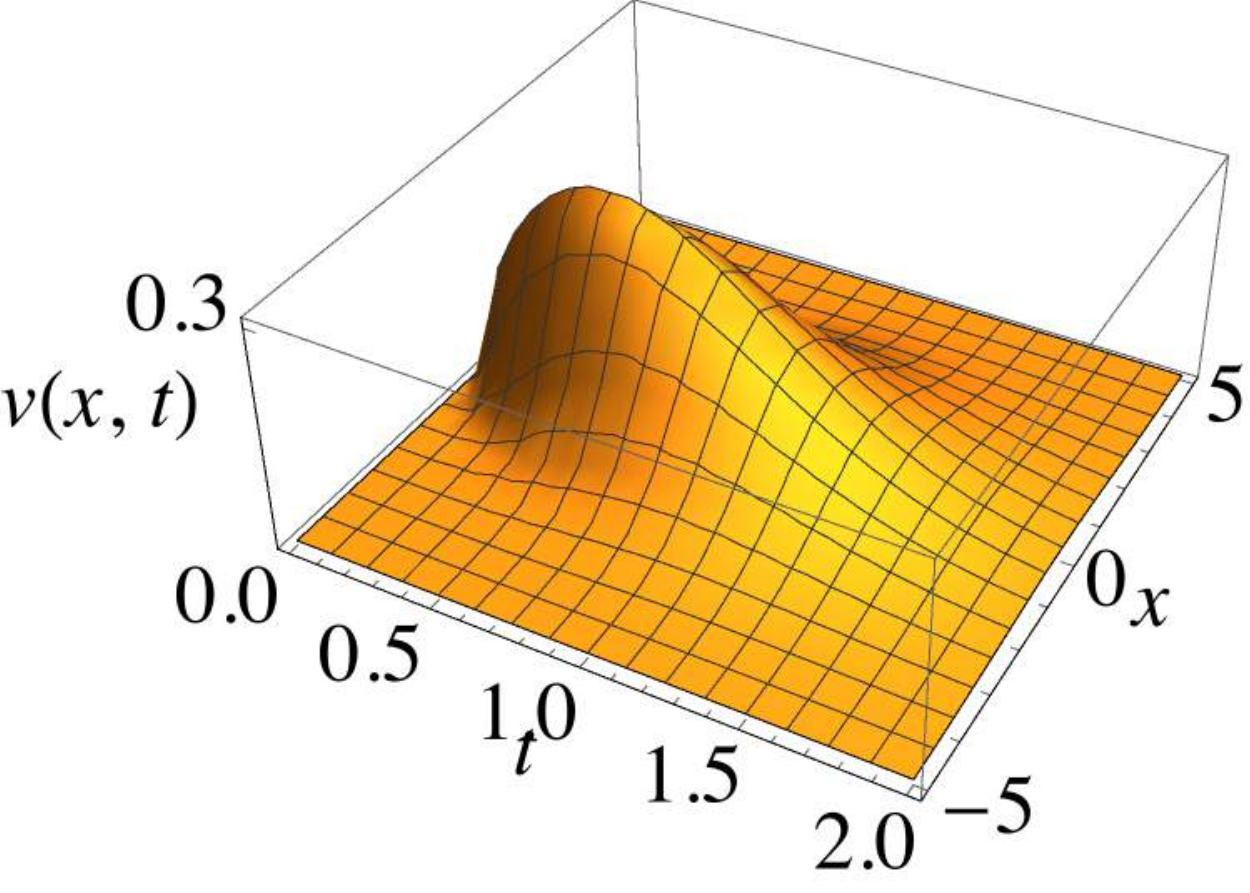} \includegraphics[width=0.25\textwidth]{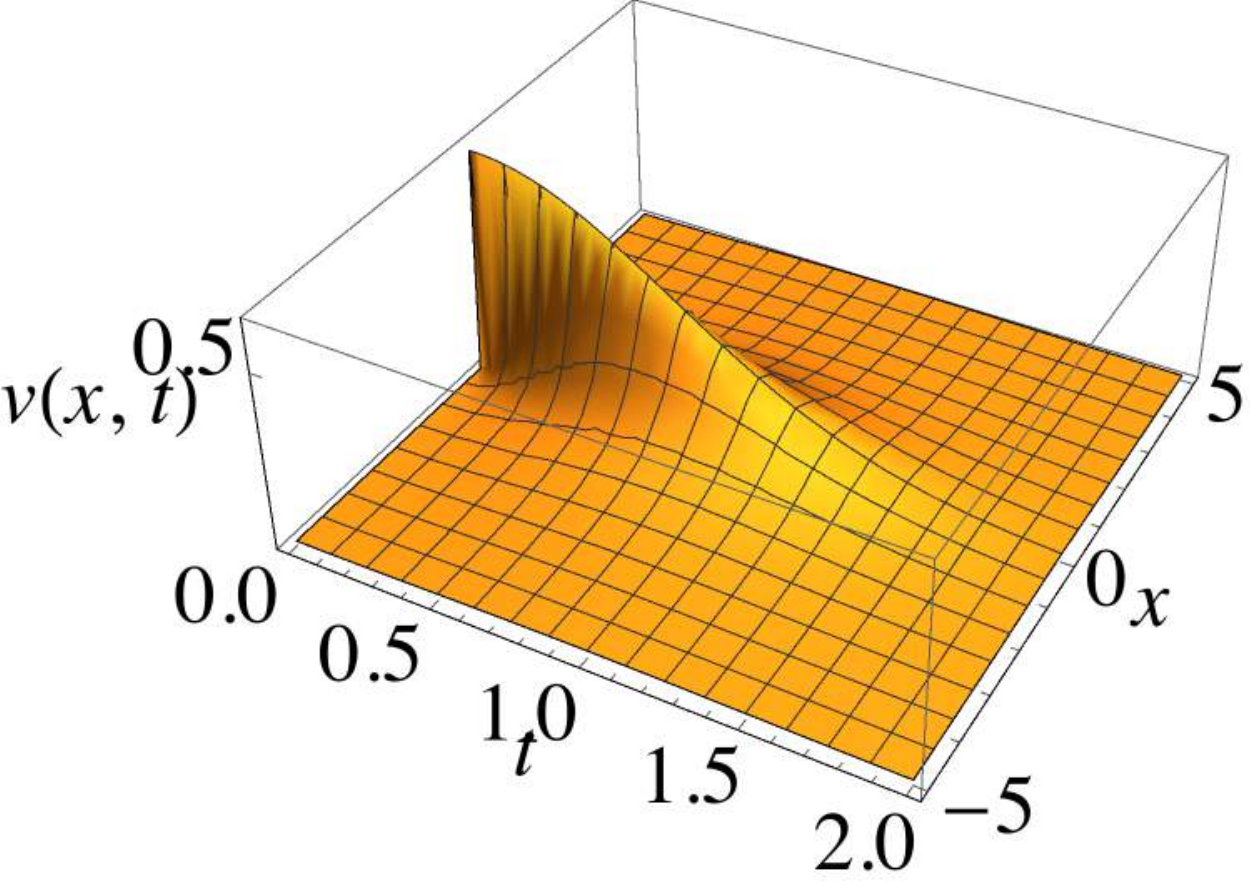} }
 \caption{Plot of the mean-field result for the space-time mean velocity profile inside an avalanche in $d=1$ for SR (left, see (\ref{SpatioTemp})) and LR elasticity (right, see (\ref{SpatioTempLR})).}
\label{fig:3D}
\end{figure}
\medskip

\begin{figure}[h]
\centerline{
\includegraphics[width=0.165\textwidth]{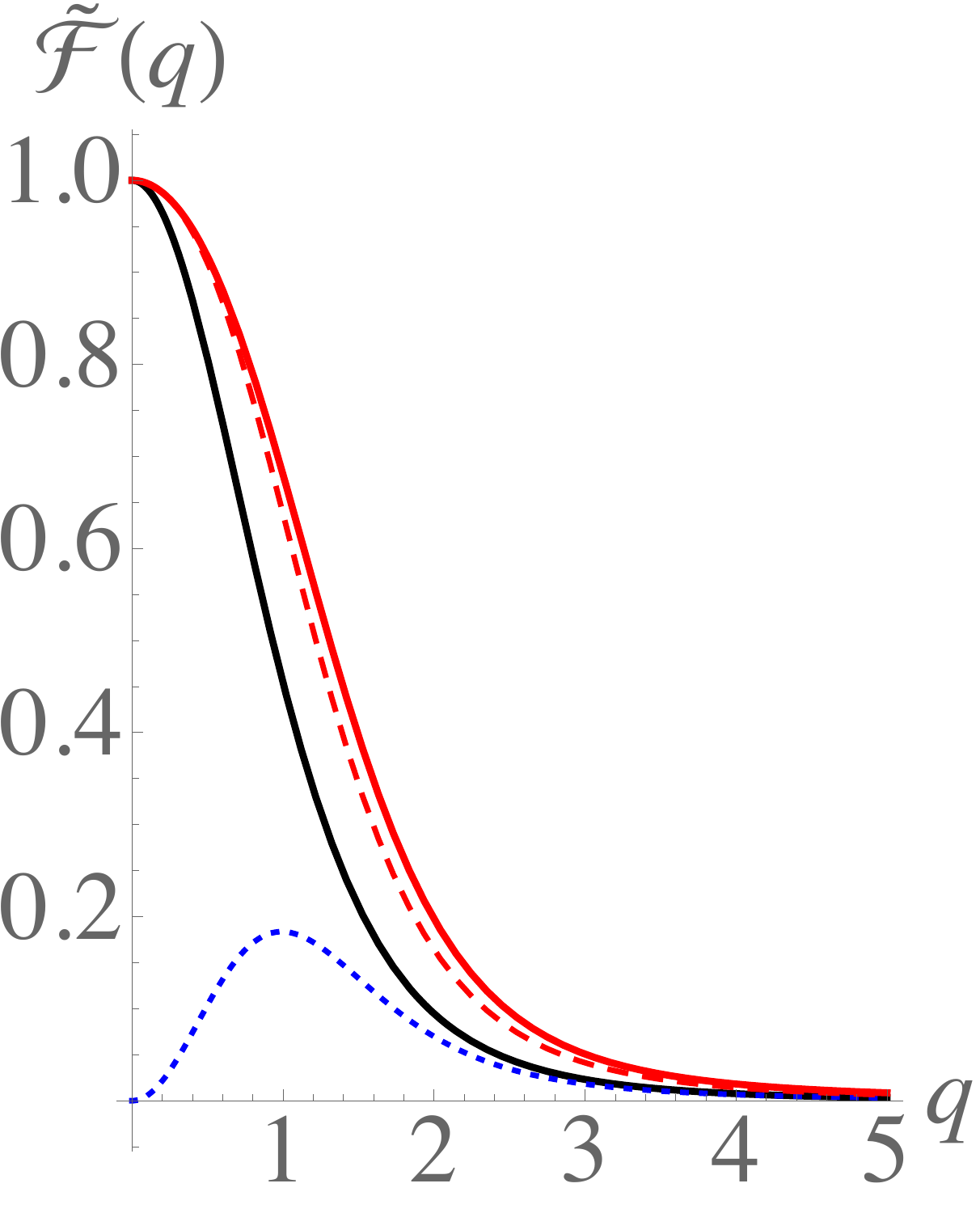} \includegraphics[width=0.165\textwidth]{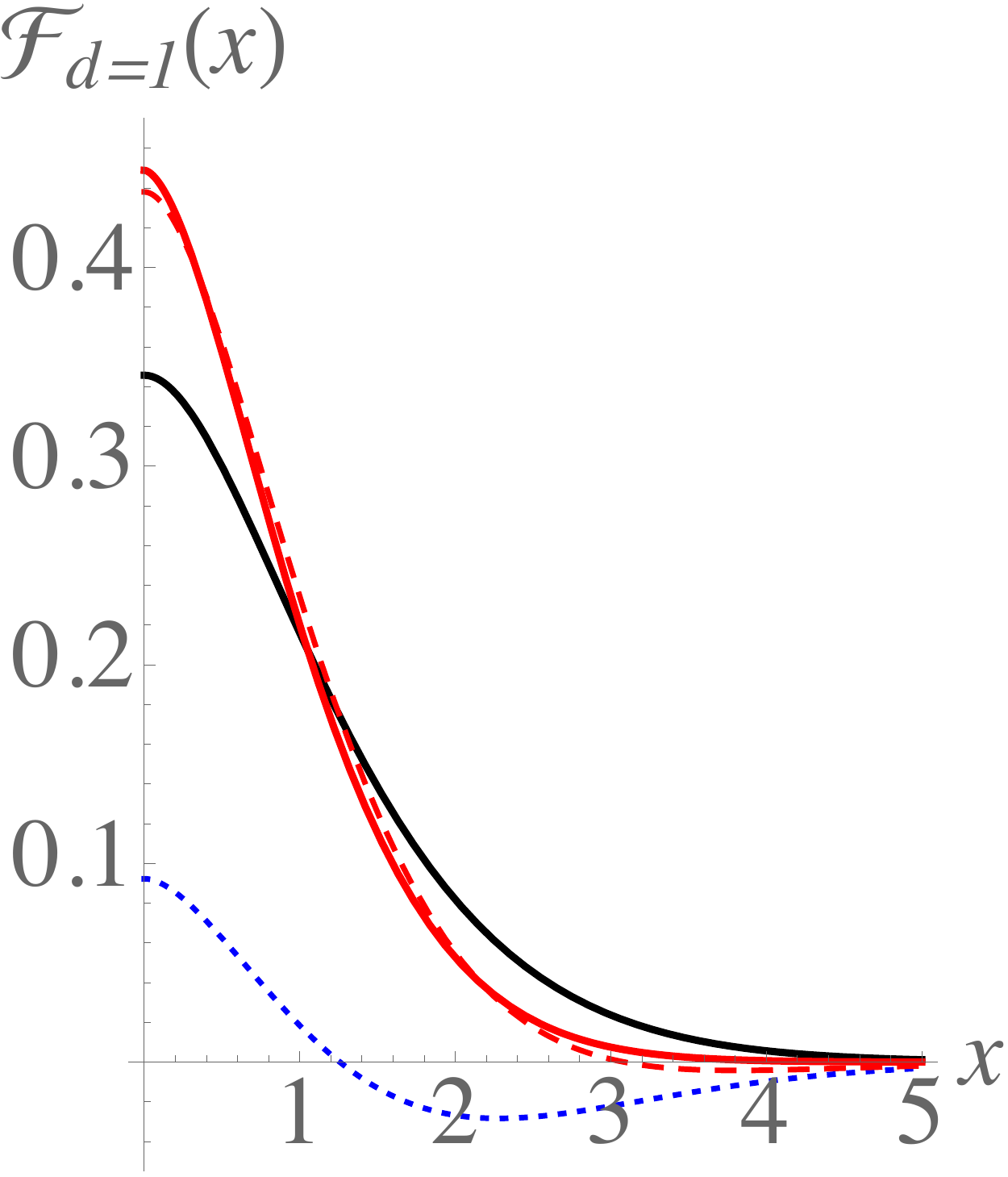} \includegraphics[width=0.165\textwidth]{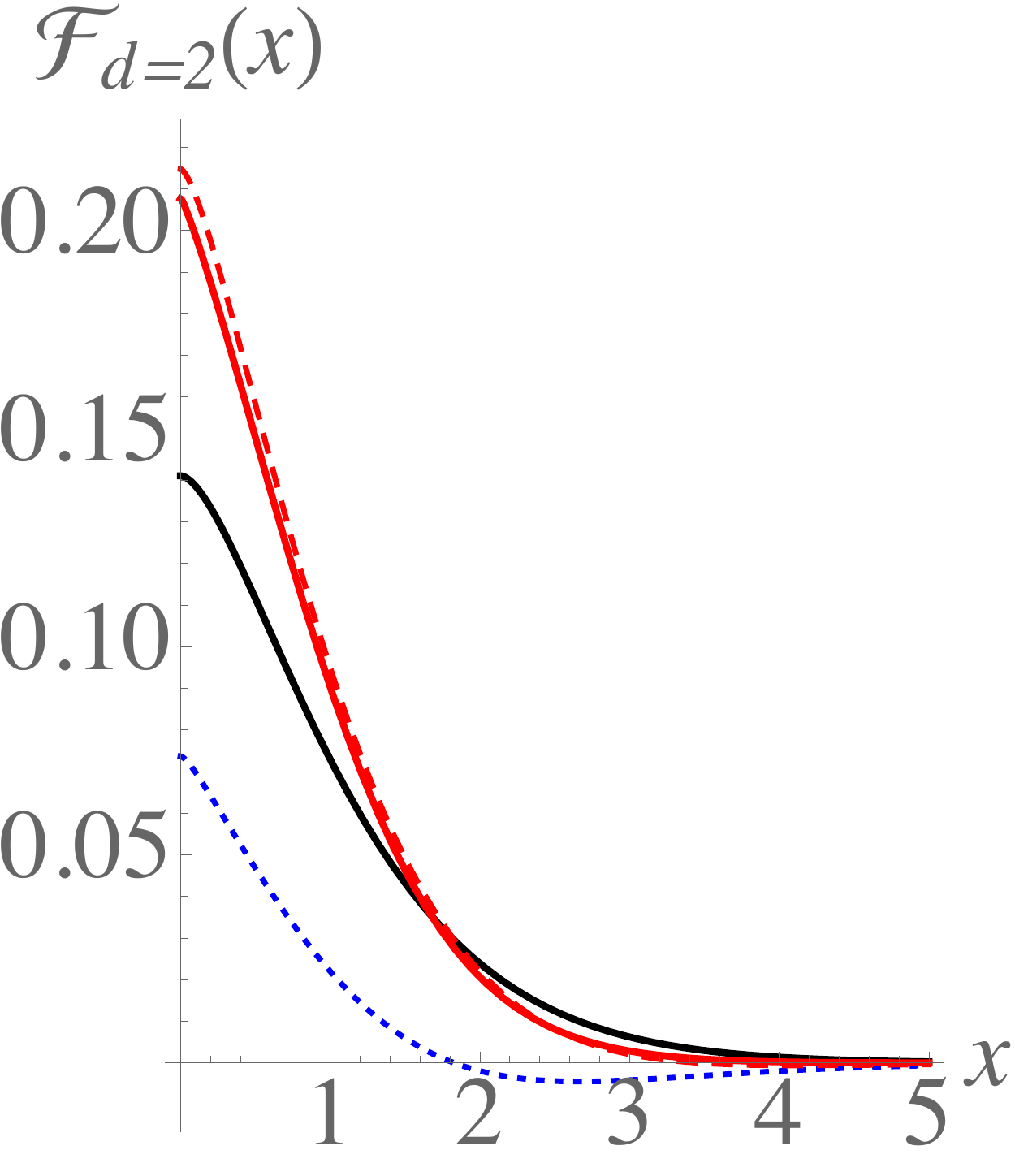}
 }
 \caption{(color online). Analytical results at MF and $O(\epsilon)$ level for the universal scaling function $\tilde{{\cal F}}_{d=1}$ in Fourier space (Left) and ${\cal F}_d$ in real space for $d=1$ (Middle) and $d=2$ (Right) for SR elasticity. Black lines: tree/mean-field results. Dotted blue lines: universal corrections, $\delta \tilde{{\cal F}}_1(q) $ (left, $O(\epsilon)$ correction in Fourier space in $d=1$), $\delta {\cal F}_1(x) $ (middle) and $\delta{\cal F}_2(x) $ (right). Red-dashed lines: $O(\epsilon)$ estimate obtained by simply adding the corrections to the MF value. 
 Red lines: improved $O(\epsilon)$ estimate, 
 which, through a re-exponentiation procedure, takes properly into account the modification of exponents (\ref{FourierFatTail}) and (\ref{Real1loopExpDecay}) (see \cite{SM}). Note that the cusp at the origin  of the avalanche shape at $O(\epsilon)$ is not obvious in this plot since the non-analyticity is rather small, but it can be emphasized using a log-log scale (and measured in numerics, see Fig.~\ref{fig:resSimuSmallxLargeq}).}
\label{fig:treeAnd1loop}
\end{figure}

{\it Results within mean-field:} The BFM can be studied analytically
in any dimension $d$.  Let us first consider the case of SR elasticity.  The exponents are $\tau_S=3/2$, $\tau_T=z=2$ and $\zeta=4-d$. The scaling function in (\ref{SpatioTemp2}) admits a very simple expression:
\bea \label{SpatioTemp}
%&& \langle v(x,t) \rangle_S = S^\frac{2-d}{4} F(t/S^{1/2},x/S^{1/4}) \nn \\
&& F(t,x)=2 t e^{-t^2} \frac{1}{(4 \pi t)^{d/2}} e^{-x^2/(4 t)} \ ,
\eea
which is plotted in Fig.~\ref{fig:3D}. Here we use 
%expressed time, space and avalanche sizes units in the dimensionless, natural units of the model. 
dimensionless units, the original units can be recovered using $x \to m x$, $t \to t/\tau_m$ and $S \to S/S_m$ where $\tau_m = \eta/m^2$ and $S_m = \sigma/m^4$ and the parameters $\eta, m$ and $\sigma$ are those 
%that appear 
in the equation of motion of the model (\ref{eqm}). Time integration of (\ref{SpatioTemp}) 
%against $t$ easily 
confirms for the BFM the general scaling law (\ref{ScalingS}) and (\ref{ScalingS2}) with ${\cal F}^{\rm MF}_d(x) = \int_0^{+\infty} dt F(t,x)$ and $\ell_{\sigma} = \sigma^{-1/4}$. The result %is most simply expressed 
is simplest in Fourier space and does not depend on the dimension: %$\tilde{{\cal F}}_d^{\rm MF}(q) = \tilde{{\cal F}}^{\rm MF}(q)$ %and reads
\bea \label{MFq}
\tilde{{\cal F}}_d^{\rm MF}(q) = \tilde{{\cal F}}^{\rm MF}(q) = 1 - \frac{ \sqrt{\pi} q^2 }{ 2} e^{\frac{q^4}{4}}  \text{erfc} \left(\frac{q^2}{2}\right) \ ,
\eea
where $\text{erfc}(z) =\frac{2}{ \sqrt{\pi}} \int_{z}^{+ \infty} e^{-t^2}$. 
In real space, ${\cal F}^{\rm MF}_{d}(x)$ depends on the dimension %(since the Fourier transform operation depends on the dimension). Its value at the origin is 
and can be expressed using hypergeometric functions \cite{SM} 
with ${\cal F}^{\rm MF}_{d \leq 4}(0) = \frac{2^{-d} \pi^{1-\frac{d}{2}}}{\Gamma(\frac{d}{4}) \sin(\frac{\pi d}{4})}$.
Both 
$\tilde{{\cal F}}^{\rm MF}(q)$ and ${\cal F}^{\rm MF}_{d=1,2}(x)$ are plotted in black in
%in black in the middle ($d=1$) and on the right ($d=2$) of 
Fig.~\ref{fig:treeAnd1loop}.
A fundamental property of $\tilde{{\cal F}}^{\rm MF}(q)$ is that it possesses an algebraic tail $\tilde{{\cal F}}^{\rm MF}(q) \sim q^{-4}$ at large $q$, which generates a 
non-analytic term $\sim |x|^{4-d}$ in the small $x$ expansion of ${\cal F}^{\rm MF}_{d}(x)$ around
the origin. 
Its behavior at large $x$ is evaluated using a saddle-point on (\ref{SpatioTemp}), leading to a stretched exponential decay with a $d$-independent exponent $4/3$:
\bea \label{MFlargex}
{\cal F}^{\rm MF}_d(x) \simeq_{x \to \infty} \frac{2^{-d/2} \pi ^{\frac{1}{2}-\frac{d}{2}} 
   }{\sqrt{3}} x^{\frac{2-d}{3}} e^{-\frac{3 x^{4/3}}{4}} \ .
\eea
%The full result can be expressed using hypergeometric functions (see SM). ${\cal F}^{\rm MF}_d(x)$ is plotted in black in the middle ($d=1$) and on the right ($d=2$) of Fig.~\ref{fig:treeAnd1loop}.
These results easily extend to LR elasticity, in which case $z=1$, $\zeta=2-d$ and the mean shape in Fourier space
is obtained replacing $q^2 \to q$ in (\ref{MFq}). Let us also give here the spatiotemporal
shape (\ref{SpatioTemp2}) for the experimentally most relevant case of $d=1$, with
\bea \label{SpatioTempLR} 
F(t,x) = \frac{2 t^2 e^{-t^2}}{\pi(x^2 + t^2)} \ .
\eea 

\medskip

{\it Results beyond mean-field for SR elasticity:} 
For realistic SR disorder, the BFM is the starting point in the $\epsilon = 4-d$ expansion. It is most clearly implemented in Fourier space,
since the mean-field result for $\tilde{{\cal F}}_d(q)$ does not depend on $d$:
\bea \label{epsilon1}
&& \tilde{{\cal F}}_d^{{\rm SR}}(q) = \tilde{{\cal F}}^{{\rm MF}}(q) +\delta \tilde{{\cal F}}_d (q) + O(\epsilon^2) \ ,
\eea
with $\delta \tilde{{\cal F}}_d (q) = \epsilon \tilde {\cal F}^{(1)}(q)$. Here
$ \tilde{{\cal F}}^{(1)} (q) = \int_{{\cal C}} 
 \frac{d\mu}{2i \pi}  e^{\mu} \tilde H(\mu,q)$ is obtained as an Inverse Laplace Transform (ILT) $\mu \to 1$: 
\bea \label{LT1loop}
&& \tilde H(\mu,q) =  \frac{4  \sqrt{\pi} }{9}
\bigg[    \frac{2-3\gamma_E}{8} \frac{1}{q^2 + 2 \sqrt{\mu}}  -
\frac{4 \sqrt{\mu}}{(q^2  + 2 \sqrt{\mu})^2}  \\
&& \times  \bigg(\frac{q^2 + 9 \sqrt{\mu}}{q  \sqrt{q^2 + 8 \sqrt{\mu}} } 
\sinh ^{-1}\left(\frac{q}{2 \sqrt{2 \sqrt{\mu}} }  \right) -1 + \frac{3}{16} \ln(4 \mu) \bigg) \bigg] \nn
\eea
where $\gamma_E$ is Euler's Gamma constant (see \cite{SM} for the choice of ${\cal C}$). 
We then define the correction to the mean shape in real
space as the d-dimensional Fourier transform $\delta {\cal F}_d (x) =  \int \frac{d^d q}{(2 \pi)^d} e^{-i q x} 
\delta \tilde{{\cal F}}_d (q)$. Hence, 
$ {\cal F}_d^{{\rm SR}}(x) = {\cal F}_d^{{\rm MF}}(x) + \delta {\cal F}_d (x) + O(\epsilon^2)$.
From the ILT expression (\ref{LT1loop}) we obtain the following analytical properties of the $O(\epsilon)$ 
corrections:

1) Its large $q$ expansion is $\delta \tilde{{\cal F}}_d (q) \simeq_{q \gg1} 
\epsilon \frac{ 8 \log (q)-\gamma_E-8 }{9 q^4} $, interpreted as a {\it change in the tail exponent $\tilde \eta_d$}:
%of $\tilde{{\cal F}}_d (q)$:}
\bea \label{FourierFatTail}
\tilde{{\cal F}}_d (q) \simeq_{q \gg1} \tilde A_d q^{- \tilde \eta_d} \quad , \quad 
\tilde \eta_d= 4 - \frac{4 \epsilon}{9}
+ O(\epsilon^2)  \ ,
\eea
with a universal prefactor $ \tilde A_d=2 (1 -  (2 + \frac{\gamma_E}{4}) \frac{2 \epsilon}{9})$. In real space this implies, 
in the expansion of ${\cal F}_d (x)$ at small $x$,
a non-analytic term $\sim |x|^{\eta_d}$ with $\eta_d  = \tilde \eta_d - d = \frac{5 \epsilon}{9}
+ O(\epsilon^2)$. Restoring the $S$ dependence from (\ref{ScalingS})
this leads to $\langle S(q) \rangle_S \sim_{q \to +\infty} S^{1 - \frac{\tilde \eta_d}{d+\zeta}} q^{- \tilde \eta_d}$
and the non-analytic part $\langle S(x) \rangle_S^{n.a} \sim_{x \to 0} S^{1 - \frac{\tilde \eta_d}{d+\zeta}} 
|x|^{\eta_d}$. Note that in the BFM the value $\tilde \eta_d=4=d+\zeta$ implies that the large $q$ behavior of $\langle S(q) \rangle_S$ {\it does not depend on $S$}. This may seem natural: in the BFM the small scales
do not know about the total size of the avalanche. A generalization of this property to the SR disorder case
would suggest the guess $\tilde \eta^{{\rm guess}}_d=d+\zeta$. Our result explicitly shows that 
this property fails with $\tilde \eta_d > d+\zeta$. Hence in the SR disorder case the {\it large avalanches tend to be more smooth than small avalanches.} Note that the predicted value of $\eta_d$ 
is smaller than $2$ in all physical dimension: {\it this non-analytic term should actually dominate the behavior of ${\cal F}_d (x)$ around $0$} (and thus lead to a cusp singularity). A possible interpretation of this cusp singularity is that around $0$ the mean shape of avalanches ${\cal F}_d (x)$ is dominated by avalanches whose largest local size is at their seed. This could correspond to the fact that such avalanches occur as a consequence of large fluctuations of the disorder that would pin a specific point of the interface for a long time. These would result in configurations of the interface with a single point well behind the rest of the interface. The depinning of such a point would then trigger an avalanche that is peaked around its seed \cite{MWprivate}.

2) At large $x$, we obtain that the stretched exponential decay exponent of the mean shape is modified from its MF behavior $\delta^{{\rm MF}}=4/3$: 
\bea \label{Real1loopExpDecay}
{\cal F}_d (x) \sim e^{ -C x^{\delta}} \quad , \quad \delta = \frac{4}{3}  + \frac{2}{27} \epsilon + O(\epsilon^2)  \ ,
\eea
with a universal prefactor $C =\frac{3}{4} + (\frac{7 \sqrt{3}}{36} -1  ) \frac{2}{9} \epsilon$. Remarkably, using $\zeta = \epsilon/3
+ O(\epsilon^2)$, this agrees to
$O(\epsilon)$ with the conjecture $\delta= \frac{d+\zeta }{d+\zeta -1}$ that we justify in
\cite{SM}. 

Furthermore, the ILT expression (\ref{LT1loop}) is easily calculated numerically. 
The corrections $\delta \tilde {\cal F}_d (q)$ and $\delta {\cal F}_d (x)$ are shown
in Fig.~\ref{fig:treeAnd1loop}, together with the resulting estimates for 
the functions ${\cal F}_d^{{\rm SR}}(x)$ and $\tilde{{\cal F}}_d^{{\rm SR}}(q)$.

\medskip

{\it Model and method:} For SR elasticity, the equation of motion for the interface position $u(x,t)$ (denoted
$u_{xt}$) is
\begin{equation} \label{eqm}
\eta \partial_t u_{xt} = \nabla^2_x u_{xt} - m^2 ( u_{xt}-w_t ) + F(u_{xt},x) \ ,
\end{equation}
where $\eta$ is the friction, $m$ is a mass cutoff which suppresses fluctuations
beyond the length $\ell_m=1/m$ and $m^2 w_{t}$ is the driving force. In the BFM, the random pinning force $F(u,x)$ is an independent Brownian motion in $u$ for each $x$ with $\overline{(F(u,x) - F(u',x))^2} = 2 \sigma |u-u'| $. For the SR disorder universality class, the second cumulant is $\overline{F(u,x)F(u',x')} = \delta^d (x-x') \Delta_0(u-u')$ with $\Delta_0(u)$ a fast decaying function. Eq. (\ref{eqm}) is analyzed using the dynamical field theory and the FRG \cite{SM}. This leads to an expression for $\langle S(x) \rangle_S$ as an ILT: $\langle S(y) \rangle_S \sim LT^{-1}_{\mu \to S} (\langle \tilde u_{x=0}^1  \rangle_{\xi}) /\rho(S)$ where $\rho(S)$ is the avalanche-size density (previously computed to $O(\epsilon)$ accuracy in \cite{LeDoussalWiese2008c,LeDoussalWiese2012a}) and $\tilde u_{x=0}^1$ is the $O(\lambda)$ term taken at $x=0$ of the solution $\tilde u_{x}$ of the following differential equation (here in dimensionless units):
\bea \label{instantonMainText}
-\mu + \lambda \delta(x-y) +(\tilde{u}_{x})^2 
+ \nabla_x^2 \tilde{u}_{x}- (1+ \xi_x)\tilde{u}_{x}= 0 \ ,
\eea
where $\xi_x$ is a white-noise of order $\sqrt{\epsilon}$ and $\langle . \rangle_{\xi}$ denotes the average over it. For the BFM, the result is thus obtained setting $\xi_x \to 0$ above. At $O(\epsilon)$ for the SR disorder universality class, it is thus sufficient to solve (\ref{instantonMainText}) perturbatively to second order in $\xi_x$. Here the fact that we are looking at the local size of avalanches at $x=y$ and whose seed is centered at $x=0$ is encoded in (\ref{instantonMainText}) as the fact that we are computing the value at $x=0$ (seed position) of the solution of (\ref{instantonMainText}) with a delta source $\lambda \delta(x-y)$ (local size position). The seed centering therefore allows analytical treatment here because $\tilde{u}_{x=0}$ only contains the contribution of avalanches starting at $0$ (see \cite{SM}). Using another type of spatial centering does not allow a similar simple treatment.

In our model (\ref{eqm}), the non-universal scale $\ell_{\sigma}$ in (\ref{ScalingS2}) is $m^{-1} S_m^{-1/(d+\zeta)}$ where
$S_m$ is defined from the ratio of the first two moments 
of the avalanche size distribution, $S_m = \langle S^2 \rangle/ (2 \langle S \rangle)$,
which can be measured in numerics and experiments. Here $ \langle  \rangle$ denotes the average with respect to the avalanche size distribution. In cases where the numerical or experimental setup corresponds to our model (as in our simulations, see below), this prediction for 
$\ell_{\sigma}$ allows unambiguous comparison between our results and the data. 
In cases where $\ell_{\sigma}$ cannot be predicted, some scale-independent features of the mean-shape still
allow comparison with the experiments. This includes the tail exponent of $\tilde{{\cal F}}_d (q)$ in (\ref{FourierFatTail}), the small and large distance behavior of ${\cal F}_d(x)$ 
in (\ref{Real1loopExpDecay}), and the {\it universal ratios} $c_{p} = \frac{\int d^d x  |x|^{2p} {\cal F}_d(x)}{\left(\int d^d x  |x|^{p} {\cal F}_d(x) \right)^2}$. In $d=1$, $(c_1,c_2) \simeq (1.6944,3.8197)$ for
the BFM while $(c_1,c_2) \simeq (1.641 \pm 0.001,3.43 \pm 0.02)$ for SR disorder to $O(\epsilon)$.

\begin{figure}[h]
\centerline{
\includegraphics[width=0.25\textwidth]{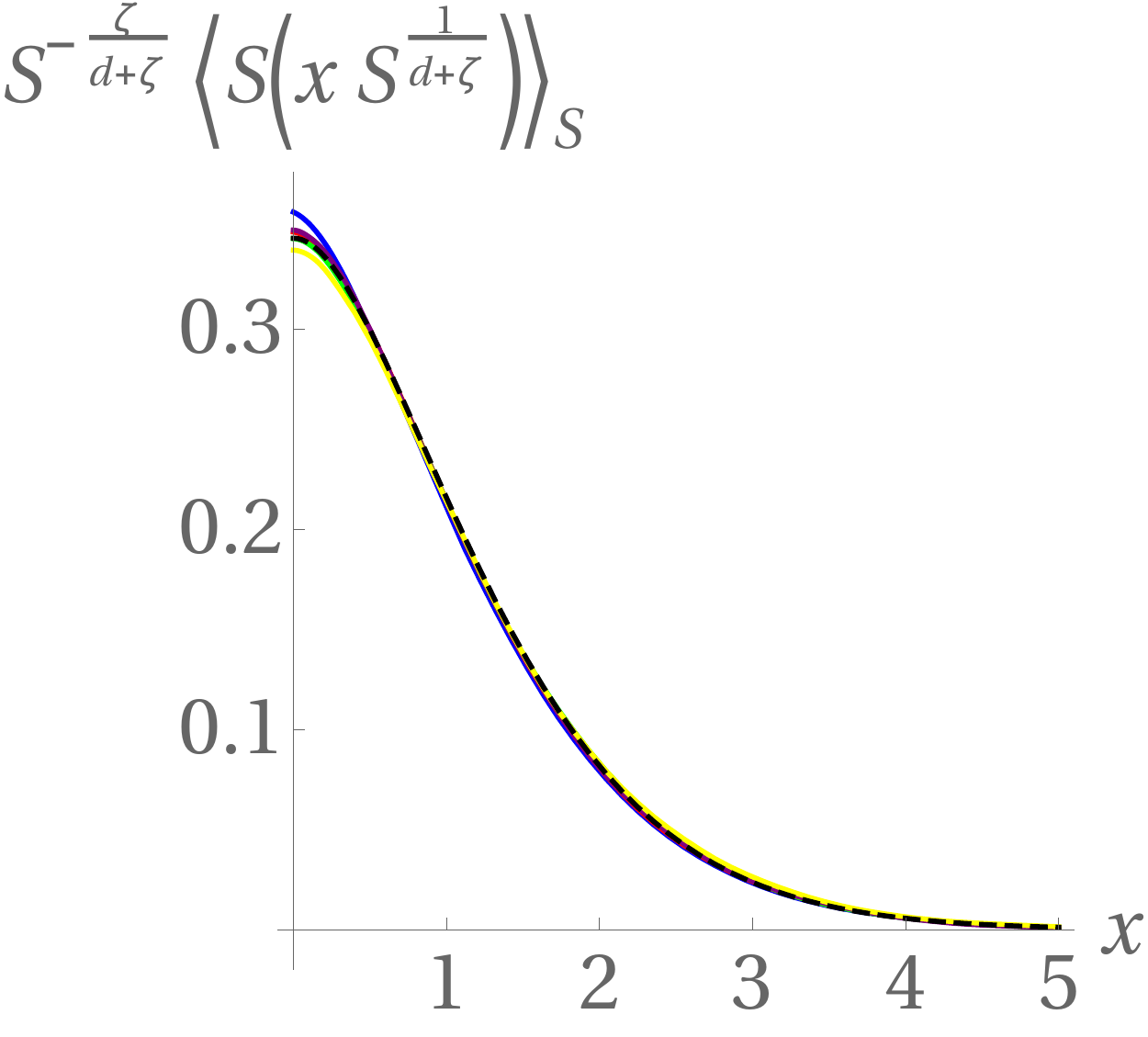} \includegraphics[width=0.25\textwidth]{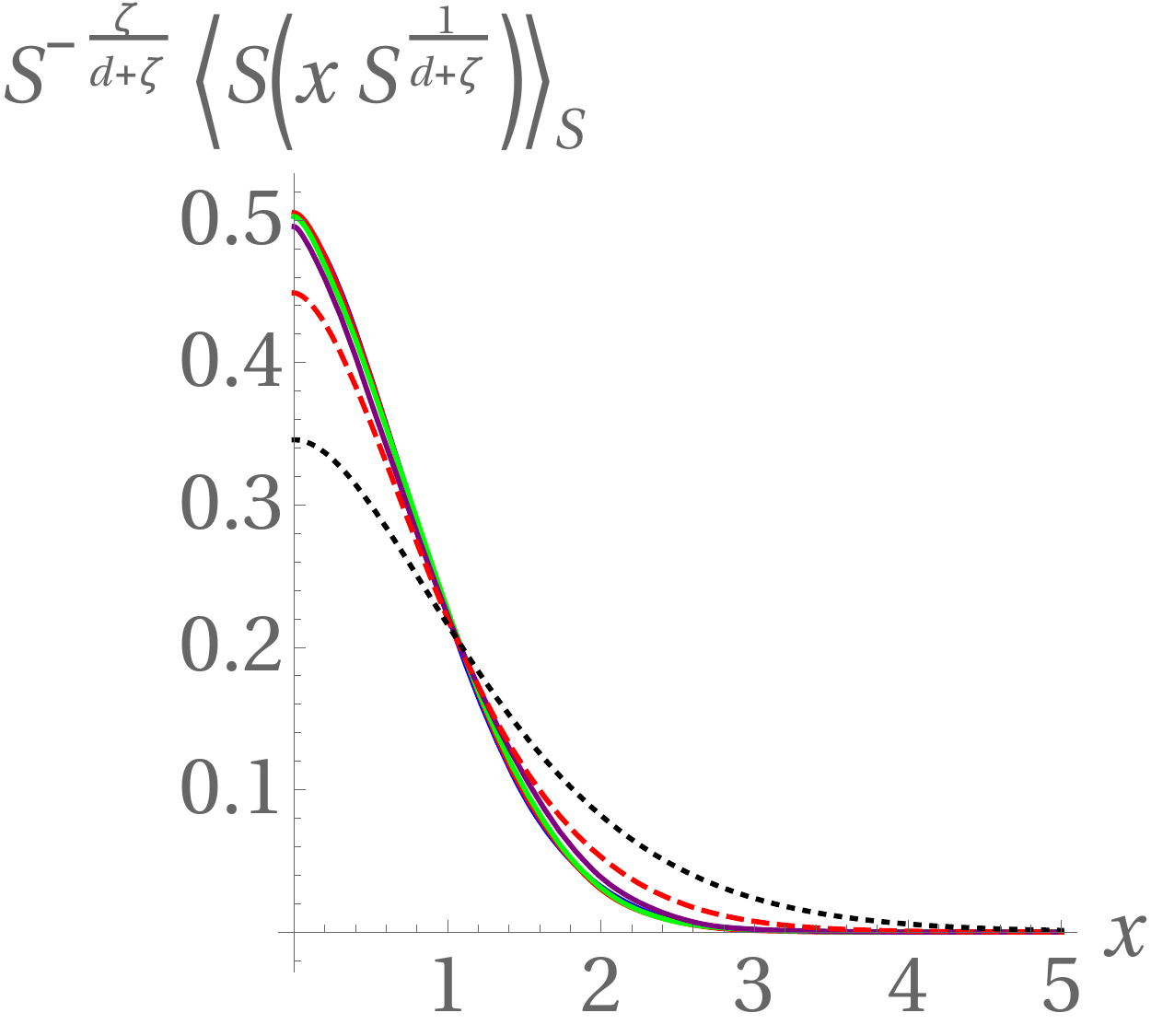}}
 \caption{(color online). Plain lines: rescaled mean shapes of avalanches at fixed size $S$ from the simulation of the BFM model (left) and of the model with SR disorder (right), in $d=1$, for $S=10$ (left only, blue), $S=50$ (right only, blue), $S=10^2$ (red), $S=10^3$ (green), $S=10^4$ (purple) and $S=10^5$ (left only, yellow). Dashed black lines: theoretical MF result. Red dashed line: $O(\epsilon)$ result. No fitting parameter.}
\label{fig:resSimuShapesd1}
\end{figure}

\medskip

{\it Numerical simulations.} A convenient choice of SR disorder, amenable to Markovian evolution,
is the Gaussian disorder $F(u,x)$ with "Ornstein-Uhlenbeck" (OU)
correlator $\Delta_0(u)=\sigma \delta u e^{- |u|/\delta u}$. It is defined by 
two coupled equations for the velocity $v_{xt} \equiv v(x,t)$ and the force
${\sf F}(x,t)$ (the first one being the time-derivative of (\ref{eqm})):
\bea \label{Simu}
&& \eta \partial_t v_{xt} = \nabla^2 v_{xt} + m^2(\dot{w}_t-v_{xt}) + \partial_t {\sf F}(x,t)  \ , \nn \\
&& \partial_t {\sf F}(x,t) = \sqrt{2 \sigma v_{xt}} \chi_{xt} - \frac{ v_{xt}}{ \delta u} {\sf F}(x,t) \ ,
\eea
with $\chi_{xt}$ a centered Gaussian white noise $\overline{\chi_{xt} \chi_{x't'}} = \delta^d(x-x') \delta(t-t')$
and initial condition $v_{xt=0}={\sf F}(x,t=0)=0$. In the stationary regime, this model is equivalent 
\cite{DobrinevskiThesis,LeDoussalWiese2014a} to
Eq. (\ref{eqm}) with $\dot u_{xt}=v_{xt}$ and ${\sf F}(x,t)=F(u_{xt},x)$ 
and initial condition $u_{xt=0}=0$. When $1/\delta u = 0$ this model becomes equivalent to the BFM. 
We discretize time in units $dt$ and space with periodic boundary conditions along $x$. To measure quasi-static avalanches, we apply a succession of kicks of sizes $\delta w$: we impose $v_{xt}  = (m^2/\eta) \delta w$ at $t=0^+$ (beginning of the avalanche), iterate (\ref{Simu}) and wait for the interface to stop before applying a new kick \cite{SM}. To identify the seed of each avalanche, we record the velocity $v(x,t)$ for the $n_{t}=10^3$ first time-steps of the avalanche. We find the position $x_{{\rm max}}(n_t)$ of maximum velocity at $t_{n_t} = n_t dt$ (or at the end of the avalanche if it has stopped before), and then successively identify at each time step $t_{n}<t_{n_t}$ the position $x_{{\rm max}}(n)$ defined as the neighbor of $x_{{\rm max}}(n+1)$ with the largest velocity at time $t_n$. $x_{{\rm max}}(n=1)$ is identified as the seed of the avalanche. The size of the kicks is chosen small enough so that the probability to trigger several macroscopic and overlapping avalanches is negligible (see \cite{SM} for details).

\begin{figure}[h]
\centerline{
\includegraphics[width=0.25\textwidth]{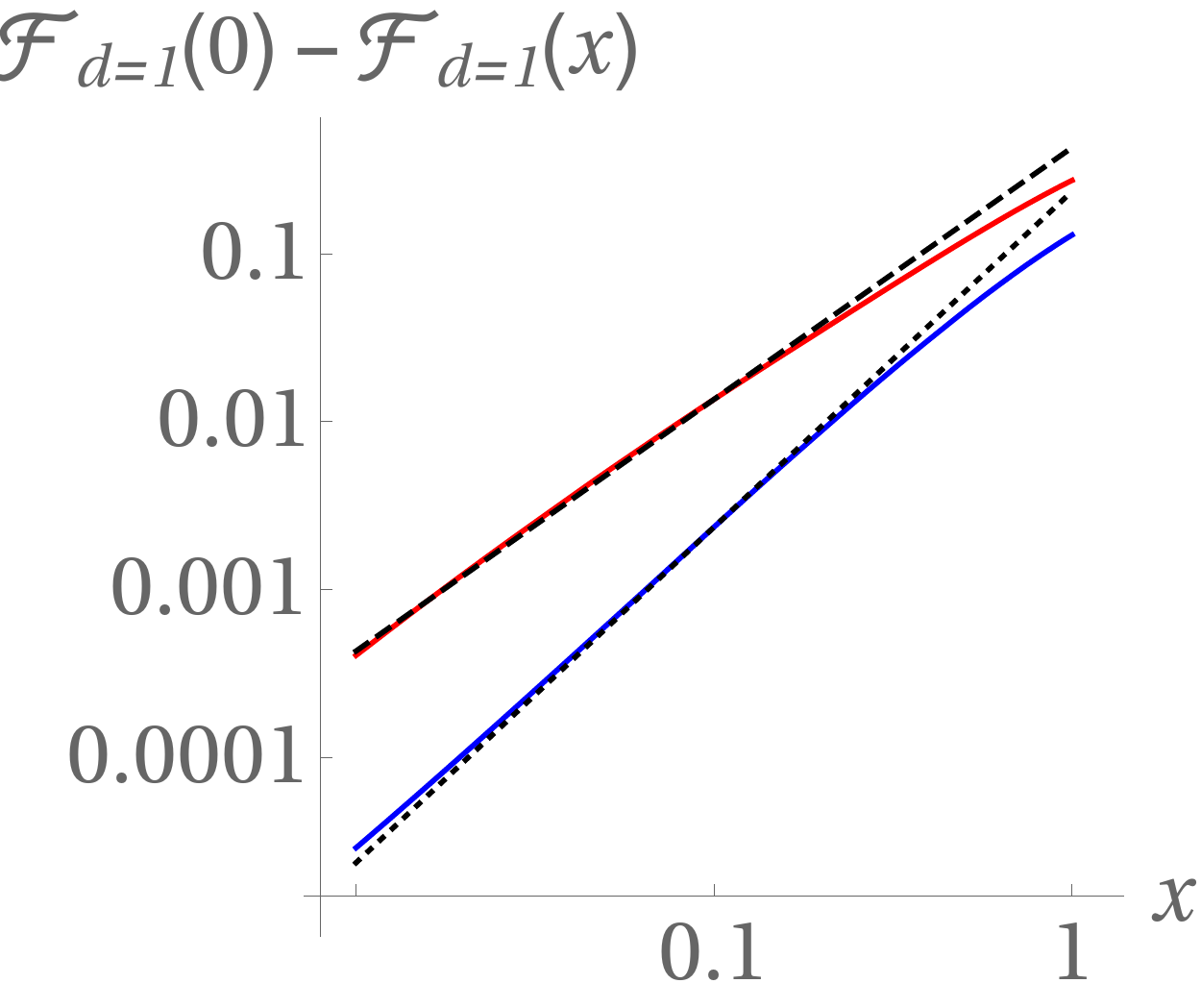} 
\includegraphics[width=0.25\textwidth]{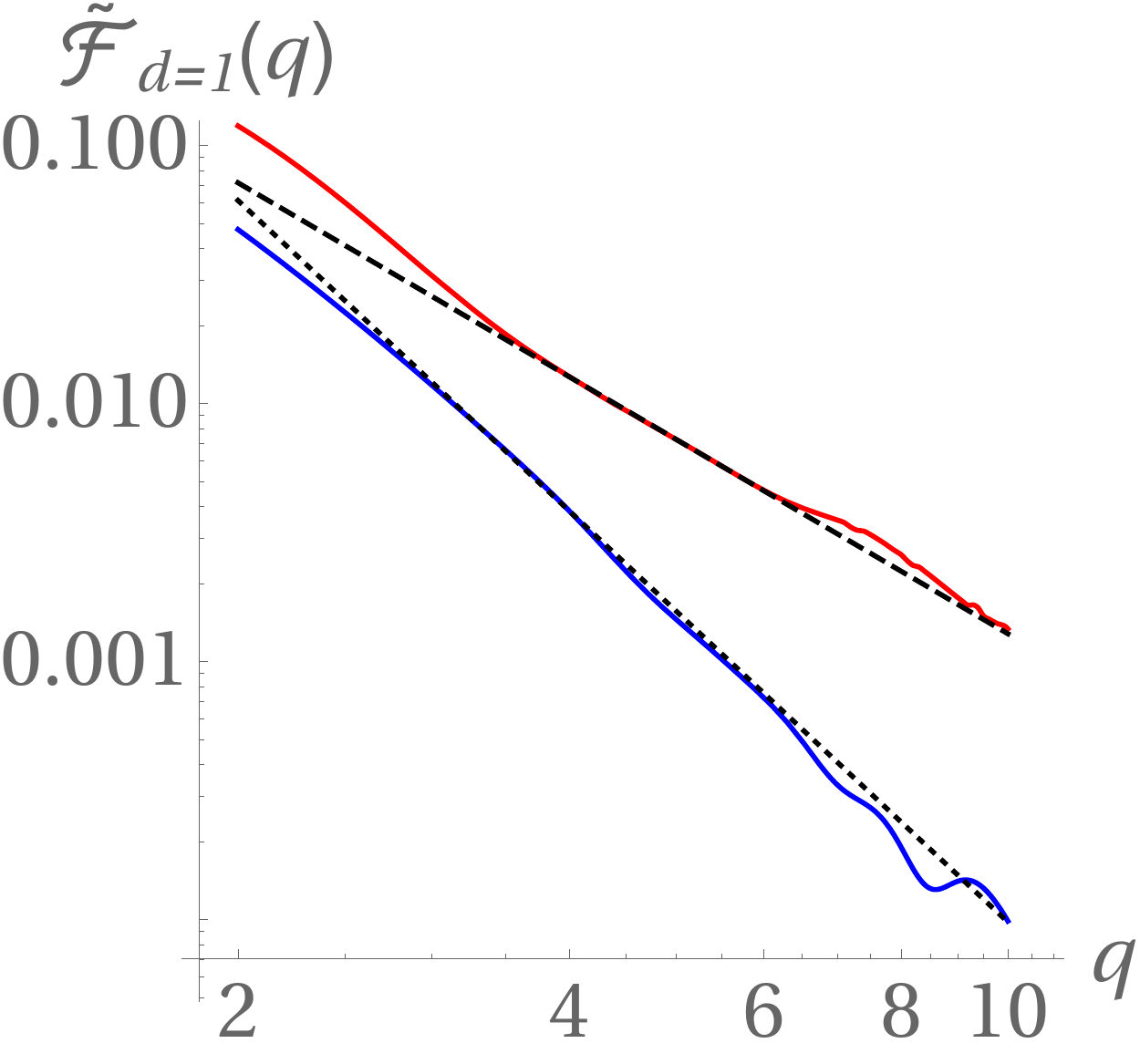}}
 \caption{(color online). Left: (resp. Right:) Log-Log plot of ${\cal F}_{d=1}(0) - {\cal F}_{d=1}(x)$ (resp. $\tilde{{\cal F}}_{d=1}(q)$) numerically obtained in the BFM model (blue) and in the model with SR disorder (red). Dotted lines: guide lines for the BFM result $x^2$ (left) and $1/q^4$ (right). Dashed lines: $x^{1.5}$ (left) and $1/q^{2.5}$ (right). These 
 results are consistent with (i) the exact result $\tilde{\eta}_{d=1} =4$ for the BFM
 (ii) $\tilde{\eta}_{d=1} \simeq 2.5$ for the SR disorder model (in between the guess $\tilde{\eta}_{d=1}^{{\rm guess}} = d+\zeta \simeq 2.25$ and our $O(\epsilon)$ prediction $\tilde{\eta}_{d=1} \simeq 8/3 \simeq 2.66$).}
\label{fig:resSimuSmallxLargeq}
\end{figure}

In dimension $d=1$ we use a system of size $L=2048$ discretized with $N=L$ points and a mass $m=10/L$.  In Fig.~\ref{fig:resSimuShapesd1} we show our results for the mean-shape for different values of $S$ and compare with our theoretical predictions using the predicted value of $\ell_{\sigma}$ (deduced from the measurement of $S_m$), hence with no fitting parameter. The results for the BFM are excellent. For the model with SR disorder, the improvement brought by the $O(\epsilon)$ correction is substantial. If one instead uses a measurement of $\ell_{\sigma}$ by e.g. setting the value of the shape at the origin, the agreement with the SR disorder model is, to the naked eye, almost perfect. We also measure properties independent of the value of $\ell_{\sigma}$: (i) in  Fig.\ref{fig:resSimuSmallxLargeq} the small $x$ and large $q$ behaviors (ii) the universal ratios $c_{p}$. We obtain $(c_1,c_2) \simeq (1.699 \pm 0.003,3.83 \pm 0.05)$ for the BFM and 
$(c_1,c_2) \simeq (1.612 \pm 0.004 , 3.16 \pm 0.03)$ for the model with SR disorder (error-bars are $3$ sigma estimates).
The above predictions are in perfect agreement for the BFM, and our $O(\epsilon)$ corrections
go in the right direction for the SR disorder case.

To conclude, we introduced an original way of characterizing the mean shape of an avalanche by centering  around its seed. We obtained theoretical predictions for this observable and confronted them to numerical simulations. We also proposed a protocol to measure it. We hope that this work stimulates measurements of this quantity in numerical setups and imaging experiments.

\acknowledgments
We thank Matthieu Wyart for interesting discussions. We acknowledge support from
PSL grant ANR-10-IDEX-0001-02-PSL.

\newpage

\newpage

\begin{widetext}

\begin{center}
{\Large Supplemental Material}
\end{center}

We give here a %succinct 
derivation of the results presented in the main text of the letter and details on the numerical simulations. 
%Additional details and results will be presented in \cite{usLongVersion}.

\begin{center}
{\bf \large Dynamical Field Theory Setting}
\end{center}

Here we first introduce the formalism used to derive the results presented in the letter.

{\bf Equation of motion and dynamical action}

As written in the main text, we consider the equation of motion for the over-damped dynamic of an elastic interface of internal dimension $d$ in a quenched random force field and driven by a parabolic well of position $w_{xt}$
\begin{equation} \label{SMeom}
\eta \partial_t u_{xt} = \nabla^2_x u_{xt} - m^2( u_{xt}-w_{xt} ) + F(u_{xt},x)
\end{equation}
where $x \in \mathbb{R}^d$, $tÃÂ \in \mathbb{R}$, $u_{xt} \in  \mathbb{R}$ (the space-time dependence is indicated by subscripts). The elastic-coefficient as been set to unity by a choice of units. In this formulation, the driving force of the parabolic well is $f_{xt} = m^2(w_{xt}-u_{xt})$. The pinning force $F(u,x)$ is chosen centered, Gaussian with second cumulant $\overline{F(u,x)F(u',x')} = \delta^d (x-x') \Delta_0(u-u')$ (the overline denotes the average over disorder) where $\Delta_0(u)$ is a short-ranged function. Higher cumulant can also exist (i.e. non Gaussian force, and are taken into account in the FRG treatment). Note that here we have written the case of short-ranged (SR) elasticity with an elastic term of the form $\nabla^2_x u_{xt}$. Other elastic kernels can also be considered, by changing
\bea
\nabla^2_x u_{xt} - m^2 u_{xt} \to \int_{x'} g_{xx'}^{-1} u_{x't}
\eea
where $g_{xx'}^{-1}$ is a translationally invariant ($g_{xx'}^{-1}= g_{x-x'}^{-1}$) elastic kernel. In particular, we will consider the following kernel (here written in Fourier space) ($ g_q^{-1} = \int_x e^{i q x} g_{x}^{-1}$, here and throughout the rest of the Supplemental Material $\int_x = \int_{x \in \mathbb{R}^d} d^d x$ and $\int_q = \int_{q \in \mathbb{R}^d} \frac{d^d q}{(2 \pi)^d}$)
\bea \label{SM:LRkernel}
g_q^{-1} = \sqrt{\mu^2 +q^2}
\eea
which is known to be relevant in the description of standard long-ranged (LR) elasticity. In this situation, the parameter $\mu$ is related to the mass $m$ as $m=\sqrt{\mu}$. In most of the following, we will deal with the SR elasticity case, and explicitly mention when we consider the LR one. Introducing a response field $\tilde{u}_{xt}$, the generating function of the velocity field $G[ \lambda_{xt} ] = \overline{ e^{\int_{xt} \lambda_{xt} \dot{u}_{xt} }}$ is computed using the dynamical action formalism for the velocity theory, that is for the time-derivative of (\ref{SMeom}) \cite{Jannsen,MSR}:
\bea \label{MSR}
&& G[ \lambda_{xt} ] = \int D[ \tilde u] D[ \dot{u}] e^{ \int_{xt} \lambda_{xt} \dot{u}_{xt}  + m^2  \int_{xt} \tilde{u}_{xt} \dot{w}_{xt}  - S_0 -S_{dis}} ÃÂ \nn \\
&& S_0 = \int_{xt} \tilde{u}_{xt} (\eta \partial_t - \nabla^2 + m^2) \dot{u}_{xt}  \quad , \quad  S_{dis} = -\frac{1}{2} \int_{xtt'} \tilde{u}_{xt} \tilde{u}_{xt'} \partial_t \partial_{t'} \Delta_{0}(u_{xt} - u_{xt'}) 
\eea
\medskip

{\bf The renormalized field theory}

 As discussed in \cite{LeDoussalWiese2012a}, in the limit of small $m$, and in the quasi-static limit $\dot{w}_{xt} = v \to 0^{+}$, universal quantities associated to the motion inside a single avalanche can be computed in an expansion in $\epsilon = 4-d$ using an effective action identical to (\ref{MSR}) with the replacement $\Delta_0(u) \to \Delta(u) = \Delta(0) - \sigma|u|- 4 \pi^2 \alpha m^{4-d} u^2 + O(\epsilon^2)$, where $\sigma$ and $\alpha = O(\epsilon)$ are renormalized quantities. $\sigma$ is a non-universal parameter whose value is related to the two first moments of the avalanche size distribution through the exact relation $2 \sigma/m^4 = \langle S^2 \rangle/\langle S \rangle$. On the other hand $\alpha$ is dimensionless and universal at the FRG fixed point
 with value $\alpha = -  2 \epsilon/9 + O(\epsilon^2)$. In terms of the action, this replacement reads $S_{dis} \to S_{dis}^{eff} = S_{tree} + \delta_{1-loop} S$ with
\bea \label{treeAction}
&& S_{tree} = -\sigma \int_{xt} \tilde{u}_{xt}^2 \dot{u}_{xt} \quad , \quad \delta_{1-loop} S = - 4 \pi^2 \alpha m^{4-d} \int_{xtt'} \tilde{u}_{xt} \dot{u}_{xt} \tilde{u}_{xt'} \dot{u}_{xt'} \label{1loopAction}
\eea
At lowest order in $\epsilon$, the action is $S_{dis}^{eff} = S_{tree}$. Using the renormalized value of $\sigma$, it gives the exact result for universal quantities in $d > 4$. In any dimension, this  tree/mean-field theory also corresponds to an interface slowly driven in a Brownian force landscape: for each $x$, $F(u,x)$ is a Brownian in $u$ independent of the others with $\overline{(F(u',x) - F(u,x))^2} = 2 \sigma |u'-u|$. This is the Brownian Force Model (BFM). The $O(\epsilon)$ corrections around the BFM are easily computed using the fact that $\delta_{1-loop} S$ can also be taken into account by introducing a fictitious Gaussian centered white noise $\xi_{xt}$ with correlations $\langle \xi_x \xi_{x'} \rangle_{\xi} = 8 \pi^2 \alpha m^{4-d}   \delta^d(x-x')$ through the identity
\bea \label{1loopTrick}
e^{ - S_0 - S_{dis}^{eff} } =  \langle e^{ - \int_{xt} \tilde{u}_{xt} (\eta \partial_t - \nabla^2 + m^2 + \xi_x) \dot{u}_{xt}  - S_{tree}  } \rangle_{\xi} 
\eea
where $\langle \rangle_{\xi}$ denotes the average over $\xi$. One-loop observables are thus rewritten as averaged tree observables in a theory with space-dependent mass $m^2 \to m^2 + \xi_x$. Since $\xi_x =O(\sqrt{\epsilon}) $, the effect of $\xi_x$ can be taken into account pertubatively up to order $O(\xi_x^2)$.

\begin{center}
{\bf \large Avalanches observables}
\end{center}

{\bf Avalanches in non-stationary driving} 

Let us first introduce our avalanche observables in a non-stationary setting. We refer the reader to \cite{DobrinevskiLeDoussalWiese2011b,LeDoussalWiese2012a, ThieryLeDoussalWiese2015} for more details on this procedure. We first prepare the interface is in its quasi-static stationary state ${\dot w}_{xt} \sim v =0^+$, 
then turn the driving off: ${\dot w}_{xt}=0$ and finally wait for the interface to stop at some metastable position. Supposing we are in such a state at $t=0$, we apply to the interface a step in the driving force localized at $x=t=0$, $\dot{f}_{xt} =  m^2 \delta w \delta(x) \delta(t)$ (local kick) and let it evolve. Information about the resulting motion of the interface is encoded in the generating functional $G[\lambda_{xt} ]= \overline{ e^{\int_{x,t>0} \lambda_{xt} \dot{u}_{xt} }}$. Remarkably, since the action (\ref{treeAction}) (written at one-loop in terms of $\xi_x$ (\ref{1loopTrick})) is linear in $\dot{u}_{xt}$, the evaluation of $G[\lambda_{xt} ]$ through the path-integral formalism simplifies. The integration on the velocity field $\dot{u}_{xt}$ leads to a delta functional and to the result:
\bea \label{SM:gen00}
G[\lambda_{xt} ] = \langle  e^{m^2 \delta w \tilde{u}_{x=t=0}^{\lambda,\xi}} \rangle_{\xi}
\eea
where $\tilde{u}_{xt}^{\lambda,\xi}$ is the solution of the so-called instanton equation:
\bea \label{SM:instanton1}
&& \partial_t \tilde{u}_{xt}  + \nabla^2 \tilde{u}_{xt}  - (1+\xi_x)\tilde{u}_{xt}  + \tilde{u}_{xt}^2 + \lambda_{xt} =0
\eea
here written in dimensionless units using the variables $\tilde{u}_{x} = \frac{m^2}{\sigma} \hat{\tilde{u}}_{\hat x} $, $x = \hat x/m$, %$\mu = \frac{m^4}{\sigma} \hat \mu$, 
$t = \frac{\eta}{m^2} \hat t$, $\lambda_{xt} = \frac{m^{4}}{\sigma}  \hat \lambda_{\hat x \hat t}$, 
%$\alpha =  m^{4-d} \hat \alpha$ (thus $\hat \alpha = -2\epsilon/9$) 
and omitting the hats in what follows, to lighten notations. The boundary conditions is $\tilde{u}_{xt}=0$ for $t=+\infty$. Here we will only be interested in single avalanche, defined as the response of the interface to an infinitesimal step in the force. We introduce the generating functional $Z[\lambda_{xt} ]$ as (expanding (\ref{SM:gen00}) in $\delta w$):
\bea \label{SM:genZ}
&& \overline{ e^{\int_{x,t>0} \lambda_{xt} \dot{u}_{xt} } -1} = \delta w Z[\lambda_{xt} ] + O(\delta w^2) \nn \\
&& Z[\lambda_{xt} ] =  m^2 \langle \tilde{u}_{x=t=0}^{\lambda,\xi} \rangle_{\xi}
\eea
In the above expansion, the $\delta w$ factor just accounts for the probability to trigger an avalanche at $t=x=0$. Introducing $\rho_{t=x=0}[\dot{u}_{xt}]$, the density of velocity field $\dot{u}_{tx}$ inside an avalanche that starts at $t=x=0$, we write
\bea
Z[\lambda_{xt} ] = \int  D[ \dot{u}]\left(  e^{\int_{xt} \lambda_{xt} \dot{u}_{xt}}  -1 \right)  \rho_{t=x=0}[\dot{u}_{xt} ] \ ,
\eea
where here this equation can actually be viewed as a definition of the density $\rho_{t=x=0}$. The fact that these definitions indeed correspond to what is usually meant by avalanches in the quasi-static limit is discussed below. This formulation is up to now completely general. Let us now focus on two types of sources: $\lambda^1_{xt} = ( -\mu  + \lambda \delta(x-y) \delta(t-s)) \theta(t)$ and  $\lambda^2_{xt} = ( -\mu  + \lambda \delta(x-y)) \theta(t)$ ($\theta(.)$ denotes the Heaviside theta function). In both cases, the $\mu$ variable probes the total size of the avalanche $S= \int_{x,t>0} \dot{u}_{xt}$. In the first case, $\lambda$ probes the local velocity at $t=s$ and $x=y$ during the avalanche. In the second case, $\lambda$ probes the local size of the avalanche at $x=y$, $S_y = \int_{t>0} \dot{u}_{yt}$. We write the associated generating function $Z^{(1)}[\lambda_{xt}^{1}] =Z^{(1)}(\mu , \lambda , y,s)$  and $Z^{(2)}[\lambda_{xt}^{2}] = Z^{(2)}(\mu , \lambda , y)$. These are obtained through the formula (\ref{SM:genZ}) by solving (\ref{SM:instanton1}) which leads to
\bea
Z^{(1)}(\mu , \lambda , y,s) = \int dS d\dot{u}_{ys} e^{ - \mu S + \lambda \dot{u}_{ys}} \rho_{t=x=0}^{(1)}(S , \dot{u}_{ys}) \quad , \quad Z^{(2)}(\mu , \lambda , y) = \int dS dS_y e^{ - \mu S + \lambda S_y } \rho_{t=x=0}^{(2)}(S , S_y),
\eea
where $\rho^{(1)}_{t=x=0}(S , \dot{u}_{ys})$ (resp. $\rho^{(2)}_{t=x=0}(S , \dot{u}_{ys})$) is the joint density of total size $S$ and velocity field $\dot{u}_{ys}$ (resp. of total size $S$ and local size $S_y$) for avalanches starting at $t=x=0$. In practice we will only be interested in computing the mean velocity-field inside avalanche of total size $S$, $\langle \dot{u}_{ys} \rangle_S$ (resp. the mean local size inside avalanche of total size $S$, $\langle S_y \rangle_S$). These are computed as
\bea \label{SM:gen2}
\langle \dot{u}_{ys} \rangle_S = \frac{LT^{-1}_{\mu \to S} \partial_{\lambda} Z^{(1)}|_{\lambda=0} }{\rho(S)/L^d} \quad , \quad \langle S_y \rangle_S = \frac{LT^{-1}_{\mu \to S} \partial_{\lambda} Z^{(2)}|_{\lambda=0} }{\rho(S)/L^d}  = \int_{s=0}^{\infty}{ds} \langle \dot{u}_{ys} \rangle_S
\eea
where $LT^{-1}_{\mu \to S}$ denotes the Inverse Laplace Transform (ILT) operation $LT^{-1}_{\mu \to S} = \frac{1}{2 i \pi} \int_{{\cal C}}d\mu e^{\mu S}$ with appropriate contour of integration, and we have introduced $\rho(S)$ the density of avalanches of total size $S$, previously computed up to one-loop in 
\cite{LeDoussalWiese2008c,LeDoussalWiese2011b,LeDoussalWiese2012a} ($\rho(S)/L^d = \int d \dot{u}_{ys} \rho_{t=x=0}^{(1)}(S , \dot{u}_{ys}) =\int dS_y \rho_{t=x=0}^{(2)}(S , S_y)$ is the density of avalanches of total size $S$ starting at $x=0$). For the observables we are interested in, we will thus only need to solve (\ref{SM:instanton1}) at first order in $\lambda$.

\medskip

{\bf Link with the stationary driving} 

Let us now present here how the precedent approach is linked to avalanches occurring in the quasi-static stationary state of the interface dynamic $\dot{w}_{xt}= v \to 0^+$. We introduce $\rho_0$ the mean density of avalanche per unit of driving and $p[\dot{u}_{tx}]$ the (functional) probability of velocity field $\dot{u}_{tx}$ inside an avalanche. At first order in $v$, the generating function $G[\lambda_{xt} ]= \overline{ e^{\int_{xt} \lambda_{xt} \dot{u}_{xt} }}$ can be written as
\bea \label{SM:gen3}
G[\lambda_{xt} ] = (1 - \rho_0 v T) + \rho_0 v T \int  D[ \dot{u}] e^{\int_{xt} \lambda_{xt} \dot{u}_{xt}} p[\dot{u}_{xt} ] + O(v^2) = 1 + v T \int  D[ \dot{u}]\left(  e^{\int_{xt} \lambda_{xt} \dot{u}_{xt}}  -1 \right)  \rho[\dot{u}_{xt} ] + O(v^2) 
\eea
where we reintroduced $\rho[\dot{u}_{xt} ] = \rho_0 p[\dot{u}_{tx}] $ the density of velocity field $\dot{u}_{tx}$ inside an avalanche. The equation (\ref{SM:gen3}) can be seen as a definition of what is meant by avalanches in the quasi-static setting. The time scale $T$ that appears in (\ref{SM:gen3}) should be much larger than the time-scale of avalanche motion (to allow the avalanche to terminate) and much smaller than the typical waiting time between avalanches. This only works if $\lambda_{xt}$ is also non-zero in a time window smaller than $T$: this ensures that the measurement made on the velocity-field is also inside a single-avalanche. On the other hand, the small velocity expansion made directly on the action (\ref{MSR}) and compared to (\ref{SM:gen3}) gives
\bea \label{SM:gen4}
G[ \lambda_{xt} ] =1+ v \langle m^2 \int_{xt } \tilde{u}_{xt}  \rangle_{\lambda_{xt}} \longrightarrow  \int  D[ \dot{u}]\left(  e^{\int_{xt} \lambda_{xt} \dot{u}_{xt}}  -1 \right)  \rho[\dot{u}_{xt} ] = \int_{xt }  \frac{m^2}{T} \langle \tilde{u}_{xt}  \rangle_{\lambda_{xt}}  , 
\eea
where here the average $\langle , \rangle_{\lambda_{xt}}$ refers to the average with respect to the dynamical action (\ref{MSR}) with source $\lambda_{xt}$. In the right of (\ref{SM:gen4}), the integral over time and space originates from the fact that we have consider the effect of avalanches starting at any point of the interface, and at any time in the time-window $T$. From a field-theory point of view, it is then natural to interpret $m^2 \langle \tilde{u}_{x=t=0}  \rangle_{\lambda_{xt}}$ as the contribution from avalanches starting at $t=x=0$ (diagrams entering into $\langle \tilde{u}_{x=t=0}  \rangle_{\lambda_{xt}}$ can only have a first non-zero $\dot{u}_{xt}$ at $x=0$). Furthermore, this is supported by the non-stationary setting in which this interpretation is immediate. In the quasi-static setting we can only a priori consider sources $\lambda_{xt}$ non-zero in time windows smaller than $T$ to make sure that only one avalanche is taken into account. However, from a practical point of view, when $T >> \tau_m$ where $\tau_m$ is the typical time scale of avalanches, both descriptions give exactly the same result as detailed in
\cite{LeDoussalWiese2012a,PLDInprep}.

\begin{center}
{\bf \large Calculation in the BFM}
\end{center}

{\bf  Mean-velocity field inside an avalanche in the BFM}

Here we present the calculations leading to the resuts Eq.(\ref{SpatioTemp}) and Eq.(\ref{SpatioTempLR}) of the letter for the mean-velocity field inside avalanche of total size $S$ in the BFM $\langle \dot{u}_{ys} \rangle_S $ (denoted $v(y,s)$ in the main text with $y=x$ and $s=t$). We have to solve to first order in $\lambda$ the instanton equation
\bea
\partial_t \tilde{u}_{xt}  + \nabla^2 \tilde{u}_{xt}  - \tilde{u}_{xt}  + \tilde{u}_{xt}^2 -\mu + \lambda \delta(x-y) \delta(t-s) =0 \ .
\eea
Note that here, in dimensionless units, time and avalanche size are measured in terms of the natural units of avalanches motion $\tau_m = \eta/m^2$ and $S_m = \sigma/m^4$. The perturbative solution is $\tilde{u}_{xt} = \tilde{u}_{xt}^0 + \tilde{u}_{xt}^1 \lambda + O(\lambda^2)$ with
\bea
&& \tilde{u}_{x}^{0} = Z(\mu)=\frac{1}{2} \left( 1 - \kappa^2( \mu) \right) \quad , \quad \kappa(\mu) = (1+4 \mu)^{\frac{1}{4}} \quad , \quad \tilde{u}_{qt}^1 = - \int_{t'=+\infty}^{t} e^{(q^2 + \kappa^2(\mu))(t-t')  + i q y } \delta(t'-s)dt'
\eea
here written in Fourier space for the $O(\lambda)$ part: $\tilde{u}_{qt}^1 = \int_x e^{iqx} \tilde{u}^1_{xt}$. This immediately gives
\bea \label{solBFM}
\tilde{u}^1_{t=x=0} = \int_q e^{iqy - (q^2 + \kappa^2(\mu))s }
\eea
Using the tree result for the avalanche size density $\rho^{{\rm MF}}(S) = \frac{L^d}{2 \sqrt{\pi} S^{3/2} } e^{-Ã¢ÂÂS/4}$ we obtain the mean velocity field inside a single avalanche using (\ref{SM:gen2}) as
\bea
< \dot u_{ys} >_{S} = 2 \sqrt{\pi} S^{3/2} e^{S/4} 
 LT^{-1}_{\mu \to S} \int_q e^{i q y - (q^2 + \sqrt{1+4 \mu}) s} = 2 s e^{- \frac{s^2}{S} } \int_q e^{i q y - q^2 s} = 2 s e^{-s^2/S} \frac{1}{(4 \pi s)^{d/2}} e^{-y^2/(4 s)} 
\eea 
In the notation of the main text, we thus obtain (\ref{SpatioTemp}) that we recall here
\bea \label{SM:SpatioTemp}
< v(x,t)>_{S} = S^\frac{2-d}{4} F(t/S^{1/2},x/S^{1/4}) \quad , \quad  F(t,x) = 2 t e^{-t^2} \frac{1}{(4 \pi t)^{d/2}} e^{-x^2/(4 t)} 
\eea 

\medskip

{\em Extension to LR elasticity}

Following the same computation, one obtains for the case of the BFM with long-ranged elasticity (with the kernel (\ref{SM:LRkernel}))
\bea \label{solBFMLR}
\tilde{u}_{t=x=0} = \int_q e^{iqy - (\sqrt{1+q^2} -1+ \kappa^2(\mu))s }
\eea
And thus
\bea 
< \dot u_{ys} >_{S} = 2 s e^{- \frac{s^2}{S} } \int_q e^{i q y - (\sqrt{1+q^2} -1 ) s}
\eea
Note that here, the spatio-temporal shape does not satisfy the expected scaling form (\ref{SpatioTemp2}), $< \dot u_{ys} >_{S} = S^\frac{2-d-1}{2} F(s/S^{\frac{1}{2}},y/S^{\frac{1}{2}})   $ for all $S$. This should not be surprising, it is known that the present theory describes scale-invariant avalanches only for $S \ll S_m$ (here $S_m =1$ in dimensionless units is the large scale cutoff $S_{\rm max}$ mentioned in the main text, and note that in our theory the low-scale cutoff on the scaling regime $S_{\rm min}$ also mentioned in the main text can effectively be taken to $0$ for shape observables). The fact that the scaling hypothesis for the mean velocity field holds $\forall S$ in the BFM with short-ranged elasticity is the true surprise. Scaling in the long-ranged model is restored at small $S$ and here
\bea \label{SM:LRspatioTemp1}
F(s,y) = \lim_{S \to 0}  S^\frac{d-1}{2} < \dot u_{S^{\frac{1}{2}} y, S^{\frac{1}{2}} s} >_{S} = 2 s e^{-s^2 } \int_q e^{i q y - |q| s }
\eea
Evaluating this integral in dimension $1$ immediately leads to the result (\ref{SpatioTempLR}).

{\bf The mean shape of avalanches in the BFM: results in Fourier space}

We now derive the result Eq.(\ref{MFq}) of the letter. Using (\ref{solBFM}), we immediately obtain the mean-shape of avalanche in Fourier space in the BFM as
\bea
\tilde{{\cal F}}^{{\rm MF}}(q) = \int_{s=0}^{\infty} 2 s e^{- s^2 - q^2 s}= 1 - \frac{ \sqrt{\pi} q^2 }{ 2} e^{\frac{q^4}{4}} \text{erfc}\left(\frac{q^2}{2}\right)
\eea
i.e. the result (\ref{MFq}) of the main text. Note that here avalanche sizes have been expressed in units of $S_m = \sigma/m^4$ and distances in units of $1/m$. Hence the non-universal scale $\ell_{\sigma}$ of the main text is indeed $\ell_{\sigma} = \frac{1}{m} S_m^{-1/4} = \sigma^{-1/4}$. Let us give here the large and small momenta behavior of $\tilde{{\cal F}}^{{\rm MF}}(q) $:
\bea
&&  \tilde{{\cal F}}^{{\rm MF}}(q) =_{q \gg 1} \frac{2}{q^4}-\frac{12}{q^8}+\frac{120}{q^{12}} + O(\frac{1}{q^{16}})\\
&&  \tilde{{\cal F}}^{{\rm MF}}(q) =_{q \ll 1} 1-\frac{\sqrt{\pi }
   q^2}{2}+\frac{q^4}{2}-\frac{\sqrt{\pi }
   q^6}{8}+O\left(q^8\right)
\eea

{\it Extension to LR elasticity}

We now compute the mean shape in real space. In particular we obtain the result Eq.(\ref{MFlargex}) of the letter. The extension of the precedent results to the case of LR elasticity is straightforward. As written in the main text and following the formula (\ref{SM:LRspatioTemp1}), the mean-shape in Fourier space in the scaling regime for LR elasticity is simply obtained from the precedent results by changing $q^2 \to |q|$:
\bea
\tilde{{\cal F}}^{{\rm MF , LR}} (q) = \tilde{{\cal F}}^{{\rm MF }} (\sqrt{q}) .
\eea
In particular it now has an algebraic tail at large $q$ with exponent $1/q^2$, $\tilde{{\cal F}}^{{\rm MF , LR}} (q)  \simeq_{q \gg 1} \frac{2}{q^2}$.

{\bf The mean shape of avalanches in the BFM: results in real space}

In real space, ${\cal F}^{{\rm MF}}_d(x)$ is most simply obtained by integration of (\ref{SM:SpatioTemp}):
\bea \label{SM:FIntegral}
{\cal F}^{{\rm MF}}_d(x) = \frac{2}{(4 \pi)^{d/2}}  \int_0^{+\infty} dt t^{1-d/2} e^{-t^2 - \frac{x^2}{4 t}} 
\eea
This integral can be expressed either as the sum of three series:
\bea \label{serFd} 
&& {\cal F}^{{\rm MF}}_d(x) = \pi^{1-\frac{d}{2}} \sum_{p=0}^\infty (-1)^p 2^{-4 p} 
 [ \frac{a_{p}}{\sin{\frac{d \pi}{4}}}  x^{4 p} -  \frac{a_{p+ \frac{1}{2}}}{4 \cos{\frac{d \pi}{4}}}
 x^{4 p+2} + \frac{b_p}{\sin{\frac{d \pi}{2}}} x^{4-d + 4 p} ]  \\
&& a_p = \frac{  2^{-d}}{(2 p)! \Gamma
   \left(\frac{d}{4}+p\right)} 
   %\quad , \quad  b_p = \frac{- 2^{-d-2}}{(2 p+1)! \Gamma  \left(\frac{d+2}{4}+p\right)} 
   \quad , \quad b_p = \frac{  2^{-3}}{p! \Gamma \left(-\frac{d}{2}+2
   p+3\right)}
\eea 
or, equivalently, as the sum of three generalized hypergeometric functions (corresponding
term by term to the series):
\bea \label{hyperFd} 
&& {\cal F}^{{\rm MF}}_d(x) = \frac{1}{8} \pi ^{1-\frac{d}{2}} \bigg(\frac{2^{3-d} \csc \left(\frac{\pi  d}{4}\right)
   \, _0F_2\left(;\frac{1}{2},\frac{d}{4};-\frac{x^4}{64}\right)}{\Gamma
   \left(\frac{d}{4}\right)} -\frac{2^{1-d} x^2 \sec \left(\frac{\pi  d}{4}\right) \,
   _0F_2\left(;\frac{3}{2},\frac{d}{4}+\frac{1}{2};-\frac{x^4}{64}\right)}{\Gamma
   \left(\frac{d+2}{4}\right)}
      \\
      && +\frac{x^{4-d} \csc \left(\frac{\pi  d}{2}\right) \,
   _0F_2\left(;\frac{3}{2}-\frac{d}{4},2-\frac{d}{4};-\frac{x^4}{64}\right)}{\Gamma
   \left(3-\frac{d}{2}\right)} \bigg) \nonumber 
\eea 
The expressions (\ref{serFd}) and (\ref{hyperFd}) are adequate for $d=1,3$. 
For $d=2,4$ one must first take the limit $d \to 2,4$ before evaluating. This is easy to do
with mathematica, and we give here only the two leading terms at small $x$:
\bea
&&{\cal F}^{{\rm MF}}_2(x) = \frac{1}{4 \sqrt{\pi }}-\frac{x^2 (-4 \log (x)-3 \gamma_E +2+\log (16))}{16 \pi }+O\left(x^3\right) \\
&& {\cal F}^{{\rm MF}}_4(x) = \frac{-4 \log (x)-3 \gamma_E +\log (16)}{16 \pi ^2}+\frac{x^2}{32 \pi
   ^{3/2}}+O\left(x^3\right)
\eea

For $d<4$ the value at zero is finite:
\bea
&& {\cal F}^{{\rm MF}}_d(0) = \frac{2^{-d} \pi ^{1-\frac{d}{2}} }{\Gamma
   \left(\frac{d}{4}\right) \sin \left(\frac{\pi  d}{4}\right)} \\
&&{\cal F}^{{\rm MF}}_1(0) %= \frac{\sqrt{\frac{\pi}{2}}}{\Gamma(\frac{1}{4})} 
\approx 0.345684
\quad , \quad  {\cal F}^{{\rm MF}}_2(0) % = \frac{1}{4 \sqrt{\pi}} 
\approx 0.141047
\quad , \quad {\cal F}^{{\rm MF}}_4(0) % = - \frac{1}{\sqrt{2 \pi} \Gamma(- \frac{1}{4})} 
\approx 0.0813891
\eea 
and ${\cal F}^{{\rm MF}}_d(0)$ diverges as $\frac{1}{4 \pi^2 \epsilon}$ as $d \to 4^-$ (it has a minimum near
$d=3.2$). For $d>4$ it diverges near zero as ${\cal F}^{{\rm MF}}_1(x) \simeq \frac{\pi ^{1-\frac{d}{2}}  \csc \left(\frac{\pi  d}{2}\right)}{8 \Gamma
   \left(3-\frac{d}{2}\right)} x^{4-d}$. The large distance behavior is easily obtained from the saddle-point method on (\ref{SM:FIntegral}). It yields 
a stretched exponential decay at large $x$ with exponent $4/3$, independent of $d$:
\bea
{\cal F}^{{\rm MF}}_d(x) \simeq \frac{2^{-d/2} \pi ^{\frac{1}{2}-\frac{d}{2}} 
   }{\sqrt{3}} x^{\frac{2-d}{3}} e^{-\frac{3 x^{4/3}}{4}}
    \eea

{\it Extension to LR elasticity}

We did not attempt to find expressions for the mean-shape in real space for LR elasticity in any $d$. In the most experimentally relevant case of $d=1$ however it takes a simple expression: integrating (\ref{SpatioTempLR}) from $t=0$ to $t= \infty$ leads
\bea
{\cal F}^{{\rm MF , LR}}_{d=1} (x) = \frac{1}{\sqrt{\pi }}-|x| e^{x^2}  \text{erfc}(|x|) \ .
\eea
We note in particular the behavior around $x=0$, ${\cal F}^{{\rm MF , LR}}_{d=1} (x)  =_{x \ll 1}  \frac{1}{\sqrt{\pi }}- |x| +O(x^2)$, reminiscent of the $2/q^2$ tail in Fourier space. At large $x$, the mean-shape now decays algebraically as $ {\cal F}^{{\rm MF , LR}}_{d=1} (x) =_{x \gg 1}  \frac{1}{2 \sqrt{\pi} x^2} + O(1/x^4)$.

\begin{center}
{\bf \large $O(\epsilon)$ corrections}
\end{center}

{\bf ``Brut'' corrections}

At $O(\epsilon)$ we focus directly on the computation of the mean-shape at fixed size $\langle S_y \rangle_S$. We need to solve
\bea
\partial_t \tilde{u}_{xt}  + \nabla^2 \tilde{u}_{xt}  - (1+\xi_x) \tilde{u}_{xt}  + \tilde{u}_{xt}^2 -\mu + \lambda \delta(x-y) =0 \ .
\eea
at order $1$ in $\lambda$ and order $2$ in $\xi_x$. When $\xi_x=0$ (corresponding to the BFM model) this equation was recently solved exactly \cite{DelormeInPrep} to study the joint distribution of total size $S$ and local size $S_y$ in the BFM. Here we will only be interested in its perturbative solution up to first order in $\lambda$ (to study the mean shape) but up to second order in $\xi_x$ (to study $O(\epsilon)$ corrections. We can look for time-independent solution and use a double expansion $\tilde{u}_{x} =\sum_{i=0}^1 \sum_{j=0}^2 \tilde{u}^i_j(x)$ where $\tilde{u}^i_j(x) = O(\lambda^i \xi^{j})$. The observable of interest is ${\cal Z}(\mu,y) = \partial_{\lambda} Z^{(2)}(\mu,y , \lambda)|_{\lambda=0}$ where $Z^{(2)}$ was introduced in (\ref{SM:genZ}). Using $Z^{(2)}(\mu,y , \lambda) = m^2 \langle \tilde{u}_{x=0}\rangle_{\xi}$ we obtain (in dimensionless units)
\bea
{\cal Z}(\mu, y) = {\cal Z}^{{\rm MF}} (\mu, y) + \delta {\cal Z} (\mu, y) \quad , \quad {\cal Z}^{{\rm MF}} (\mu, y) = \tilde{u}^1_0 (x=0) ÃÂ \quad, \quad \delta {\cal Z} (\mu, y) = \langle \tilde{u}^1_2 (x=0) \rangle_{\xi} 
\eea
These are most simply expressed in Fourier space $\tilde{{\cal Z}}(\mu, q) = \int_x e^{i q y} {\cal Z}(\mu, y) $ and we find
\bea \label{SM:1loop1}
&& \tilde{{\cal Z}}^{{\rm MF}}(\mu, q) = G_q(\mu) = \frac{1}{q^2 + \kappa^2(\mu) } \nn \\
&& \delta \tilde{{\cal Z}} (\mu, q)  = 8 \pi^2 \alpha (G_q(\mu))^2  \left( \int_{p } G_p(\mu) \left(    1 +2 Z(\mu)  G_{p-q}(\mu)  \right)^2  + 2 G_0(\mu)  \int_{p }( 1 +   Z(\mu)  G_{p}(\mu))  Z(\mu)  G_{p}(\mu) \right)
\eea
where we have introduced the response function $G_q(\mu)$, a dressed version of the elastic kernel $g_q = \frac{1}{m^2 + q^2} $.

\medskip

{\bf Counter-terms}

The result for $ \delta \tilde{Z} (\mu, q) $ is not yet complete: the integrals present in (\ref{SM:1loop1}) diverge
at large $q$ for $d<4$. This is a usual feature of one-loop computations in field theory. As detailed in \cite{LeDoussalWiese2012a}, when doing a pertubative calculation in (\ref{1loopAction}), one has to take into account a renormalization of $\sigma$ and $m^2$ (the latter being in fact an artifact due to the utilization of the oversimplified one-loop action (\ref{1loopAction})). For clarity let us now denotes $\sigma_0$ and $m_0^2$ the parameters used so far in the perturbative calculation. These are renormalized as $\sigma_0 \to \sigma = \sigma_0 + \delta \sigma$ and $m^2_0 \to m^2 = m^2_0 + \delta m^2$ with
\begin{eqnarray} \label{ct}
\delta \sigma = 24 \pi^2 \alpha  \int_{k} g_{k}^2 \quad , \quad \delta m^2 = - 8 \pi^2 \alpha \int_k g_{k}
\end{eqnarray}
where $g_k = \frac{1}{k^2 + m_0^2}$ is the bare propagator. The parameters entering in (\ref{ct}) are either the bare parameters or the renormalized parameters (these choices differ from a term of order $O(\epsilon^2)$). The fact that the theory is renormalizable imply that divergences present in (\ref{SM:1loop1}) should disappear when expressing the results in terms of renormalized parameters. Let us thus denote $\{ K_0 \} := \{ \sigma_0 , m^2_0 \}$ the set of important couplings and emphasize the dependance of $\tilde{{\cal Z}}(\mu ,q)$ by momentarily adopting the simple notation $\tilde{{\cal Z}}(\{K_0 \})$. Rewriting the result $\tilde{{\cal Z}}(\{K_0 \})$ in terms of the renormalized coupling $\{K \}$ leads to the definition of the counter-terms $\delta_{c.t.} \tilde{Z}(\{K\})$ as
\bea
\tilde{{\cal Z}}(\{K_0 \}) = \tilde{{\cal Z}}(\{K - \delta K \} ) = \tilde{{\cal Z}}^{{\rm MF}}(\{K \}) + \delta_{c.t.} \tilde{{\cal Z}}(\{K\}) + \delta \tilde{{\cal Z}} (\{KÃÂ \})  +O(\epsilon^2)
\eea
and thus $\delta_{c.t.} \tilde{{\cal Z}}(\{K\}) =  -\frac{\partial  \tilde{{\cal Z}}^{{\rm MF}}(\{ K \} )  }{ \partial K_{\alpha} } \delta K_{\alpha} $. To compute these partial derivatives, we reintroduce the original units of the problem in $\tilde{{\cal Z}}^{{\rm MF}}(\{K \}) $:
\bea
\tilde{{\cal Z}}^{{\rm MF}}(\{K \}) %= \frac{m^2}{\sigma} m^{-d}  \frac{\sigma}{m^{4-d}}\frac{e^{i q y}}{ \kappa^2(\mu \frac{\sigma}{m^4} ) + (\frac{q}{m})^2 } 
= \frac{e^{i q y}}{ q^2 + \sqrt{1+ 4 \sigma \mu/m^4}}
\eea
The $\frac{m^2}{\sigma}$ comes from the rescaling of $\tilde{u}$, the $m^{-d}$ from the rescaling of the Fourier Transform and the $ \frac{\sigma}{m^{4-d}}$ from the rescaling of $\lambda$. Computing the derivatives with respect to $\sigma$ and $m^2$ and going back to dimensionless units leads to the following expression for the counter terms:

\bea \label{oneloopct}
\delta_{c.t.} \tilde{{\cal Z}}(\mu ,q) =  8 \pi^2 \alpha  e^{i q y} G_{q=0}(\mu) G_q(\mu)^2  (6  \mu 
  \int_{k} g_{k}^2   -  \int_k g_{k} )
\eea

%\bea \label{oneloopct}
%\delta_{c.t.}^{\sigma} \tilde{Z}(\{K\}) =   2 \mu e^{i q y} G_{q=0}(\mu)  G_q(\mu)^2 \delta \sigma  \quad , \quad 
%\delta_{c.t.}^{m^2} \tilde{Z}(\{K\}) = e^{i q y} G_{q=0}(\mu)  G_q(\mu)^2   \delta m^2
%\eea

It is then easy to check that adding (\ref{oneloopct}) to (\ref{SM:1loop1}) indeed regularizes the result. The computation of the resulting, convergent integrals in $d=4$ leads to the full result for the one loop correction $\delta \tilde Z(\mu,q) \to \delta \tilde Z(\mu,q)+ \delta_{c.t.} \tilde Z(\mu,q)$ with

\bea \label{SM:1loop3}
\delta \tilde {\cal Z}(\mu,q)= \alpha (G_q(\mu))^2  \left( \frac{\left(1+6\mu \right) \log (1-2 Z)+2 Z}{2 (1 - 2 Z)}  + 4 Z \left( 1 + \sinh^{-1}( \frac{q}{2 \sqrt{1-2 Z}}) \frac{Z - (q^2+4(1-2Z))}{q \sqrt{q^2+4(1-2Z)}}  \right) \right) 
\eea
and $Z \equiv Z(\mu)$.

{\bf The mean-shape at $O(\epsilon)$: Laplace transform in Fourier}

We now obtain the result Eq.(\ref{LT1loop}) presented in the letter. Using (\ref{SM:gen2}), the mean-shape in Fourier space is computed as $\langle S(q) \rangle_S = \frac{L^d}{\rho^{{\rm MF}}(S) } LT^{-1}_{\mu \to S} \left( \tilde{{\cal Z}}^{{\rm MF}}(\mu, q) \right)$. To order $O(\epsilon)$, we have $\tilde{{\cal Z}}(\mu, q)  = \tilde{{\cal Z}}^{{\rm MF}}(\mu, q) + \delta \tilde{{\cal Z}}(\mu, q)$. The density $\rho$ was computed to $O(\epsilon)$ in \cite{LeDoussalWiese2011b} with the result $\rho(S) = \rho^{{\rm MF}}(S) + \delta \rho(S)$ with
\bea
\delta \rho(S) = \alpha \rho^{{\rm MF}}(S)  \times \frac{\gamma_E  (S-6)+4 S-8 \sqrt{\pi } \sqrt{S}+(S-6) \log (S)+4}{16
  } 
\eea
$\langle S(q) \rangle_S $ can thus be computed to $O(\epsilon)$ as
\bea \label{SM:Sofq1}
\langle S(q) \rangle_S = \frac{L^d}{\rho^{{\rm MF}}(S) } LT^{-1}_{\mu \to S} \left( \tilde{{\cal Z}}^{{\rm MF}}(\mu, q) \right) - \frac{L^d \delta \rho(S)}{(\rho^{{\rm MF}}(S))^2} LT^{-1}_{\mu \to S} \left( \tilde{{\cal Z}}^{{\rm MF}}(\mu, q) \right) +  \frac{L^d}{\rho^{{\rm MF}}(S) } LT^{-1}_{\mu \to S} \left(\delta \tilde{{\cal Z}}(\mu, q) \right) + O(\epsilon^2) \ .
\eea
One can check that the $O(\epsilon^0)$ part of this result allows to retrieve directly the result of the precedent section for the mean-shape (i.e. without computing $\langle v(x,t) \rangle_S$ first), so that everything is consistent. A new difficulty (compared to the BFM case), is that $\langle S(q) \rangle_S$ defined in (\ref{SM:Sofq1}) does not satisfy the scaling form $\langle S(q) \rangle_S = S \tilde{{\cal F}}_d (qS^\frac{1}{d + \zeta} )$ $\forall S$. This is natural: the scaling regime of the problem is for $S \ll S_m$ (here $S_m=1$ in dimensionless units) and the universal shape of avalanches is the one obtained from (\ref{SM:Sofq1}) as $S \to 0$. It is thus obtained here as
\bea \label{SM:Sofq2}
\tilde{{\cal F}}_d (q) = \lim_{S \to 0} \frac{\langle S(q S^\frac{-1}{d + \zeta}) \rangle_S}{S}
\eea
We now compute the $\epsilon$ expansion of (\ref{SM:Sofq2}) using (\ref{SM:Sofq1}). By definition $\tilde{{\cal F}}_d (q) = \tilde{{\cal F}}^{{\rm MF}} (q) + \delta \tilde{{\cal F}}_d (q)$. We also use the one-loop value of $\zeta = \zeta_1 \epsilon$ ($\zeta_1 = 1/3$) and obtain
\bea \label{SM:Sofq3}
\delta \tilde{{\cal F}}_d (q) = \lim_{S \to 0} \epsilon \frac{\zeta_1 -1}{16}  q \log(S) \frac{\partial \tilde{{\cal F}}^{{\rm MF}}}{\partial q} ( q) + L^d \frac{ LT^{-1}_{\mu \to S} \delta \tilde {\cal Z}(\mu,q S^{- \frac{1}{4}} ) }{S \rho^{{\rm MF}}(S)} - \tilde {\cal F}^{{\rm MF}}( q ) \frac{\delta \rho(S)}{  \rho^{{\rm MF}}(S)}
\eea
Let us first look at the second term in (\ref{SM:Sofq3}):
\bea 
L^d \frac{ LT^{-1}_{\mu \to S} \delta \tilde {\cal Z}(\mu,q S^{- \frac{1}{4}} ) }{S \rho^{{\rm MF}}(S)}  && =\frac{L^d}{ S\rho^{{\rm MF}}(S)} \int_{c-i\infty}^{c+i\infty} \frac{d \mu}{2 i \pi } e^{ \mu S}  \delta \tilde {\cal Z}(\mu,q S^{- \frac{1}{4}} ) \nn \\
&&= \frac{L^d e^{-S/4}}{  S \rho^{{\rm MF}}(S)} \int_{c'-i\infty}^{c'+i\infty} \frac{d \mu}{2 i \pi S } e^{ \mu}  \delta \tilde {\cal Z}(-1/4 + \mu/S ,q S^{- \frac{1}{4}} ) \nn \\
&& \simeq_{S << 1}  \alpha LT^{-1}_{\mu \to 1} \left( H(q,\mu)- \frac{3\sqrt{\pi}}{2} \frac{ \sqrt{\mu} \log(S)}{\left(2 \sqrt{\mu }+q^2\right)^2}  + O(S)\right) 
\eea
Where here from the first to the second line we used a change of variables $\mu \to -1/4 + \mu/S$ and then took the limit $S \to 0^+$ of (\ref{SM:1loop3}) to define
\bea \label{SM:H}
H(q,\mu) =  \frac{\sqrt{\mu } \sqrt{\pi} \left(q \left(6 \log \left(2 \sqrt{\mu }\right)-16\right) \sqrt{8 \sqrt{\mu }+q^2}+16 \left(9 \sqrt{\mu }+q^2\right) \sinh ^{-1}\left(\frac{q}{2 \sqrt{2} \sqrt[4]{\mu }}\right)\right)}{2 q \left(2 \sqrt{\mu }+q^2\right)^2 \sqrt{8 \sqrt{\mu }+q^2}}
\eea
Using similar manipulations, the other terms are inserted inside the ILT using the representation
\bea \label{SM:LTMF}
\tilde{{\cal F}}^{{\rm MF}} (q)= LT^{-1}_{\mu \to 1} \left( \frac{2 \sqrt{\pi}}{2 \sqrt{\mu}  + q^2 } \right) \quad , \quad \frac{\partial \tilde{{\cal F}}^{{\rm MF}}}{\partial q} ( q) = LT^{-1}_{\mu \to 1} \left( \frac{-4  \sqrt{\pi} q}{(2 \sqrt{\mu}  + q^2 )^2} \right)  \quad , \quad \frac{\zeta_1 -1}{16} = \frac{3 \alpha}{16 \epsilon}
\eea
This representation shows that the $O(\log(S))$ terms present in (\ref{SM:Sofq3}) cancel and we obtain the result 
\bea \label{SM:LTFq}
 \delta \tilde{{\cal F}}_d (q)  =   \alpha LT^{-1}_{\mu \to 1} \left( -\frac{\sqrt{\pi}}{8} \frac{(4 - 6 \gamma_E)}{\left(2 \sqrt{\mu }+\tilde{q}^2\right)}   +  H(q,\mu)\right)
\eea
which leads to the result (\ref{LT1loop}) in the main text. Note that the result satisfies,
as required from normalization
\bea
\tilde{{\cal F}}_d (q=0) = 1 \quad , \quad \delta \tilde{{\cal F}}_d (q=0) = 0 
\eea 
which can be checked explicitly from the above expressions using that 
$LT^{-1}_{\mu \to 1}  \frac{\gamma_E + \ln(4 \mu)}{\sqrt{\mu}} = 0$. Equivalently, the total shape in Fourier takes the form
\bea  \label{eq:res1} 
\tilde{{\cal F}}_d (q)  =  LT^{-1}_{\mu \to 1} \left( (1+ \alpha \frac{3 \gamma_E -2}{8})  \frac{2 \sqrt{\pi}}{q^2 + 2 \sqrt{\mu} + \Sigma(q,\mu)} \right) + O(\alpha^2) 
\eea 
where the "self-energy" correction reads, to lowest order
\bea \label{eq:res2} 
\Sigma(q,\mu) = - 4 \alpha  \sqrt{\mu} 
\bigg(\frac{q^2 + 9 \sqrt{\mu}}{q \sqrt{q^2 + 8 \sqrt{\mu}}} 
\sinh ^{-1}\left(\frac{q}{2 \sqrt{2 \sqrt{\mu}} }\right) - 1 + \frac{3}{16} \ln(4 \mu) \bigg) 
\eea

\medskip
{\it Units and scales:}
Let us mention here that, since this result was obtained in dimensionless units, the universal scale $\ell_{\sigma}$ appearing in the main text is here given by $\ell_{\sigma} = \frac{1}{m} \left(\frac{1}{S_m}\right)^{\frac{1}{d+\zeta}}$. $S_m$ can always be measured as $S_m= \frac{\langle S^2 \rangle}{ 2 \langle S \rangle} $ and is exactly given in terms of the parameters of the model by $S_m = \frac{\sigma}{m^4}$. As $m\to 0$, the dependence of $\sigma$ on $m$ is universal: $\sigma \sim m^{4 - d - \zeta} \sigma*$ with $\sigma*$ a dimensionless number. Thus $\ell_{\sigma} \simeq (\sigma*)^\frac{-1}{d + \zeta}$. The number $\sigma*$ is non-universal and depends on the microscopic disorder. Thus the scale $\ell_{\sigma}$ is non-universal and depends on {\it microscopic} properties of the disorder. Note also that using (\ref{SM:Sofq1}) one can also study the dependence of the mean-shape when $S$ gets close to the cutoff avalanche size $S_m$. This dependence is expected to be non-universal and in our model we find that the amplitude of the $O(\epsilon)$ corrections decrease as $S$ increases close to $S_m$.

\medskip

{\bf Small and large $q$ expansion of the mean-shape in Fourier space}

We now derive the result Eq.(\ref{FourierFatTail}) of the letter. The small $q$ expansion of $\delta \tilde{{\cal F}}_d (q) $ is obtained from (\ref{SM:LTFq}) at any order. The first terms are:
\bea \label{SM:smallq}
\delta \tilde{{\cal F}}_d (q) && \simeq_{q \ll 1} \alpha \left(-\frac{1}{16} \sqrt{\pi }   (-3 \gamma_E +1+\log (4096))  q^2  + \frac{1}{240} (299-90 \gamma_E )  q^4+\frac{\sqrt{\pi } \alpha  (1890 \gamma_E -3121-5040 \log (2))}{13440} q^6  \right. \nn\\
&& \left. + \left( \frac{2299}{5040}-\frac{\gamma_E }{8} \right)q^8 + O(q^{10})\right) \nn \\
&& \simeq_{q \ll 1} \alpha \left( -0.840378 q^2  + 1.02938q^4  -0.728437 q^6 +0.383999 q^8 + O(q^{10}) \right) 
\eea

For the large $q$ expansion, the expansion at large $q$ of $\delta \tilde{{\cal F}}_d (q) $ cannot be naively ILT. However, since we compute the ILT from $\mu$ to $1$, one can derive the result with respect to $\mu$ an arbitrary number of times $m$ to make the ILT convergent before taking the ILT since this just multiplies the end result by an innocent $(-1)^m$ factor). This leads to
\bea \label{SM:largeq}
\delta \tilde{{\cal F}}_d (q) &&  \simeq_{q \gg 1} \alpha \left( \frac{\frac{\gamma_E  \ }{2}+4   -4\log(q) }{q^4} -\frac{8 \sqrt{\pi }  }{q^6}+\frac{-48 \gamma_E   +23  -120\log(q)}{q^8} +  \frac{624 \sqrt{\pi }  }{q^{10}}  + O(\frac{1}{q^{12}} )\right)ÃÂ \nn \\
&& \simeq_{q \gg 1} \alpha \left(\frac{-4 \log(q)+4.28861}{q^4} -\frac{14.1796}{q^6} +\frac{-120 \log(q)-4.70635}{q^8} + \frac{1106.01}{q^{10}}  + O(\frac{1}{q^{12}} ) \right)
\eea
And as explained in the main text, the first term of this expansion is interpreted as a modification of the power-law behavior of $\tilde{{\cal F}}_d (q)$, $ \tilde{{\cal F}}_d (q) \simeq_{q \gg1} 2 (1 + (2 + \frac{\gamma_E}{4}) \alpha) q^{- 4 - 2 \alpha} + O(\epsilon^2)$. 

\newpage

{\bf Dominant non-analyticity at small $x$} 

Let us now understand more precisely how the large $q$ behavior %$(2 + \frac{\gamma}{4}) \alpha) q^{- 4 - 2 \alpha}$ 
of $\tilde{{\cal F}}_d (q)$ generates a non-analyticity in ${\cal F}_d(x)$ at small $x$. We consider the effect of a fat tail $q^{-2 \beta}$ in a Fourier transform. We write
\bea \label{SM:smallx1}
\int \frac{d^dq}{(2 \pi)^d} \frac{e^{i q_1 x}}{q^{2\beta}} && = \frac{1}{\Gamma(\beta)} \int_{t>0} \frac{dt}{t} t^{\beta} \int  \frac{d^dq}{(2 \pi)^d} e^{i q_1 x - q^2 t } = |x|^{2 \beta -d} \int dt \frac{2^{-d} \pi ^{-\frac{d}{2}} e^{-\frac{1}{4 t}} \left(t \right)^{\beta -\frac{d}{2}}}{t \Gamma (\beta )} \nn \\
&& \sim |x|^{2 \beta -d} \frac{2^{-2 \beta } \pi ^{-\frac{d}{2}} \Gamma \left(\frac{d}{2}-\beta \right)}{\Gamma (\beta )}
\eea
The above derivation is formal since e.g. the first integral on $q$ on the left-hand side of (\ref{SM:smallx1}) do not converge but we notice that (\ref{SM:smallx1}) indeed gives, for $\beta=2$, the dominant non-analyticity in the expansion (\ref{serFd}) (i.e. the $b_{p=0}$ term). The above calculation indicates that the leading non-analyticity present in the small $x$ expansion of ${\cal F}_d(x)$ is a term of the form
\bea
 {\cal F}^{sing}_d (x) \simeq 2 (1 + (2 + \frac{\gamma_E}{4}) \alpha) |x|^{4+2\alpha-d} \frac{ 2^{-4- 2 \alpha} \pi^{-d/2} \Gamma(d/2 - 2 - \alpha) }{\Gamma(2+\alpha)}
\eea
Expanding this result in $\alpha$, it implies the existence of a term 
\bea \label{SM:smallx2}
 \delta{\cal F}^{sing}_d (x) \simeq  \frac{\alpha}{32} \pi ^{-\frac{d}{2}} x^{4-d} \Gamma \left(\frac{d}{2}-2\right) \left(-4 \psi \left(\frac{d}{2}-2\right)+8 \log (x)+5 \gamma_E +4-8 \log (2)\right)
\eea
in the small $x$ expansion of $\delta {\cal F}_d(x)$ ($\psi = \frac{\Gamma'}{\Gamma}$ is the diGamma function). For $d=1,3$ this result correctly gives the dominant non-analyticity in $\delta{\cal F}_d (x)$. For $d=2$, one has to look at the expansion of (\ref{SM:smallx2}) around $d=2$. In doing so, one obtains terms (i) regular in $x$ (proportional to $x^2$) that diverge as $d\to 2$: these terms are unimportant and would be cancelled by other regular terms present in $\delta{\cal F}_d (x)$, and (ii) a singular term which admit a well defined $d\to 2$ limit and read:
\bea \label{SM:smallx2d2}
 \delta{\cal F}^{sing}_{d=2} (x) \simeq  \frac{\alpha}{16 \pi} (9 \gamma_E - 8 \log(2) + 4 \log(x) ) x^2 \log(x) \ . 
\eea
This term is the dominant non analyticity present in $\delta{\cal F}_{d=2} (x)$.

\medskip

{\bf Large $x$ expansion of the mean-shape in real space}

We now obtain the modification of the large $x$ behavior of ${\cal F}_d (x)$, and derive Eq.(\ref{Real1loopExpDecay}) of the letter.
The mean shape in real space is obtained by Fourier transform and ILT from (i) the expressions
$\tilde{{\cal F}}^{{\rm MF}}(q)$ (\ref{SM:LTMF}), $\delta \tilde{{\cal F}}_d (q)$ (\ref{SM:LTFq}) and the definition of $H(q,\mu)$, (\ref{SM:H}), or, equivalenty to lowest order in $\alpha$, (ii) from the expressions (\ref{eq:res1}, \ref{eq:res2}).
We use the latter here:
\bea
{\cal F}_d (x) && =  \int \frac{d^d q}{(2 \pi)^d} e^{ - iq_1 x} \int_{{\cal C}}  \frac{d\mu}{2 i \pi}  e^{\mu} 
 \left(   \frac{2 \sqrt{\pi} c}{q^2 + 2 \sqrt{\mu} + \Sigma(q,\mu)} \right) \quad , \quad
 c = (1+ \alpha \frac{3 \gamma_E -2}{8}) 
\eea 
where here the contour ${\cal C}$ can be chosen as a wedge around the branch cut $\mu<0$ of the integrand, such as e.g. ${\cal C} = (1 + e^{- \frac{3 i \pi}{4} }  \mathbb{R}_+ ) \cup (1 + e^{ \frac{3 i \pi}{4} }  \mathbb{R}_+ )$. 
To compute this radial Fourier transform, we chose $x>0$ oriented along the first axis. 
The integration over the other components 
%The integral on momenta is splitted between the integration on $q_1$ and on the other components $q_2 \dots q_d$. The integral 
$q_2 \dots q_d$ %is simplified by the fact that it depends only on 
depends only on $q = \sqrt{ q_2^2 + \cdots q_d^2}$: the change of variable brings out a factor $S_{d-1} = \frac{ 2(\pi)^{\frac{d-1}{2}} }{\Gamma(\frac{d-1}{2})}$. 
Performing the rescaling $(q_1 , q) \to \sqrt{2}(q_1 , q)$ we obtain the more convenient form
\bea
 {\cal F}_d (x/\sqrt{2})  && =  2^{\frac{d}{2}} \sqrt{\pi} c\int_{ -\infty}^{\infty} \frac{d q_1}{2 \pi } \int_{0}^{\infty}  \frac{dq S_{d-1}}{(2\pi)^{d-1}} q^{d-2} e^{ - iq_1 x} \int_{{\cal C}}  \frac{d\mu}{2 i \pi}  e^{\mu} \frac{1}{q_1^2 +q^2 + \sqrt{\mu} -\frac{ \alpha}{2} h(\sqrt{2}\sqrt{q_1^2 + q^2},\mu)}
\eea
where we denote $\Sigma(q,\mu)  = - \alpha~ h(q, \mu)$.

At the mean-field level, i.e. $\alpha=0$, the integral on $q_1$ can be performed by closing the contour of integration in the upper half plane (the integrand is then analytic in $q_1$), and taking into account the contribution of the pole at $q_1(\mu) = i \sqrt{ q^2 + \sqrt{\mu}}$. The scaling of this pole with $\mu$, $q_1 \sim \mu^{\frac{1}{4}}$ notably leads to the stretched exponential decay of the shape at large $x$ with exponent $4/3$. Here, at $O(\epsilon)$ we cannot a priori performs this residue calculation since the integrand is non analytic in $q_1$. It seems however reasonable to assume that the behavior of ${\cal F}_d (x)$ at large $|x|$ will still be dominated by this pole in the integration on $q_1$. At first order in $O(\epsilon)$ the position of this pole is shifted as
\bea
q_1(\mu) %&& \simeq i \sqrt{q^2 + \sqrt{\mu}  -  \alpha \frac{h(i \sqrt{2 } \mu^{1/4} , \mu)}{2}} 
\simeq i \left( \sqrt{q^2 + \sqrt{\mu}}  - \alpha \frac{ \delta q(\mu)}{\sqrt{q^2 + \sqrt{\mu}}}ÃÂ  \right)
%\eea
%where we introduced the notation
%\bea
\quad , \quad \delta q(\mu) = \frac{1}{4} h(i \sqrt{2 } \mu^{1/4} , \mu)    = \frac{1}{72} \sqrt{\mu } \left(27 \log \left(2 \sqrt{\mu }\right)+14 \pi  \sqrt{3}-72\right)
\eea
And for the saddle-point calculation of the integral on $q_1$, we can approximate
\bea
&& \frac{1}{q_1^2 +q^2 + \sqrt{\mu} -\frac{ \alpha}{2} h(\sqrt{2}\sqrt{q_1^2 + q^2},\mu)}  
%\nn \\
%&&\simeq  \frac{1}{q_1^2 +q^2 + \sqrt{\mu} - \frac{\alpha}{2} ( h (i \sqrt{2} \mu^{\frac{1}{4}} , \mu) +  \sqrt{2} \sqrt{ q^2 + \sqrt{\mu}}(\mu)^{-\frac{1}{4}}\partial_1 h(i \sqrt{2} \mu^{\frac{1}{4}} , \mu)    (q_1 -i \sqrt{q^2 + \sqrt{\mu}}) ) } \nn \\
%&&  
\simeq \frac{1}{(q_1 - q_1(\mu)) (q_1 + q_1(\mu) - \frac{\alpha}{2} \Delta q(\mu))}
\eea
With
\bea
\Delta q(\mu) = \sqrt{2} \sqrt{ q^2 + \sqrt{\mu}}(\mu)^{-\frac{1}{4}}\partial_1 h(i \sqrt{2} \mu^{\frac{1}{4}} , \mu) = \frac{2 i }{27} (13 \sqrt{3} \pi - 63 ) \sqrt{q^2 + \sqrt{\mu}}
\eea
 (Through rescaling one shows that higher order terms in the series expansion of $h(\sqrt{2}\sqrt{q_1^2 + q^2},\mu)$ around $q_1 = i\sqrt{q^2 + \sqrt{\mu}}$ do not contribute). Hence we have 
\bea
{\cal F}_{d}(x/\sqrt{2}) && = c 2^{d/2} \sqrt{\pi} \int \frac{d \mu}{2 i \pi }  e^{\mu} \int_{0}^{+\infty} \frac{S_{d-1}}{(2\pi)^{d-1}} q^{d-2} dq   \frac{2 i \pi}{2 \pi} \frac{e^{ -x \left( \sqrt{q^2 + \sqrt{\mu}} - \frac{\alpha \delta q}{\sqrt{q^2 + \sqrt{\mu}}} \right) }}{2 q_1(\mu) - \frac{\alpha}{2} \Delta q(\mu)} \nn  \\ 
&& \simeq  c 2^{d/2-1} \sqrt{\pi} \frac{S_{d-1}}{(2\pi)^{d-1}}  \int \frac{d \mu}{2 i \pi } e^{\mu - x (\mu^\frac{1}{4} - \alpha \frac{\delta q}{\mu^{\frac{1}{4}}}) }  \int_{0}^{+\infty} q^{d-2} \frac{1}{ (1-\frac{\alpha}{54}  (13 \sqrt{3} \pi - 63 ) )\mu^{\frac{1}{4}} - \alpha \frac{\delta q}{\mu^{\frac{1}{4}}}   }  e^{ -x \frac{q^2 \left(\delta q \alpha+\sqrt{\mu }\right)}{2 \mu ^{3/4}}} \nn \\
&&  \simeq  c \frac{\pi ^{1-\frac{d}{2}}}{\sqrt{2}}   \int \frac{d \mu}{2 i \pi } e^{\mu - x a \mu^b } \frac{1}{a'  \mu^b } (\frac{a \mu^b}{x})^{\frac{d-1}{2}}  
\eea 
Where we have used the fact that the dominant behavior of the integral on $q$ is given by $q \simeq 0$, and we have introduced the notation
\bea
&& a = 1 + \frac{-14 \sqrt{3} \pi +72-9 \log (8)}{72} \alpha \quad , \quad b = \frac{1}{4} - \frac{3}{16} \alpha  \quad , \quad  %a' = 1 + \frac{468-42 \sqrt{3} \pi - 52 \sqrt{3 \pi} - 81 \log(2)}{216} \alpha \\
 a' = 1 + \frac{468-94 \pi \sqrt{3}  - 81 \log(2)}{216} \alpha
\eea
So that $ a \mu^b = \mu^\frac{1}{4} - \alpha \frac{\delta q}{\mu^{\frac{1}{4}}} + O(\epsilon^2)$ and $a' \mu^b= (1-\frac{\alpha}{54}  (13 \sqrt{3} \pi - 63 ) )\mu^{\frac{1}{4}} - \alpha \frac{\delta q}{\mu^{\frac{1}{4}}} + O(\alpha^2) $. Note that, using $\zeta = \frac{1}{3} \epsilon$ and $\alpha = - 2\epsilon/9$, the $O(\epsilon)$ value of $b$ is consistent with the conjecture $b = \frac{1}{d + \zeta}$ which is quite natural: the exponent $b$ gives the scaling with $\mu$ of the pole $q_1(\mu) \sim \mu^b$. We know that momenta inside avalanches of sizes $S$ scale with $S$ as $S^\frac{-1}{d + \zeta}$. On the other hand, $\mu$ is conjugate to $S$: $\mu \sim S^{-1}$, hence the conjecture $q_1(\mu) \sim \mu^\frac{1}{d + \zeta}$. At large $x$, the integral on $\mu$ can now be evaluated using a saddle-point calculation. It leads to, at first order in $\epsilon$,
\bea \label{SM:largex1loop}
&& {\cal F}_{d}(x) \simeq A x^B e^{-C x^\delta}  \nn \\
&&   A = \frac{2^{-d/2} \pi ^{\frac{1}{2}-\frac{d}{2}}}{\sqrt{3}} 
( 1+\frac{1}{216} \alpha  \left(4 \sqrt{3} \pi  (27-7 d)+9 (13 d+9 (\gamma_E -8))\right) )  \nn \\
&&  B = - \frac{d-2}{2}  \frac{1-2 b}{1-b} =\frac{2-d}{3} (1 + \frac{1}{2} \alpha) \nn \\
&& C =\frac{3}{4} + \alpha \frac{\left(36-7 \sqrt{3} \pi \right) }{36} \quad , \quad  \delta  = \frac{1}{1-b}  = \frac{4}{3}- \frac{\alpha}{3} \nn 
\eea
Following the conjecture on the value of $b$ we can also conjecture
\bea \label{SM:Dconj}
B = -\frac{(d-2) (d+\zeta -2)}{2 (d+\zeta -1)} \quad , \quad \delta = \frac{d+\zeta }{d+\zeta -1}
\eea
Setting $\alpha = 0$ in the above result, we retrieve the large $x$ behavior of ${\cal F}^{{\rm MF}}_{d}(x)$ using here a totally different route. Lets us warn the reader that there is some uncertainty on the values of $A$ and $B$ since %we were not allowed to close the contour of integration on $q_1$ in the first place and 
additional contributions could come from the branch cut in $q_1$. 
The values of $C$ and $\delta$ however should be correct. 
%if one assumes that the dominant behavior of ${\cal F}_{d}(x)$ at large $x$ is indeed given this simple deformation of the dominant behavior of ${\cal F}^{{\rm MF}}_{d}(x)$ (through the shift of the pole $q_1(\mu)$). 
The resulting numerical values of the exponents $B$ and $\delta$ 
are summarized in Table~\ref{tab:SM:delta}.

 Note that (\ref{SM:largex1loop}) can also be expanded in $\alpha$ and gives the prediction 
 \bea\label{SM:largex1loop2}
 \delta{\cal F}_{d}(x) \simeq_{x>>1}&&   \alpha \frac{  2^{-\frac{d}{2}-3} \pi ^{\frac{1}{2}-\frac{d}{2}} e^{-\frac{3}{4} x^{4/3} } x^{\frac{2}{3}-\frac{d}{3}}}{27 \sqrt{3}} \left(   2 \pi  \sqrt{3} \left(-14 d+21 x^{4/3}+54\right) \right. \nn \\
 && \left. +9 \left(\left(-4 d+6 x^{4/3}+8\right) \log (x)+13 d-24 \left(x^{4/3}+3\right)+9 \gamma \right) \right)
 \eea

 \begin{center}
\begin{table}
\vspace{0.1cm}
\begin{tabular}{|l | l | l | l | l |}
 \hline			
  & $\epsilon=0$ & $\epsilon=1$ & $\epsilon=2$ & $\epsilon=3$  \\
   \hline
 $B$ at $O(\epsilon)$  & $-2/3$ & $-0.298 \pm 0.002$ & $0$ & $0.235 \pm 0.014$ \\
    \hline
 $B$ conjecture & $-2/3$ & $-0.2876 \pm 0.0001$ & $0$ & $0.100 \pm 0.002$ \\
    \hline
 $\delta $ at $O(\epsilon)$ & $4/3$ & $1.410 \pm 0.002$ & $1.49 \pm 0.01$ & $1.58 \pm 0.02$ \\
     \hline
 $\delta $ conjecture& $4/3$ & $1.4246 \pm 0.0002$ & $1.570 \pm 0.001 $ & $1.800 \pm 0.004$\\ 
 \hline  
 \end{tabular}
\caption{Predicted values for the exponents $B$ and $\delta$ from the $O(\epsilon)$ calculation,
and from the conjecture (\ref{SM:Dconj}) (the values are averaged over the two Pade, and the
spread is indicated), and compared to the conjecture (\ref{SM:Dconj}) using the value of $\zeta$ determined numerically in \cite{RossoHartmannKrauth2002} ($\zeta = 0.355 \pm 0.001$ for $d=3$ and $\zeta = 0.753 \pm 0.002$ for $d=2$) and \cite{FerreroBustingorryKolton2012} ($\zeta = 1.250 \pm 0.005$ in $d=1$).}
\label{tab:SM:delta}
\end{table}
\end{center}

\medskip

{\bf Numerical obtention of the mean shape}

We now explain how our analytical results are used to obtain numerically the mean shape computed at $O(\epsilon)$. In particular we explain how we obtain the theoretical curves presented in Fig.~\ref{fig:treeAnd1loop} and Fig.~\ref{fig:resSimuShapesd1} of the letter. The correction $\delta \tilde{{\cal F}}_d (q)$ can easily be obtained numerically using a numerical integration on the formula (\ref{SM:LTFq}) and choosing a contour of integration for $\mu$ as ${\cal C} = (1 + e^{- \frac{3 i \pi}{4} }  \mathbb{R}_+ ) \cup (1 + e^{ \frac{3 i \pi}{4} }  \mathbb{R}_+ ) $. The precision of the numerical integration can be tested against the exact results at small and large $q$, (see Fig.~\ref{fig:SM:Correc}). It can easily be Fourier transformed in any dimension to find the correction  $\delta {\cal F}_d (x)$: 
\bea \label{SM:dFreal}
\delta {\cal F}_{d=1} (x) = 2 \int_{0}^{\infty} \frac{dq}{2 \pi} \cos(q x) \delta \tilde{{\cal F}}_d (q) \quad , \quad  \delta {\cal F}_{d} (x) =  \frac{1}{(2 \pi)^{\frac{d}{2}} x^{\frac{d-2}{2}} } \int_{0}^{\infty} dq  J_{\frac{d-2}{2} } ( q x ) q^{\frac{d}{2}}\delta \tilde{{\cal F}}_d (q) 
\eea
where $J_n(x)$ denotes the Bessel function of the first kind. The large $x$ behavior of these corrections agrees with our prediction (\ref{SM:largex1loop2}), to a surprisingly large extent (see Fig.~\ref{fig:SM:Correc}). Some properties of these corrections are their values at the origin $\delta {\cal F}_{d=1} (0) = 0.09227$, $\delta {\cal F}_{d=2} (0) = 0.04912$, the position where they cross $0$, $x_0 = 1.2567$ ($d=1$), $x_0 = 1.8286$ ($d=2$), the position of their minimum and minimal value, $x_{min} = 2.2783$, ${\cal F}_{1}(x_{min})=-0.02835$, $x_{min}=2.6634$; ${\cal F}_{2}(x_{min})=-0.002980$ ($d=2$). We also investigate the presence of non-analyticities in the form of logarithm in the short-distance behavior of the result. In dimension $1$, the correction $\delta {\cal F}_{1} (0)$ has a second derivative at $0$ evaluated as $a_0 = \delta {\cal F}''_{1} (0) \simeq -0.512$. By plotting $\frac{1}{x^3} \left( \delta {\cal F}_{1} (x) - \delta {\cal F}_{1} (0) - \frac{a_0}{2} x^2 \right)$, we shed the light on the non analyticity present in $\delta {\cal F}_{1}(x)$ at small $x$, which is found to be in very good agreement with (\ref{SM:smallx2}) (see Fig.~\ref{fig:SM:Correc}). In dimension $2$, the dominant non-analyticity predicted in (\ref{SM:smallx2d2}) compares very well with the plot of $\frac{\delta {\cal F}_{2} (x) -\delta {\cal F}_{2} (0) + 0.06 x^2}{x^2}$ at small $x$ (the $0.06 x^2$ term is a regular term which was not predicted by our calculations).

\begin{figure}
\centerline{
   \fig{.3\textwidth}{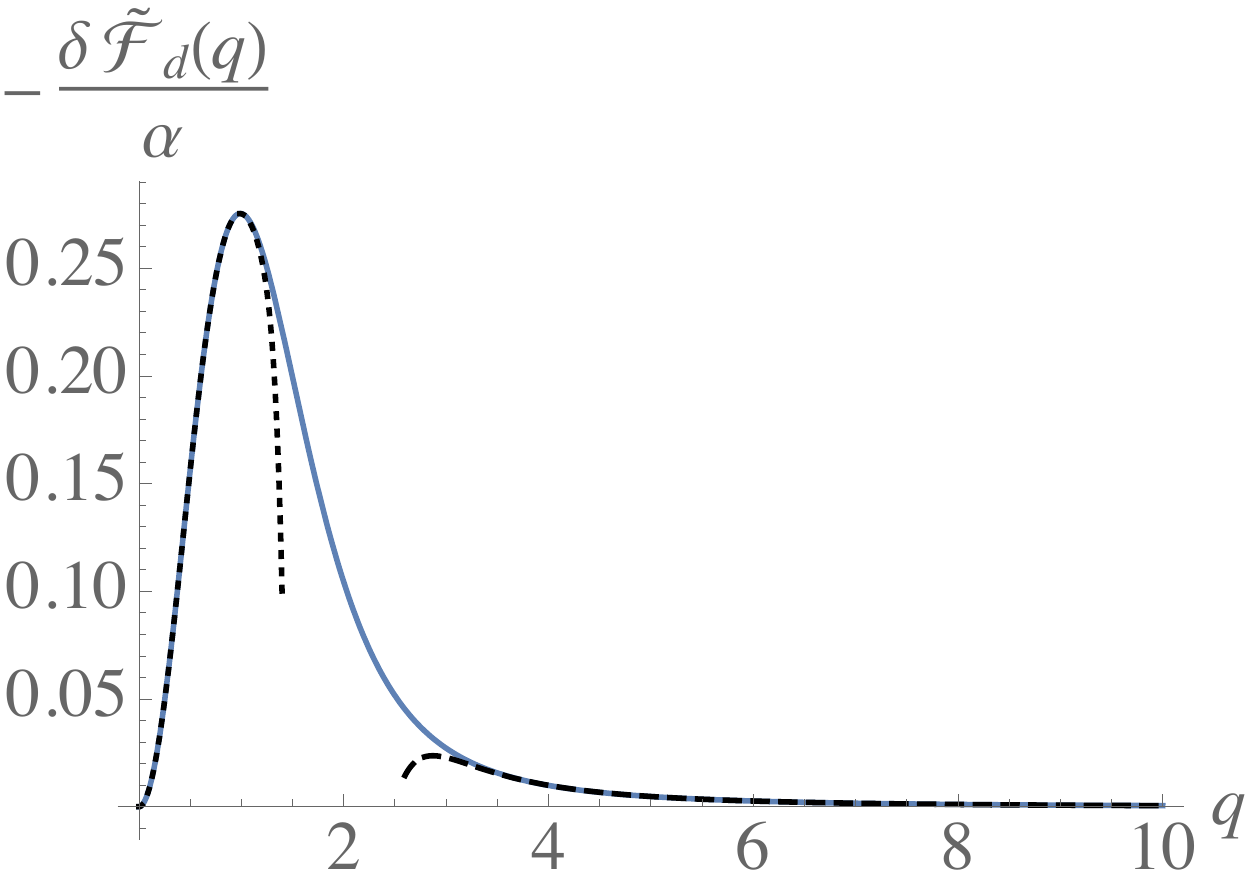} \fig{.3\textwidth}{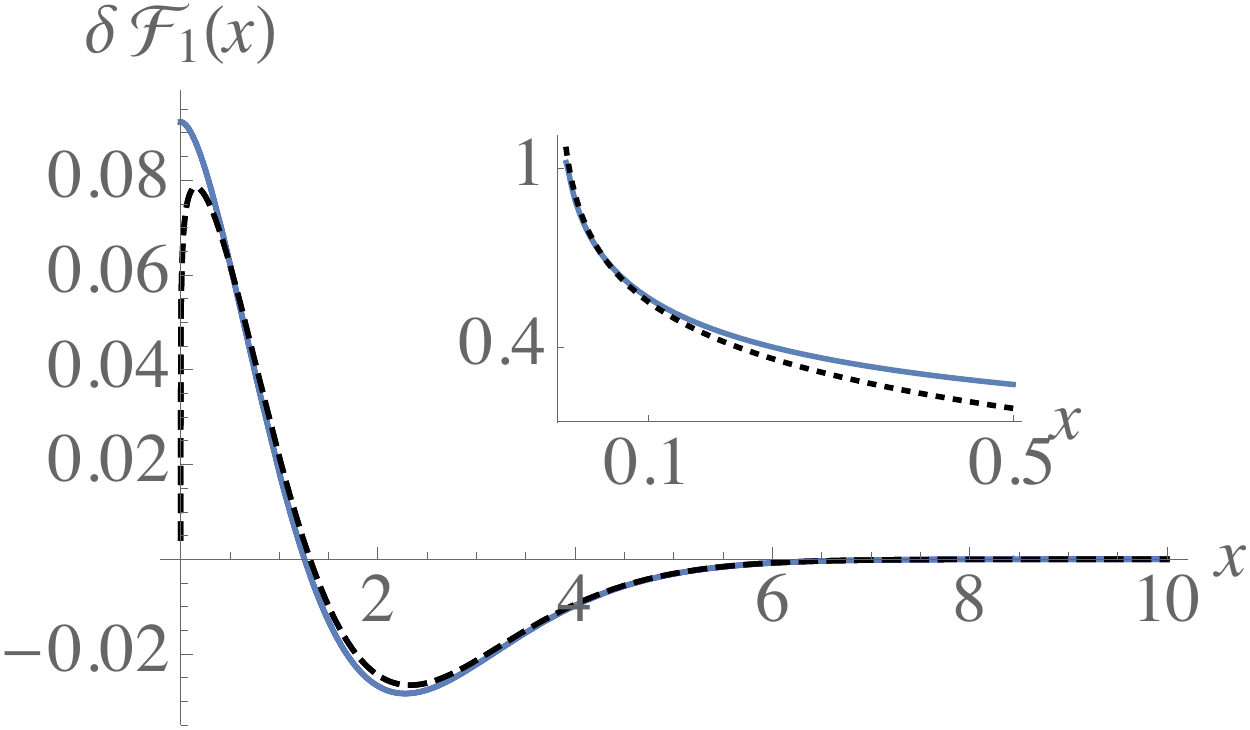} \fig{.3\textwidth}{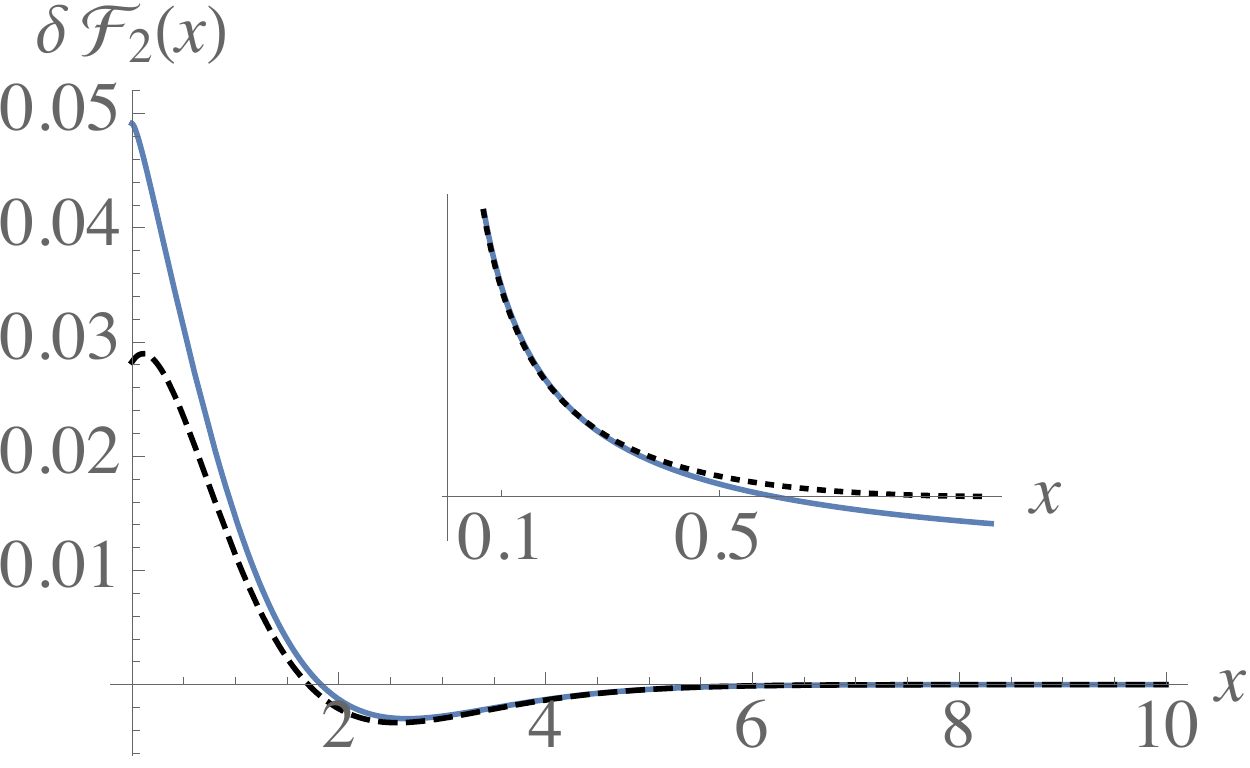}  }
 \caption{In blue from left to right: $O(\epsilon)$ correction to the mean-shape in Fourier space divided by $-\alpha$, $- \frac{\delta \tilde{{\cal F}}_d(q)}{\alpha}$, in real space in $d=1$, $\delta {\cal F}_1(x)$ and in $d=2$, $\delta {\cal F}_2(x)$. The dotted line on the left is the theoretical small $q$ expansion (\ref{SM:smallq}) up to $O(q^{20})$ and the dashed line is the large $q$ expansion (\ref{SM:largeq}). The dashed line in the middle and on the right are the theoretical large $x$ expansion (\ref{SM:largex1loop2}). Middle inset: plot of $\frac{1}{x^3} \left( \delta {\cal F}_{1} (x) - \delta {\cal F}_{1} (0) - \frac{a_0}{2} x^2 \right)$ (plain line), compared with the prediction $(\ref{SM:smallx2})$ (dashed line). Right inset: plot of $- \frac{\delta {\cal F}_{2} (x) -\delta {\cal F}_{2} (0) + 0.06 x^2}{x^2}$ (plain line), compared with the prediction $(\ref{SM:smallx2d2})$ (dashed line).}
\label{fig:SM:Correc}
\end{figure}
\medskip

 Adding naively these corrections to the mean-field result $ {\cal F}_{d} (x) = {\cal F}^{{\rm MF}}_{d} (x) + \delta {\cal F}_{d} (x)$ then gives a result which suffers from several problems. At large $x$ it becomes slightly negative in $d=1$ and does not have the right non-analytic behavior at small $x$. The second problem can be cured by considering the reexponentiated Fourier result
\bea \label{SM:regFourier}
\tilde{{\cal F}}^{{\rm reg}}_d(q) = \tilde{{\cal F}}^{{\rm MF}}_d (q) \exp\left( \frac{\delta \tilde{{\cal F}}_d (q)}{\tilde{{\cal F}}^{{\rm MF}}_d (q)} \right)
\eea
This result is still correct to first order in $\epsilon$ and has the advantage of having the correct behavior at large $q$, $\tilde{{\cal F}}^{{\rm reg}}_d(q) \simeq 2 (1 + (2 + \frac{\gamma_E}{4}) \alpha) q^{- 4 - 2 \alpha} + O(\epsilon^2)$. It is plotted in plain red in Fig.~\ref{fig:SM:Regularization}. Taking the Fourier transform of this result we obtain a function ${\cal F}_{d}^{{\rm reg} 1}(x) $ which has now the correct behavior at small $x$ but is still slightly negative at large $x$. On the other hand the function
\bea 
{\cal F}_{d}^{{\rm reg} 2}(x)   =\frac{1}{{\cal N}}  \exp\left( - \exp\left(  \log (-\log({\cal F}^{{\rm MF}}_d (x))) + \frac{\delta {\cal F}_d (x)}{{\cal F}^{{\rm MF}}_d (x) \log({\cal F}^{{\rm MF}}_d (x))} \right)\right)
\eea
where ${\cal N}$ is a normalization constant ensuring that $\int d^d x{\cal F}_{d}^{{\rm reg} 2}(x)  =1$, is correct to $O(\epsilon)$ and takes properly into account the change of exponent in the exponential decay of the shape at $x = \infty$ and is everywhere positive. However, it doesn't have the correct behavior at small $x$. Since ${\cal F}_{d}^{{\rm reg} 1}(x)$ and ${\cal F}_{d}^{{\rm reg} 2}(x)$ intersect themselves at some $x_c$, we construct the function
\bea \label{SM:reg3}
{\cal F}_{d}^{{\rm reg}}(x) = \frac{1}{{\cal N}}\left( r(x){\cal F}_{d}^{{\rm reg} 1}(x) +  (1-r(x)){\cal F}_{d}^{{\rm reg} 2}(x) \right)
\eea
where ${\cal N}$ is a normalization factor and $r(x)$ is a function that interpolates smoothly between $r(1)=1$ and $r(\infty)=0$ sufficiently fast to obtain a positive result everywhere. Here we have chosen $r(x) = e^{-x^2/xc^2}$ but this choice does not matter drastically since all these functions are close to each others (see Fig.~\ref{fig:SM:Regularization}). The result (\ref{SM:reg3}) is still correct to $O(\epsilon)$ and has the right behavior at small and large $x$. It is plotted for $d=1$ and $d=2$ in plain red in (\ref{fig:SM:Regularization}) and used for comparison to numerical simulations.

\begin{figure}
\centerline{
   \fig{.3\textwidth}{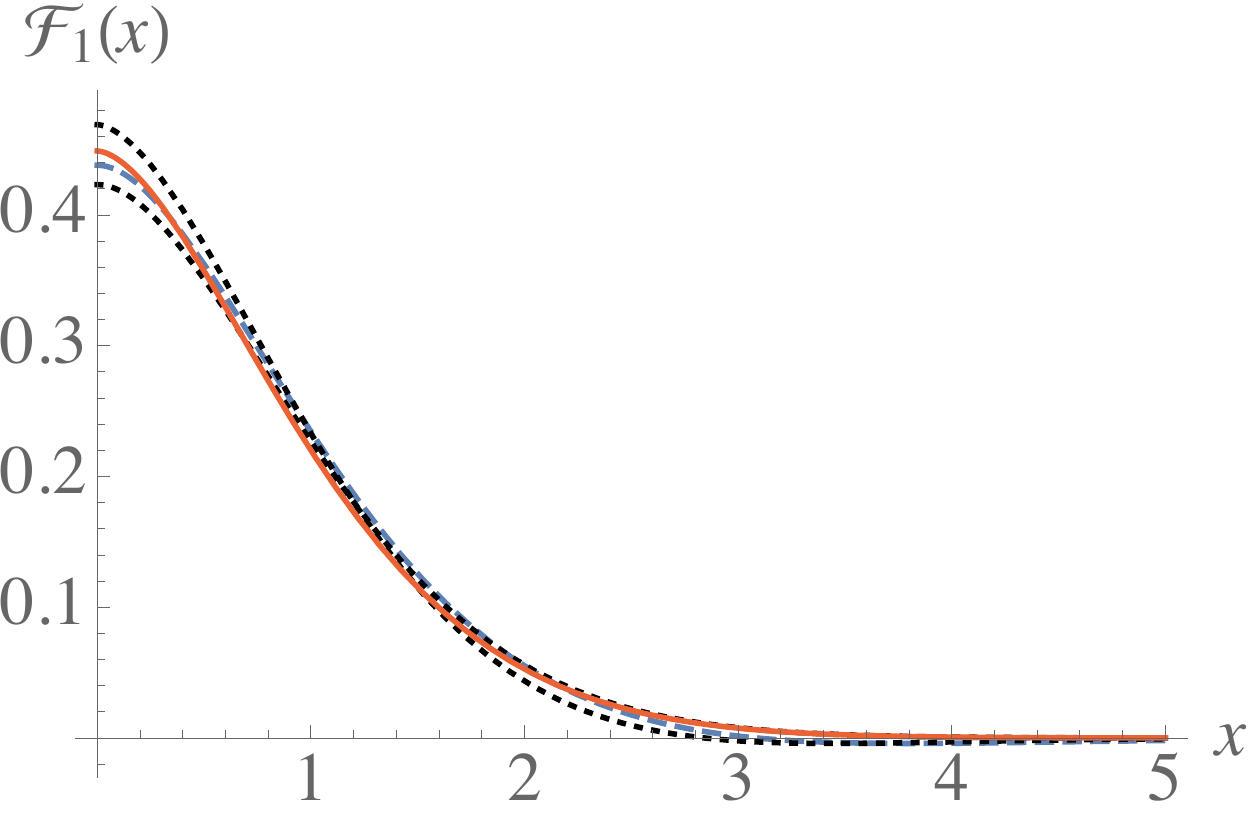} \fig{.3\textwidth}{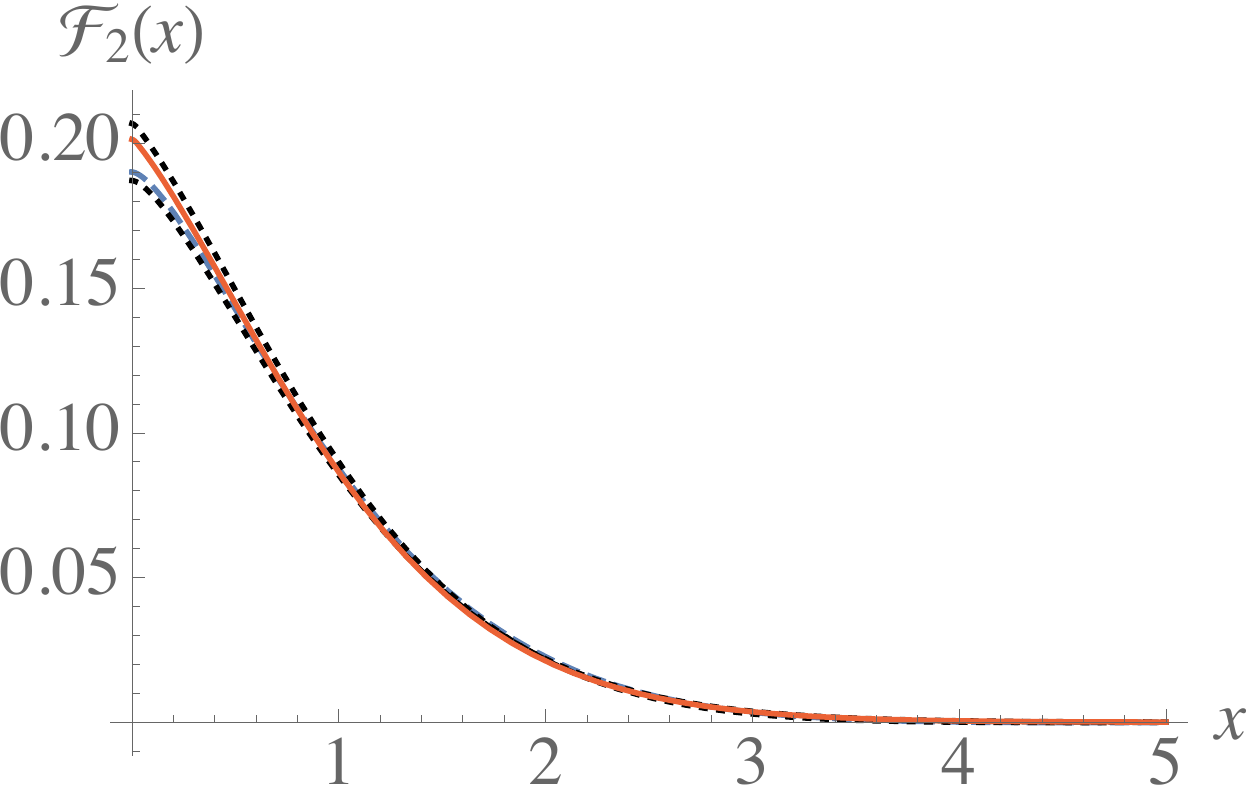}}
 \caption{Different mean shape ${\cal F}_{d}(x)$ correct at $O(\epsilon)$ for $d=1$ (left) and $d=2$ (right). Dashed-blue lines: naive result ${\cal F}_{d}(x) = {\cal F}^{{\rm MF}}_{d}(x) + \delta {\cal F}_{d}(x) $. Dotted lines: ${\cal F}_{d}^{{\rm reg1}}(x) $ (largest at the origin) and ${\cal F}_{d}^{{\rm reg2}}(x) $ (smallest at the origin). Red line: regularized result ${\cal F}_{d}^{{\rm reg}}(x) $ used for comparison with numerics.}
\label{fig:SM:Regularization}
\end{figure}

\medskip

{\bf Universal ratios}

Here we compute the universal ratios in dimension $1$ and $2$ of the various mean-shapes. These are defined as $c_{p} = \frac{\int d^d x  |x|^{2p} {\cal F}_d(x)}{\left(\int d^d x  |x|^{p} {\cal F}_d(x) \right)^2}$. In dimension 1 and for $p$ even they are exactly obtained as $c_p = 
\frac{\tilde{{\cal F}}_d^{(2p)}(0) }{\left(  \tilde{{\cal F}}_d^{(p)}(0)    \right)^2}$. For $p$ odd and in dimension $d=2$ one has to rely on direct numerical integration techniques. Fortunately, the exponential decay of the shape at large $x$ (which is known analytically) allows us to obtain an excellent numerical precision, we compute them pertubatively in $O(\epsilon)$ using
\bea \label{SM:Ratios}
c_p %\frac{\int d^d x  |x|^{2p} {\cal F}_d(x)}{\left(\int d^d x  |x|^{p} {\cal F}_d(x) \right)^2} 
\simeq \frac{\int d^d x  |x|^{2p} {\cal F}^{{\rm MF}}_d(x)}{\left(\int d^d x  |x|^{p} {\cal F}^{{\rm MF}}_d(x) \right)^2}  + \alpha \left( \frac{\int d^d x  |x|^{2p} \delta{\cal F}_d(x)}{\left(\int d^d x  |x|^{p} {\cal F}_d^{{\rm MF}}(x) \right)^2 }- 2 \frac{\int d^d x  |x|^{2p} {\cal F}^{{\rm MF}}_d(x) \int d^d x  |x|^{p} \delta{\cal F}_d(x) }{\left(\int d^d x  |x|^{p} {\cal F}^{{\rm MF}}_d(x) \right)^3}          \right) 
\eea

Table \ref{tab:SM:UniversalRatios} contains our results in $d=1$ and $d=2$. The even values in $d=1$ are exact for both the BFM and (to $O(\epsilon)$) the SR case. The odd values are results of numerical integration. The uncertainty on the numerical integration is evaluated in $d=1$ by comparing the result obtained using numerical integrations for even ratios to the exact ones. The values in $d=2$ are results of numerical integrations. We also give for reference in Table \ref{tab:SM:UniversalRatios} the value of the universal ratios for a Gaussian shape function ($ {\cal F}_{d=1}^{{\rm Gauss}} (x) = \frac{e^{-x^2}}{\sqrt{\pi}} $ and $ {\cal F}_{d=2}^{{\rm Gauss}} (x) = \frac{e^{-x^2}}{\pi}  $)

\begin{center}
\begin{table}
\vspace{0.1cm}
\begin{tabular}{|l | l | l | l | l | l | l |}
 \hline			
  & $c_1$  \hspace{1.5 cm} & $c_2$ \hspace{1.5 cm} & $c_3$ \hspace{1.5 cm} & $c_4$ \hspace{1.5 cm} & $c_5$ \hspace{1.5 cm} & $c_6$  \hspace{1.5 cm} \\
   \hline
 Gaussian $d=1$ & $1.5708$ & $3$ & $5.8905$ & $11.67$ & $29.1938$ & $46.2$ \\
    \hline
 BFM $d=1$: Theory & $1.6944$ & $3.8197$ & $9.2703$ & $23.3333$ & $60.045$ & $156.863$ \\
    \hline
 SR $d=1$: Theory & $1.6944$ & $3.8197$ & $9.2703$ & $23.3333$ & $60.045$ & $156.863$ \\
  & $+0.0798\alpha$ & $+0.6196 \alpha$ & $+2.8\alpha$ & $+11.4444\alpha$ & $+37\alpha$ & $+138.296 \alpha$ \\
     & $\simeq 1.641 $ & $\simeq 3.43  $ & $\simeq 7.53 $ & $\simeq 16.6  $ & $\simeq 38.5  $ & $\simeq 81  $ \\
     & $\pm 0.001$ &  $\pm 0.02$ &  $ \pm 0.16$ & $ \pm 0.9 $ & $ \pm 3.7 $ & $ \pm 17$ \\
  \hline
 Gaussian $d=2$ & $1.27324$ & $2$ & $3.3953$ & $6$ & $10.865$ & $20$ \\
    \hline
 BFM $d=2$: Theory & $1.3734$ & $2.5464$ & $5.3435$ & $12$ & $28.1289$ & $67.9111$ \\
    \hline
 SR $d=2$: Theory & $1.3734$ & $2.5464$ & $5.3435$ & $12$ & $28.1289$ & $67.9111$ \\
  & $+ 0.06482 \alpha $ & $+ 0.4110 \alpha $ & $+ 1.6647 \alpha $ & $+ 5.7758 \alpha $ & $+ 18.6579 \alpha  $ & $+ 58.0856 \alpha $ \\
   & $\simeq 1.3449 $ & $\simeq 2.369  $ & $\simeq 4.65 $ & $\simeq 9.6   $ & $\simeq 20.8 $ & $\simeq 45.7 $ \\
   & $\pm 0.0002$  & $ \pm 0.006$  & $\pm 0.05$ & $ \pm 0.2$ & $ \pm 0.9$ &   $ \pm 3.6$  \\
    \hline
 \end{tabular}
\caption{Prediction for the universal ratios in dimension 1 ($\epsilon=3$) and 2 ($\epsilon=2$). 
Here $\alpha=-2 \epsilon/9$. The values displayed are the average over the two Pade
and their spread is indicated (as an indication of the uncertainty).}
\label{tab:SM:UniversalRatios}
\end{table}
\end{center}

\newpage

\begin{center}
{\bf \large Details on numerical simulations}
\end{center}
We now give details on the numerical simulations leading to the results presented in Fig.~\ref{fig:resSimuShapesd1} and Fig.~\ref{fig:resSimuSmallxLargeq} in the letter.
{\bf Parameters of the simulations}

For our simulations we have used $\sigma =1$ and $dt=0.02$. The discretization in time is handled using an algorithm similar to the one presented in \cite{Chate}. The used values of $\delta w$ and number of simulated kicks $n_{kicks}$ are: $\delta w = 0.1$  and $n_{kicks} = 40 \times 10^6$ for the SR model; $\delta w = 1$  and $n_{kicks} = 100 \times 10^6$ for the BFM model. As discussed in the main text, these simulations are performed in $d=1$ for a line of size $L=2048$ discretized with $N=L$ points.
%
%$\delta w=0.01$ and $n_{kicks} = 2 \times 10^6$ for the SR model in $d=2$ with $128^2$ points; $\delta w=0.01$ and $n_{kicks} = 8 \times 10^6$ for the BFM model in $d=2$ with $128^2$ points; $\delta w = 1$ and $n_{kicks} = 20 \times 10^6$ for the SR model in $d=1$ with $128$ points; $\delta w = 1$ and $n_{kicks} = 20 \times 10^6$ for the BFM model in $d=1$ with $128$ points.
For the SR model, $\delta u $ is chosen as $\delta u = 5 \delta w$.

{\bf PDF of avalanche sizes and measurement of $S_m$}

The measurement of the PDF $P(S)$ (plotted in Fig.~\ref{fig:SM:pofS}) shows that the avalanche size distribution of both models have a lower cutoff $S_{\delta w} \simeq \frac{(L^d \delta w)^2}{S_m^{{\rm BFM}}}$ where $S_m^{{\rm BFM}}$ is always given by $\sigma/m^4$. In the BFM model, we observe a scaling regime $P(S) \sim S^{-\tau_S^{{\rm BFM}}}$ with $\tau_S^{{\rm BFM}} = 3/2 = 2-\frac{d}{d + \zeta^{{\rm BFM}}}$ ($\zeta^{{\rm BFM}} = 4-d$) for $S_{\delta w} \ll S \ll S_m^{{\rm BFM}}$. In the SR model, for $S_{\delta w} \leq S \leq S_{\delta u} = S_{\delta u} \simeq (\delta u)^{\frac{d + \zeta^{BFM}}{\zeta^{BFM}}}$, the interface does not feel the short-ranged nature of the disorder and we observe a first scaling regime coherent with the BFM, $P(S) \sim S^{-\tau_S^{{\rm BFM}}}$. In the SR model, $S_m^{{\rm SR}}$ is measured as $\langle S^2 \rangle/(2 \langle S\rangle)$ with the result $S_m^{{\rm SR}} = (1.40 \pm 0.05) \times 10^5$ (statistical uncertainty given with $3$ sigma estimation).  For $S_{\delta u } \ll S \ll S_m^{{\rm SR}}$, we observe a second scaling regime coherent with the known features of the SR fixed point: $P(S) \sim S^{-\tau_S^{{\rm SR}}} $ with $\tau_S^{{\rm SR}} = 2-\frac{d}{d + \zeta^{{\rm SR}}}$ and our data are consistent with the value of $\zeta$ numerically estimated in \cite{FerreroBustingorryKolton2012}, $\zeta^{{\rm SR}} \simeq 1.250 \pm 0.005$ (see Fig.~\ref{fig:SM:pofS}). These measurements allows us  to identify the desired scaling regime and compare our simulations with known features of the BFM and SR fixed point.

\begin{figure}
\centerline{
   \fig{.3\textwidth}{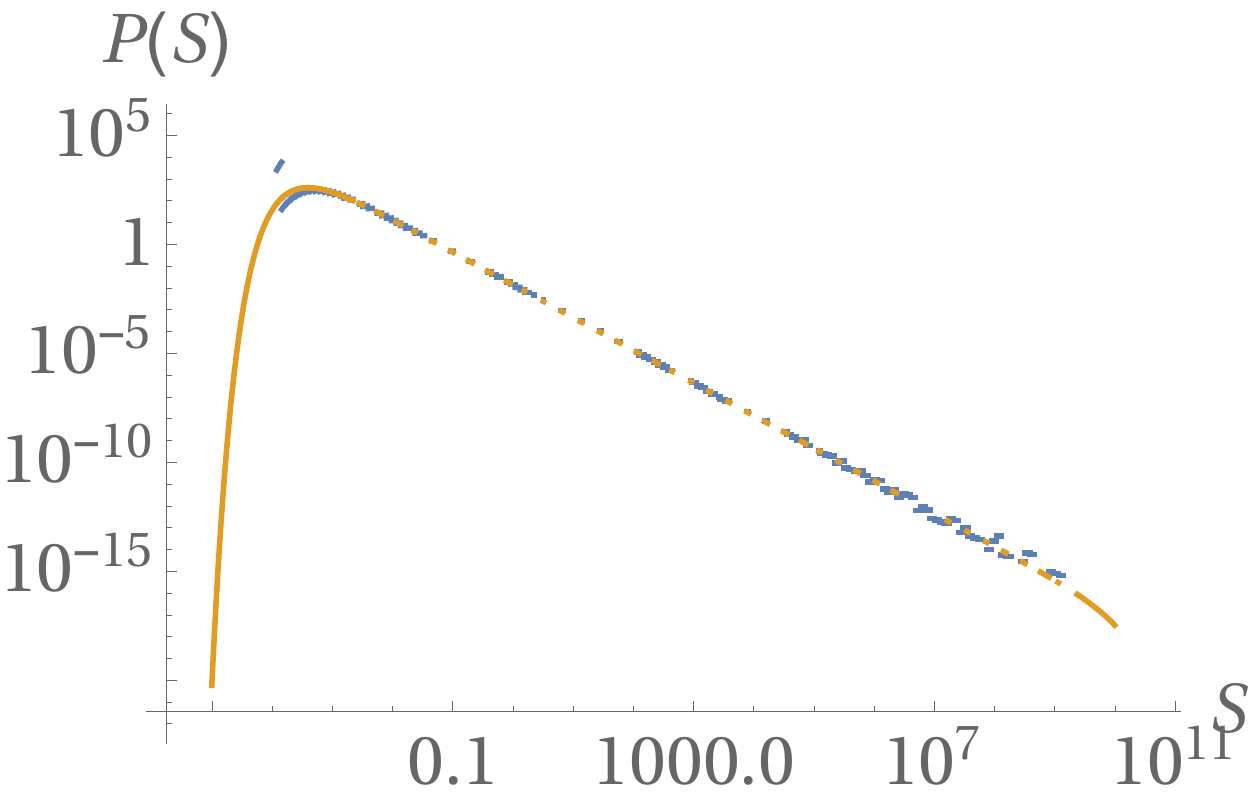} \fig{.3\textwidth}{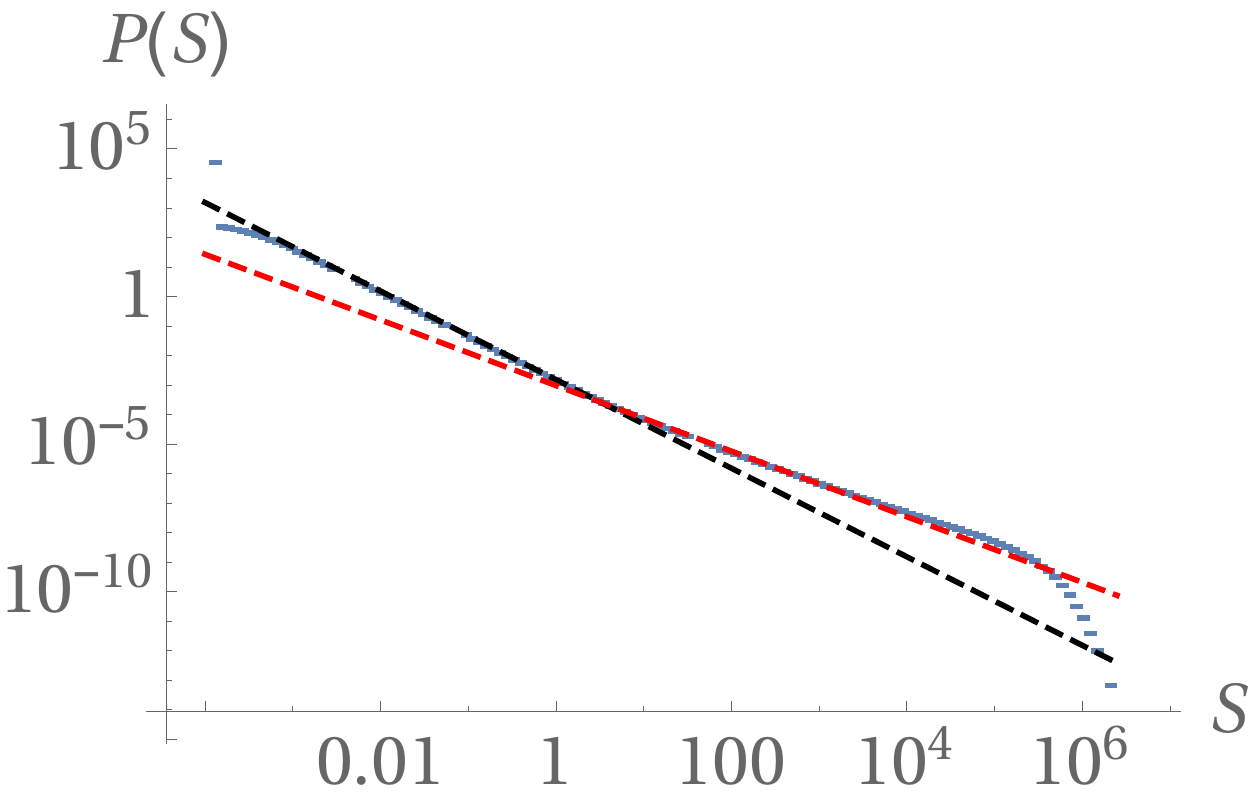}}
 \caption{Blue: Measurement of the avalanche size distribution in the BFM model (left) and the SR model (right). Yellow curve on the left: theoretical prediction for $P(S) = p^{{\rm MF}}(S)$ (no scaling parameter). The excess of small avalanches is an artifact due to the discretization and does not affect the statistics of larger avalanches. Black dashed line on the right: power-law $S^{-\tau_S^{{\rm BFM}}}$ with $\tau_S^{{\rm BFM}}=3/2$. Red dashed line on the right: power-law $S^{-\tau_S^{{\rm SR}}}$ with $\tau_S^{{\rm SR}}\simeq 2 - \frac{2}{1+1.250} \simeq 1.11$.}
\label{fig:SM:pofS}
\end{figure}

{\bf Details on the search for the seed}

Let us now make a few comments on some subtle points and emphasize the importance of the algorithm used in the main text to retrieve the seed of each avalanche. When we apply a uniform kick of size $\delta w$ to the system, the interface always moves from a small amount. As seen above and in Fig.~\ref{fig:SM:pofS}, avalanches of size much smaller than $S_{\delta w}$ are very unlikely (note that the discretization procedure introduces another sharp, artificial, small scale cutoff on the avalanches size: since each points moves at least during the first iteration of the algorithm with velocity $m^2 \delta w/\eta$, the avalanche cannot be smaller than $L^d dt m^2 \delta w/\eta$). After the first iteration, it is actually highly probable that several points along the interface are still moving, each of them being the seed of an avalanche. With a high probability, these small avalanches have sizes of order $S_{\delta w}$ and quickly perish, hence we do not analyze their shapes (they are 'microscopic avalanches'). In the following we are only interested in the shape of avalanches of total size $S > 1 \gg S_{\delta w}$ ('macroscopic avalanches'), which only occur with a small probability. When such an avalanche occurs, since there is a large separation of scales with the small avalanches of order $S_{\delta w}$, we expect its shape to be only very weakly perturbed by the fact that other small avalanches could have been triggered after the kick. We neglect the small probability that more than one macroscopic avalanche have been triggered by the kick. A crucial step is to unambiguously identify, from the set of points still moving during the second iteration of the algorithm, which one is the true seed of the observed macroscopic avalanche. This is what is accomplished by the algorithm explained in the text: after $n_t$ iterations of the algorithm, all the small avalanches triggered at the beginning of the avalanche have already stopped (thus in general $n_t$ has to be chosen sufficiently large). Identifying the maximum velocity inside the avalanche at time $n_t$, we are sure to have identified a point which is inside the macroscopic avalanche. The algorithm is then devised to run within the history of the avalanche backward in time and always identify a point moving along the interface {\it which is in the correct cluster of moving points defining the macroscopic avalanche}. This is illustrated in Fig.~\ref{fig:SM:NicePic}

\begin{figure}
\centerline{
 \fig{.4\textwidth}{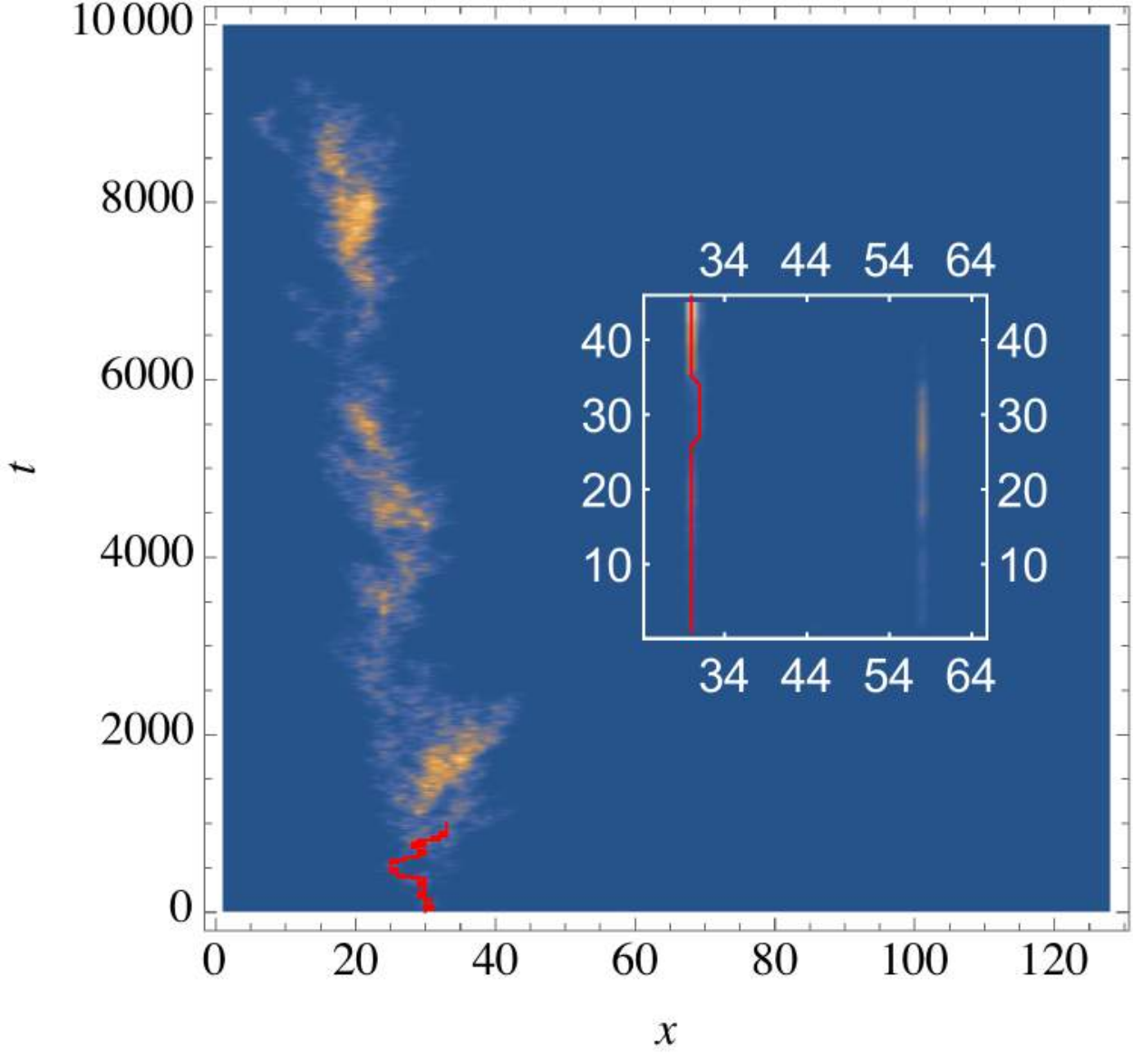}}
 \caption{Density plot of the velocity field $v(x,t)$ inside an avalanche of size $S=1760$ in the mean-field model (BFM) for $d=1$ discretized with $N=128$ points. Line in red: backward path produced by the algorithm to find the seed of the avalanche. The inset illustrates the efficiently of the algorithm to identify, from the set of moving points of the interface just after the kick, the true seed of the observed macroscopic avalanche. In this avalanche (at least) two points (at $x=32$ and $x=57$) still moves at $t=2 dt$, but only the point at $x=32$ is inside the cluster of moving points of the macroscopic avalanche and can be its seed.}
\label{fig:SM:NicePic}
\end{figure}

{\bf Measurement of the mean-shape}

We always only measure mean-shape with values of $S$ well inside the desired scaling regime. The binning on the values of the total size $S$ is of $0.05$, we construct a grid of total sizes with the values $S_i = 1 \times (\frac{1,05}{0.95})^{i-1}$ and avalanches with total size $S$ such that $0.95 S_i <S< 1,05 S_i$ are rescaled as $S \to S_i$. The difference between $S_m^{{\rm SR}}$ and $S_m^{{\rm BFM}}$ and $\tau_S^{{\rm SR}}$ and $\tau_S^{{\rm BFM}}$ explains the difference between the chosen values of $\delta w$ and $n_{kicks}$ for each model: these parameters are adjusted so as to give a comparable numerical precision for the measurement of the mean-shape of interest (i.e. large avalanches which provide a good spatial precision - for the same $\delta w$, one observes more large avalanches in the SR model than in the BFM model). The shapes are rescaled onto one another using the value of $\zeta$ given above and determined numerically in \cite{FerreroBustingorryKolton2012}. The fact that they collapse (see Fig.~\ref{fig:resSimuShapesd1}) using this value is another check that our simulations are correct since they appear in agreement with the high-precision simulations performed in \cite{FerreroBustingorryKolton2012}. Let us also present here the results analogous to Fig.~\ref{fig:resSimuShapesd1} in Fourier space: see Fig.~\ref{fig:SM:resd1Fourier}.
\begin{figure}
\centerline{
   \fig{.3\textwidth}{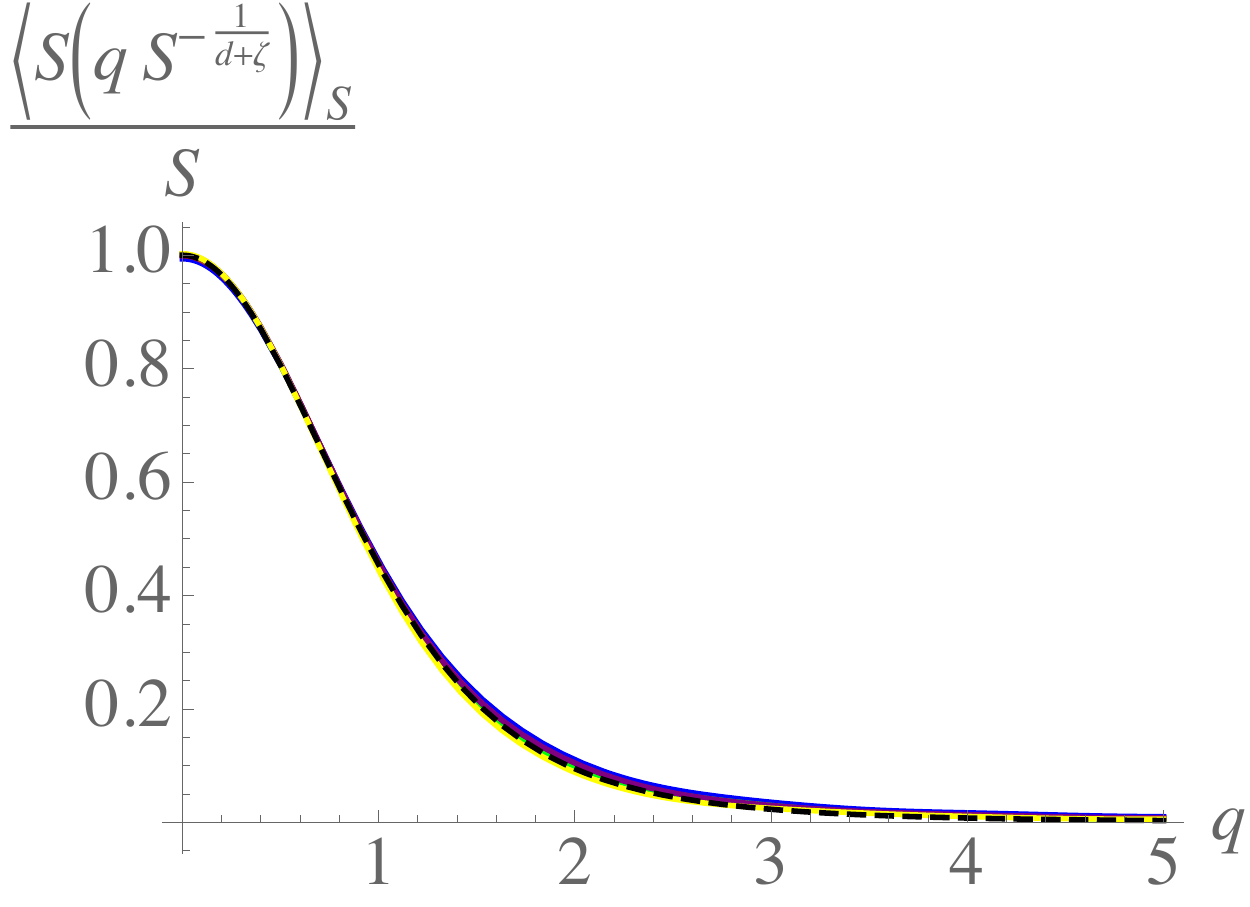} \fig{.3\textwidth}{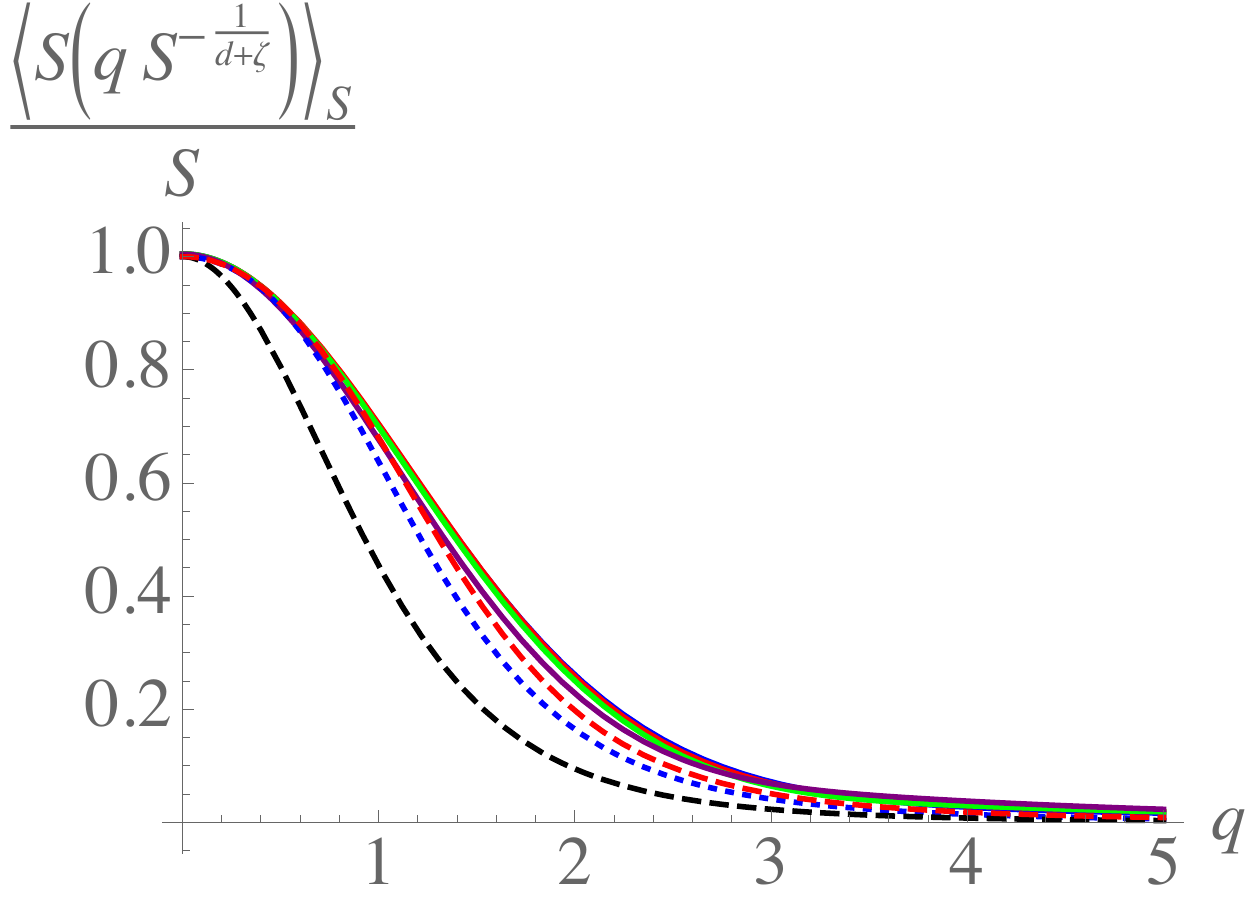} }
 \caption{The mean shape in Fourier space measured in simulations (left: BFM and right: SR), (plain lines, same color code as Fig.~\ref{fig:resSimuShapesd1}) and compared to the theoretical predictions (dashed-black: BFM result, dotted-blue: naive $O(\epsilon)$ result and dashed-red: improved $O(\epsilon)$ result (\ref{SM:regFourier}).}
\label{fig:SM:resd1Fourier}
\end{figure}

\begin{figure}
\centerline{
   \fig{.3\textwidth}{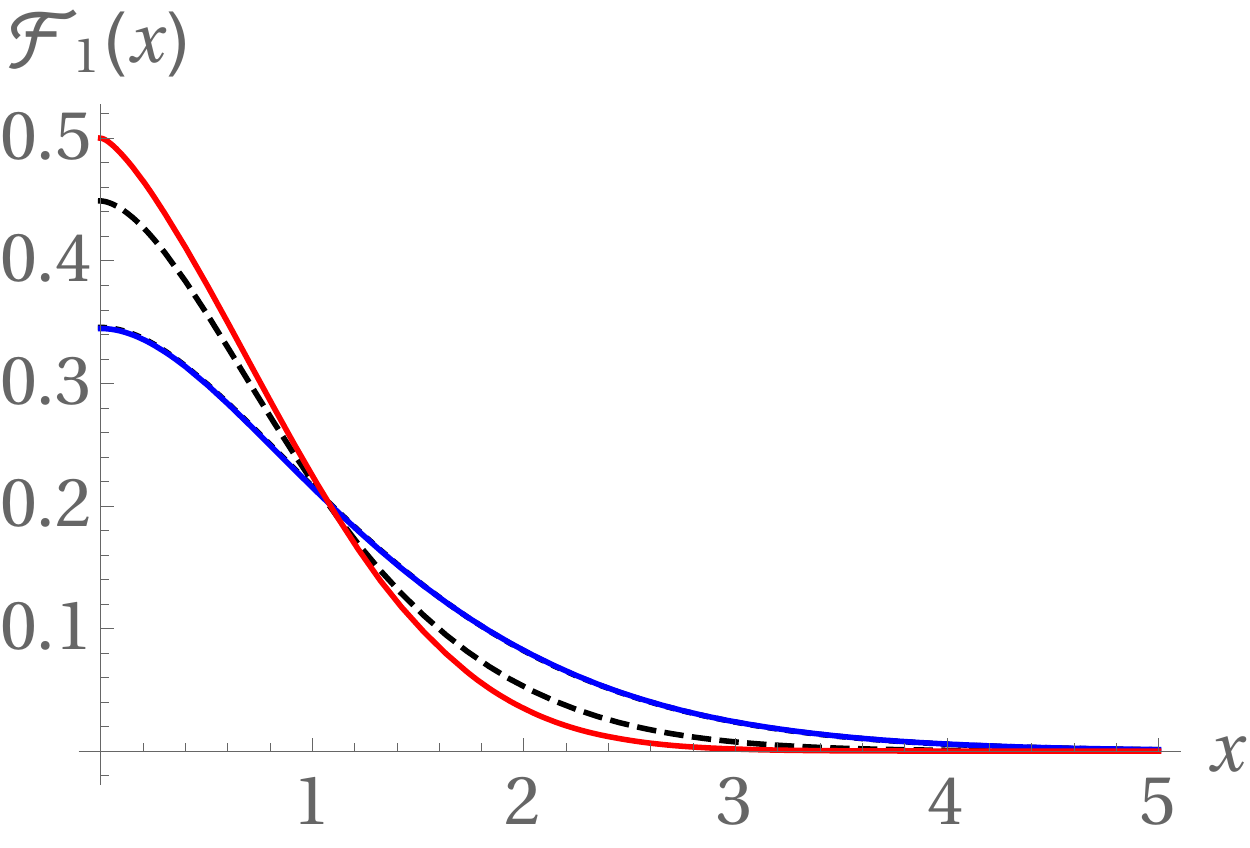} \fig{.3\textwidth}{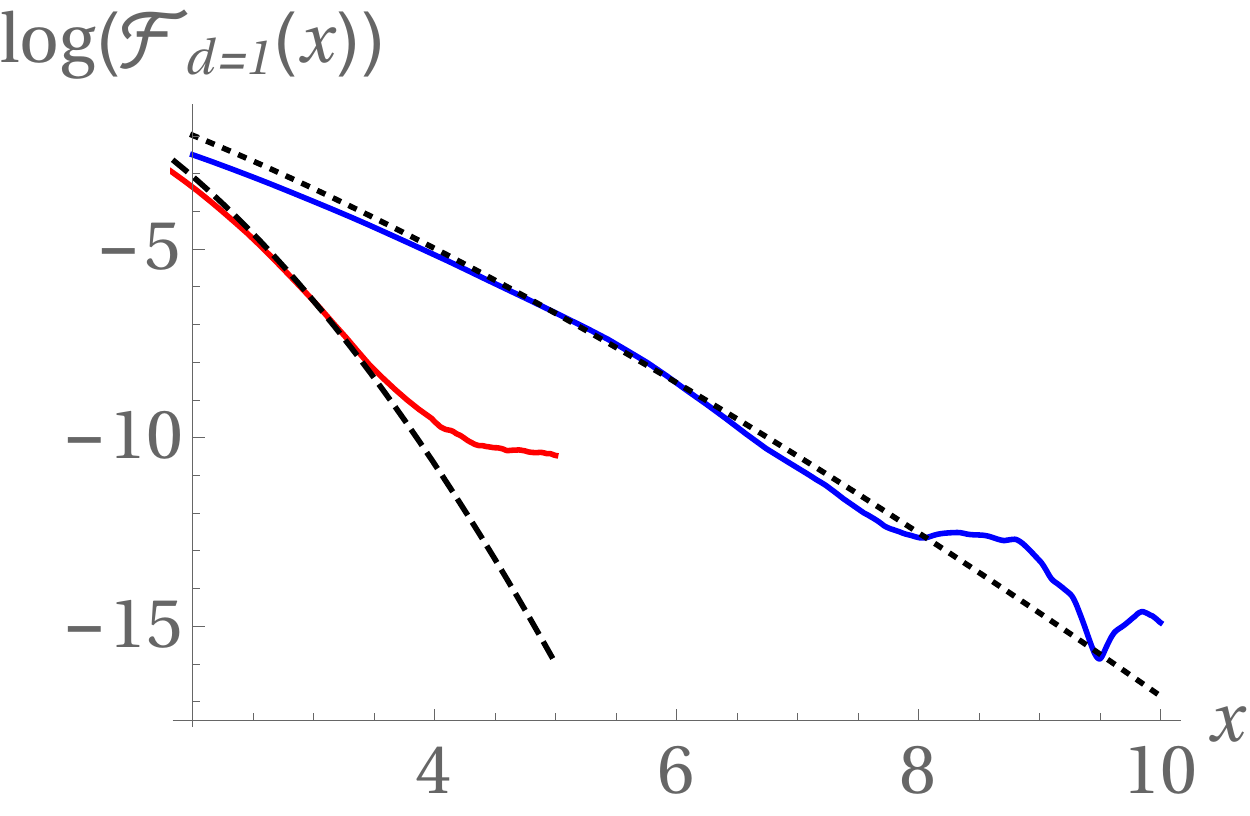}}
 \caption{Left: mean shapes obtained in the simulations of the SR model (red) and of the BFM model (blue) compared with the $O(\epsilon)$ result (dashed, black) and BFM result (dotted black). Right: blue (resp. red) large $x$ behavior of the mean shape measured in the BFM model (resp. SR model). To avoid the noise present at large $x$ to dominate the large $q$ behavior of the mean shape, we smooth our result at large $x$ using an exponential ansatz as explained below.}
\label{fig:SM:meanShapeEtc}
\end{figure}

\begin{figure}
\centerline{
   \fig{.3\textwidth}{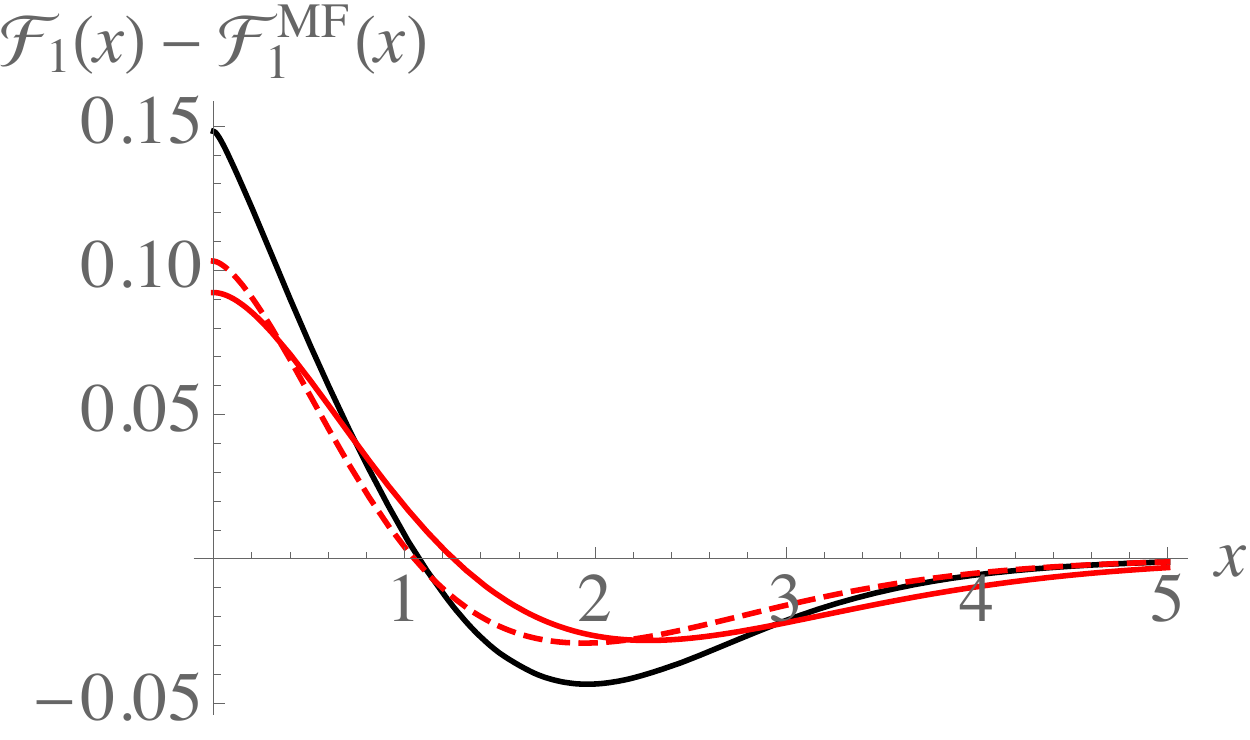} \fig{.3\textwidth}{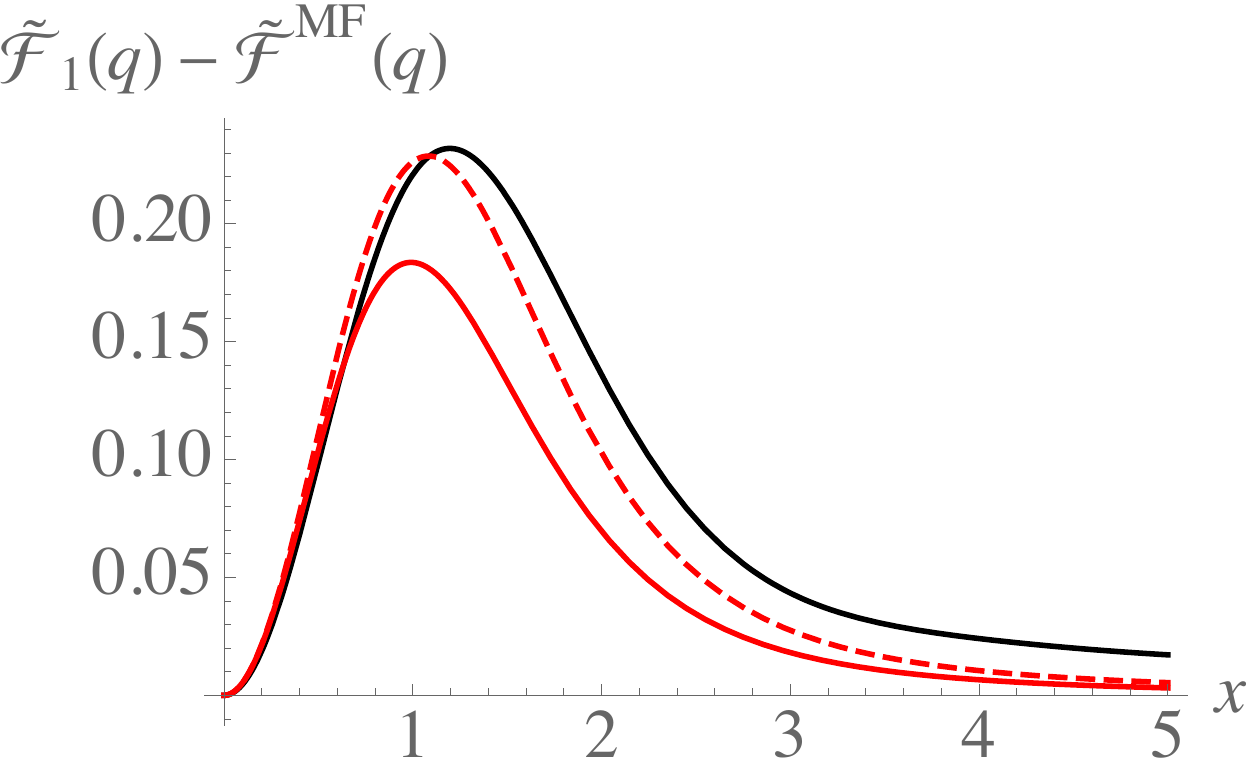}}
 \caption{ Left: (resp. Right:) Black line: Difference between the mean shape measured in the numerical simulations of the SR model in real space ${\cal F}_1(x)$ (resp. in Fourier space $\tilde{{\cal F}}_1(q)$) and the theoretical mean field result ${\cal F}^{\rm MF}_1(x)$ (\ref{hyperFd}) (resp. $\tilde{{\cal F}}^{\rm MF}(q)$ (\ref{MFq})). Red line: theoretical $O(\epsilon)$ result $\delta {\cal F}_1(x)$ (\ref{SM:dFreal}) (resp. $\delta \tilde{{\cal F}}_1(q)$ (\ref{SM:LTFq})). Red-dashed line: improved (through the reexponentiation procedure) theoretical $O(\epsilon)$ result ${\cal F}^{{\rm reg}}_1(x) -{\cal F}^{\rm MF}_1(x) $ (\ref{SM:reg3}) (resp. $\tilde{{\cal F}}^{{\rm reg}}_1(q) -\tilde{{\cal F}}^{\rm MF}(q) $ (\ref{SM:regFourier})). The reexponentiation procedure chosen in Fourier space sensibly improves the accuracy of the result. Nevertheless, higher loop corrections will be necessary to account for the remaining difference.}
\label{fig:SM:DeltameanShape}
\end{figure}

\newpage 
{\bf Measurement of the non-analyticity at small $x$ and fat tail at large $q$}

To measure these observables with a good precision in $d=1$, we use the models discretized using $2048$ points. We first obtain a smooth numerical mean-shape for the BFM and SR model by taking the average of several mean-shapes obtained for various sizes (taken large to obtain a good spatial precision: for the BFM we use $20$ shapes with $13575<S<100478$, for the SR model we use $10$ shapes with $7386<S<20095$). The resulting shapes are shown on the left of Fig.~\ref{fig:SM:meanShapeEtc}. We also plot in Fig.~\ref{fig:SM:DeltameanShape} the difference between the mean shape measured in our numerical
simulations of the SR model and the theoretical mean-field result in $d=1$ and compare it with our theoretical $O(\epsilon)$ predictions. This notably highlights the efficiency of the reexponentiation procedure discussed previously. We then directly study the small $x$ behavior of these shapes, leading to the results presented on the left of Fig.~\ref{fig:resSimuSmallxLargeq}. The study of the large $q$ behavior is more tedious: at large $x$ the mean shapes we obtained start to be dominated by the noise present in our numerical results. This noise blurs the analysis of the large frequency content of the mean-shape. We thus first smooth our results at large $x$ result by using an exponential fit $e^{-ÂÂC x^\delta}$ with the theoretical value of $\delta$ previously obtained exactly for the BFM and using our conjecture (\ref{SM:Dconj}) for the SR model (see Table~\ref{tab:SM:delta}). This fitting procedure is illustrated in Fig.~\ref{fig:SM:meanShapeEtc}. By Fourier transform, we then obtain the results presented on the right of Fig.~\ref{fig:resSimuSmallxLargeq}.

\begin{figure}
\centerline{
   \fig{.3\textwidth}{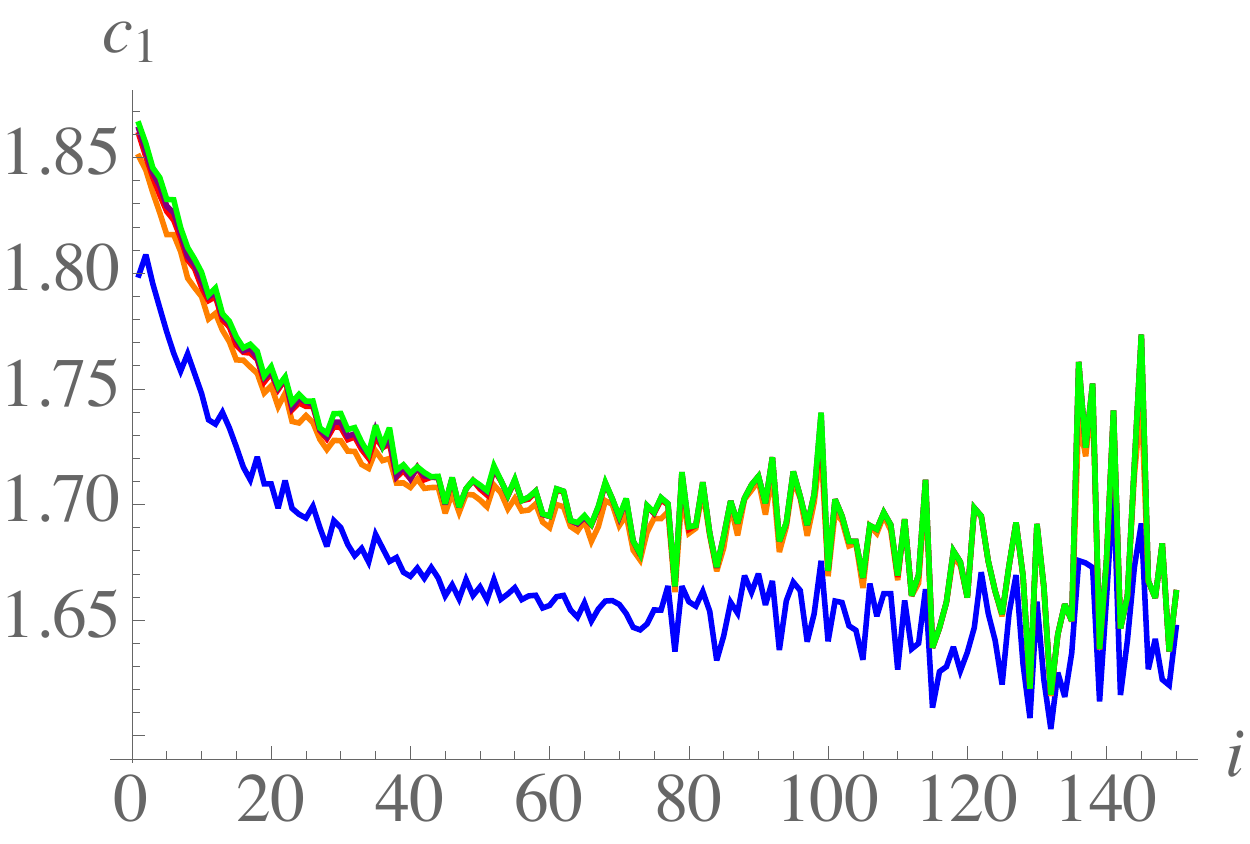} \fig{.3\textwidth}{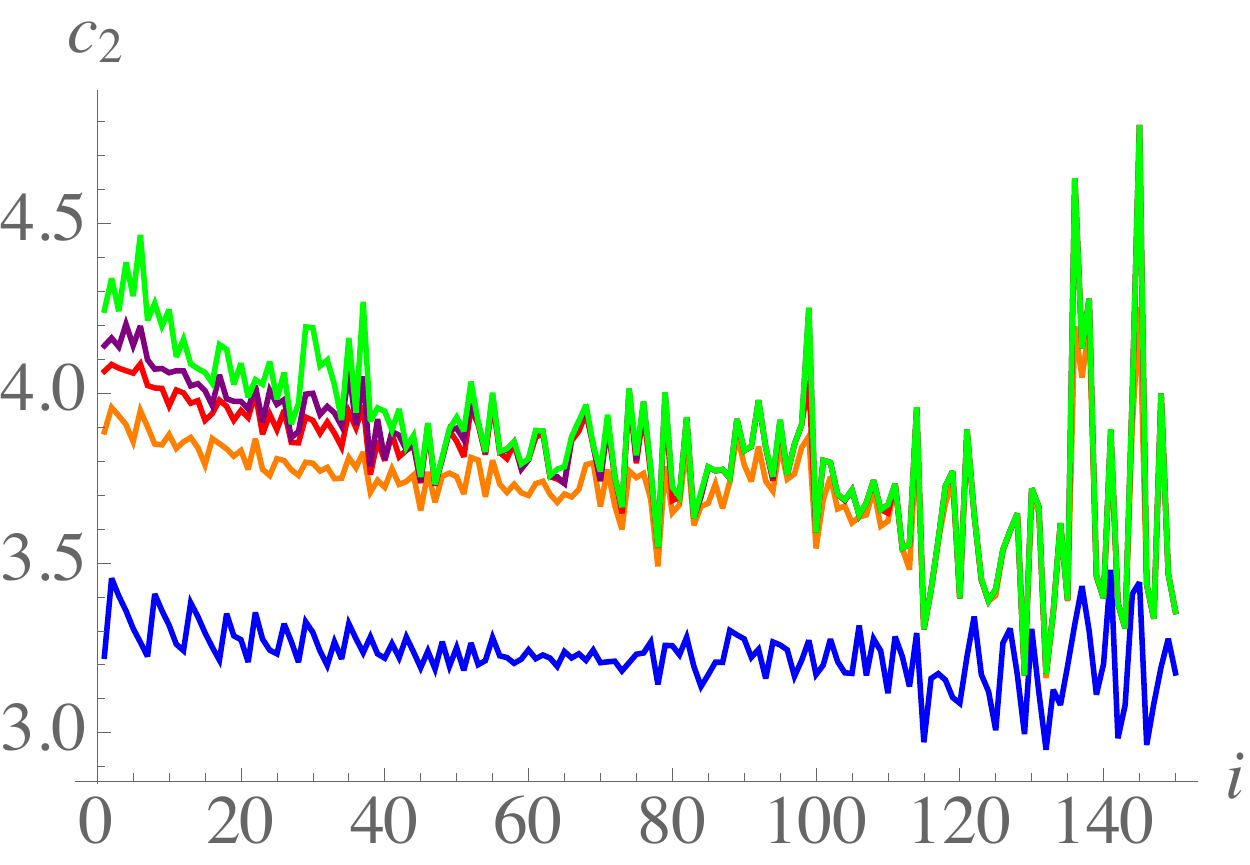}}
 \caption{Universal ratios $c_1(\ell_{cut})$ (left) and $c_2(\ell_{cut})$ (right) measured in the BFM for various cutoff length $\ell_{cut} = 4,6,8,10,12$ (Blue, Orange, Red, Purple and Green) as a function of the total sizes $S=S_i= 1 \times (\frac{1.05}{0.95})^{i-1}$. For the BFM, as a consequence of these plots, the results presented in Table~\ref{SM:tab:UniversalRatios} are averages on the universal ratios obtained for $S>S_i$ with $i=60$ and $\ell_{cut} = 8$ to obtain a result that do not depend on $\ell_{cut}$ and is free of discretization artifacts as explained in the text. A similar procedure is used for the SR model. Note that the important variations observed here for large $i$ are just a consequence of the fact that only a few avalanches with the largests $S_i$ have been measured, hence the statistical uncertainty on the measurements of $c_i(\ell_{cut})$ increases when $S_i$ increases.}
\label{fig:SM:univ}
\end{figure}

{\bf Measurement of the universal ratios}

Here we describe the protocol used to measure the universal ratios. We measure the universal ratios defined in (\ref{SM:Ratios}) using severall cutoff length $\ell_{cut}$ for the integral on $x$ (i.e. we consider different approximations of the universal ratios $c_j(\ell_{cut}) = \frac{\int_{-\ell_{cut}}^{\ell_{cut}} d x  |x|^{2j} {\cal F}_1(x)}{\left( \int_{-\ell_{cut}}^{\ell_{cut}} dx |x|^{j} {\cal F}_1(x)  \right)^2}$ that should converge to the true universal ratios $c_j$ as $\ell_{cut} \to \infty$). These are measured on the mean-shape ${\cal F}_1(x)$ numerically obtained for each possible total size $S_i$ (see above for the definition of the binning procedure). Using these measurements we make sure that $\ell_{cut}$ is chosen large enough so that the results are not sensitive to its finite value. We also control discretization artifacts by studying the dependence of the measured universal ratios $c_j(\ell_{cut})$ on the total size $S_i$: for small $S_i$, the avalanches extend only over a few sites and the mean shape deduced from them is different from the one of the continuum theory, a difference that is seen in the universal ratios. For large enough $S_i$, the universal ratios become 
size independent 
%do not depend anymore on the total size when 
and we reach the continuum regime. This is illustrated for the two first universal ratios in the BFM model in Fig.~\ref{fig:SM:univ}. In the end, the universal ratios are measured by performing an average over various, large enough total sizes $S_i$, leading to the values presented in Table~\ref{SM:tab:UniversalRatios}.

\begin{table}
\vspace{0.1cm}
\begin{tabular}{|l | l | l | l | l | l | l |}
 \hline			
  & $c_1$ & $c_2$ & $c_3$ & $c_4$ & $c_5$ & $c_6$ \\
    \hline
 BFM $d=1$: Theory  & $1.694$ & $3.819$ & $9.270$ & $23.334$ & $59.255$ & $156.863$ \\
    \hline
 SR $d=1$: Theory  & $\simeq 1.641 $ & $\simeq 3.43  $ & $\simeq 7.53 $ & $\simeq 16.6  $ & $\simeq 38.5  $ & $\simeq 81  $ \\
     & $\pm 0.001$ &  $\pm 0.02$ &  $ \pm 0.16$ & $ \pm 0.9 $ & $ \pm 3.7 $ & $ \pm 17$ \\
  \hline

 BFM $d=1$: Numerics & $1.699$ & $3.83$ & $9.3$ & $23$ & $59$ & $143 $ \\
 & $\pm0.003$ & $\pm 0.05$ & $\pm0.3 $ & $\pm 7$ & $\pm 26$ & $\pm 41$ \\
    \hline
  SR $d=1$: Numerics & $1.612 $ & $3.16$ & $6.4 $ & $ 13.6$ & $27$ & $57 $ \\
  & $\pm 0.004$ & $\pm 0.03$ & $\pm 0.3$ & $\pm 0.2$ & $\pm 2$ & $\pm 9$ \\
 \hline  
 \end{tabular}
\caption{Universal ratios in dimension $1$. First two lines: theoretical result for the BFM and $O(\epsilon)$ theoretical result for the SR universality class. Last two lines: numerical measurement in the simulations of the BFM and SR model. Error-bars for the numerics are $3$-sigma estimates. Note that the statistical uncertainty on the numerical measurements of the universal ratios $c_j$ increases with $j$ since these quantities become more and more sensitive to the presence of noise in the large $x$ tail of the measured shapes of avalanches.}
\label{SM:tab:UniversalRatios}
\end{table}

\end{widetext}

\end{document}